\documentclass[twocolumn]{aastex63}
\usepackage{graphicx,times}    
\usepackage{amssymb,amsmath}
\usepackage{amsfonts}
\usepackage{subfigure}
\usepackage{natbib}
\shorttitle{Sample article}
\shortauthors{Zhang et al.}
\usepackage{multirow}

\newcommand{\teff}{\ensuremath{T_{\mathrm {eff}}\,}}
\newcommand{\logg}{\ensuremath{{\mathrm {log}\, } g\,}}
\newcommand{\feh}{\ensuremath{[{\mathrm {Fe/H}}]\,}}

\DeclareUnicodeCharacter{02BC}{'}
\DeclareUnicodeCharacter{2212}{-}
\usepackage{hyperref}
\hypersetup{
    colorlinks=true,      
    linkcolor=blue,       
    urlcolor=blue,        
    citecolor=blue        
    }

\begin{document}
\title{Homogeneous Stellar Atmospheric Parameters and 22 Elemental Abundances for FGK Stars Derived From LAMOST Low-resolution Spectra with {\sc DD-Payne}}
\author{Meng Zhang}
\affiliation{National Astronomical Observatories, Chinese Academy of Sciences, Beijing 100101, Peopleʼs Republic of China\\}
\author{Maosheng Xiang}
\affiliation{National Astronomical Observatories, Chinese Academy of Sciences, Beijing 100101, Peopleʼs Republic of China\\}
\affiliation{Institute for Frontiers in Astronomy and Astrophysics, Beijing Normal University, Beijing 102206, Peopleʼs Republic of China\\}
\author{Yuan-Sen Ting}
\affiliation{Department of Astronomy, The Ohio State University, Columbus, OH 45701, USA.}
\affiliation{Center for Cosmology and AstroParticle Physics (CCAPP), The Ohio State University, Columbus, OH 43210, USA.\\}
\affiliation{Research School of Astronomy \& Astrophysics, Australian National University, Cotter Rd., Weston, ACT 2611, Australia\\}
\affiliation{School of Computing, Australian National University, Acton, ACT 2601, Australia\\}
\author{Anish Mayur Amarsi}
\affiliation{Theoretical Astrophysics, Department of Physics and Astronomy, Uppsala University, Box 516, SE-751 20, Uppsala, Sweden\\}
\author{Hua-Wei Zhang}
\affiliation{Department of Astronomy, School of Physics, Peking University, Beijing 100871, Peopleʼs Republic of China\\}
\affiliation{Kavli institute of Astronomy and Astrophysics, Peking University, Beijing 100871, Peopleʼs Republic of China\\}
\author{Jianrong Shi}
\affiliation{National Astronomical Observatories, Chinese Academy of Sciences, Beijing 100101, Peopleʼs Republic of China\\}
\affiliation{School of Astronomy and Space Science, University of Chinese Academy of Sciences, Beijing 100049, Peopleʼs Republic of China\\}
\author{Haibo Yuan}
\affiliation{Institute for Frontiers in Astronomy and Astrophysics, Beijing Normal University, Beijing 102206, Peopleʼs Republic of China\\}
\affiliation{Department of Astronomy, Beijing Normal University, Beijing 100875, Peopleʼs Republic of China\\}
\author{Haining Li}
\affiliation{National Astronomical Observatories, Chinese Academy of Sciences, Beijing 100101, Peopleʼs Republic of China\\}
\author{Jiahui Wang}
\affiliation{National Astronomical Observatories, Chinese Academy of Sciences, Beijing 100101, Peopleʼs Republic of China\\}
\affiliation{School of Astronomy and Space Science, University of Chinese Academy of Sciences, Beijing 100049, Peopleʼs Republic of China\\}
\author{Yaqian Wu}
\affiliation{National Astronomical Observatories, Chinese Academy of Sciences, Beijing 100101, Peopleʼs Republic of China\\}
\author{Tianmin Wu}
\affiliation{National Astronomical Observatories, Chinese Academy of Sciences, Beijing 100101, Peopleʼs Republic of China\\}
\affiliation{School of Astronomy and Space Science, University of Chinese Academy of Sciences, Beijing 100049, Peopleʼs Republic of China\\}
\author{Lanya Mou}
\affiliation{National Astronomical Observatories, Chinese Academy of Sciences, Beijing 100101, Peopleʼs Republic of China\\}
\affiliation{School of Astronomy and Space Science, University of Chinese Academy of Sciences, Beijing 100049, Peopleʼs Republic of China\\}
\author{Hong-Liang Yan}
\affiliation{National Astronomical Observatories, Chinese Academy of Sciences, Beijing 100101, Peopleʼs Republic of China\\}
\affiliation{Institute for Frontiers in Astronomy and Astrophysics, Beijing Normal University, Beijing 102206, Peopleʼs Republic of China\\}
\affiliation{School of Astronomy and Space Science, University of Chinese Academy of Sciences, Beijing 100049, Peopleʼs Republic of China\\}
\author{Jifeng Liu}
\affiliation{National Astronomical Observatories, Chinese Academy of Sciences, Beijing 100101, Peopleʼs Republic of China\\}
\affiliation{School of Astronomy and Space Science, University of Chinese Academy of Sciences, Beijing 100049, Peopleʼs Republic of China\\}
\affiliation{Institute for Frontiers in Astronomy and Astrophysics, Beijing Normal University, Beijing 102206, Peopleʼs Republic of China\\}
\affiliation{ WHU-NAOC Joint Center for Astronomy, Wuhan University, Wuhan, Hubei 430072, Peopleʼs Republic of China\\}

\correspondingauthor{Maosheng Xiang, Meng Zhang}
\email{msxiang@nao.cas.cn, zhangm@nao.cas.cn}


\begin{abstract}
A deep understanding of our Galaxy desires detailed decomposition of its stellar populations via their chemical fingerprints. This requires precise stellar abundances of many elements for a large number of stars. Here we present an updated catalog of stellar labels derived from LAMOST low-resolution spectra in a physics-sensible and rigorous manner with {\sc DD-Payne}, taking labels from high-resolution spectroscopy as training set. The catalog contains atmospheric parameters for 6.4 million stars released in LAMOST DR9, and abundances for 22 elements, namely, C, N, O, Na, Mg, Al, Si, Ca, Ti, Cr, Mn, Fe, Ni, Sr, Y, Zr, Ba, La, Ce, Nd, Sm, and Eu, for nearly 3.6 million stars with spectral signal-to-noise ratio (S/N) higher than 20. The $\feh$ is valid down to $\simeq-4.0$, while elemental abundance ratios [X/Fe] are mostly valid for stars with $\feh\gtrsim-2.0$. Measurement errors in these labels are sensitive to and almost inversely proportional with S/N. For stars with ${\mathrm S/N}>50$, we achieved a typical error of 30~K in $\teff$, 0.07\,dex in $\log~g$, $\sim0.05$~dex in abundances for most elements with atomic number smaller than Sr, and 0.1--0.2~dex for heavier elements. Homogenization to the label estimates is carried out via dedicated internal and external calibration. In particular, the non-local thermal equilibrium effect is corrected for the [Fe/H] estimates, the $\teff$ is calibrated to the infrared flux method scale, and the $\log~g$ is validated with asteroseismic measurements. The elemental abundances are internally calibrated using wide binaries, eliminating systematic trend with effective temperature. The catalog is publicly available.
\end{abstract}

\keywords{Surveys; Spectroscopy; Stellar parameters}

\section{introduction}

Galactic archaeology has experienced wonderful progress in the past decade, driven by the Gaia mission and ground-based large surveys. Extensive new discoveries are reached by investigating the kinematics, age, and the chemical abundances of only a few elements for a massive set of stars. However, one of the bottlenecks in achieving further understanding of our Galaxy is the limited ability to finely decompose the stellar populations, which requires the element abundances in high dimensions for a vast amount of stars.

For this purpose, a number of large, high-resolution spectroscopic surveys have been conducted. The 17th data release (DR17) of Apache Point Observatory Galactic Evolution Experiment \citep[APOGEE;][]{Majewski2017} provided precise chemical abundances for 20 species
derived from $R\simeq22,500$ infrared spectra for 0.7 million stars in the full sky \citep{Garcia2016, Abdurrouf2022}. The Gaia-ESO survey \citep{Gilmore2022,Hourihane2023} provided homogeneous abundances of more than 30 elements derived from
high-resolution ($R\simeq 50,000$ or $R\simeq 20,000$) spectra for 111,400 stars. 
The recent fourth data release (DR4) of Galactic Archaeology with HERMES (GALAH) survey provided chemical abundances of 32 elements, including s- and r-process elements,
derived from $R\simeq24,000$ optical spectra for a million stars in the southern sky \citep{Buder2024}. Upcoming high-resolution spectroscopic surveys, such as the SDSS-V \citep{Kollmeier2017}, 4MOST \citep{deJong2019}, WEAVE \citep{Dalton2014} and others will further increasing the sample significantly. Attempts have been also made to deliver abundances from medium-resolution spectroscopic surveys, such as the Gaia-RVS \citep{Recio-Blanco2016, Recio-Blanco2023}, the LAMOST-MRS \citep{LiuC2020, WangR2020}.  

On the other hand, spectra from low-resolution surveys of $R\simeq2000$ also have been demonstrated to be able to deliver abundances for many ($\gtrsim15$) elements \citep{TingYS2017a, TingYS2017b, XiangMS2019, WangZX2022, ZhangM2024}. This is mainly attribute to the fact that the low-resolution survey spectra usually have broad wavelength coverage thus contain a wealth of spectral features for these elements. As tens of millions of stars have been sampled by low-resolution spectroscopic surveys, such as the LAMOST \citep{YanHL2022}, the DESI \citep{DESI2016}, and spectra of a even larger amount of stars will be collected, e.g. 4MOST/LRS, SDSS-V/BOSS \citep{Chiappini2019, Kollmeier2017}. This offers a great synergy with $Gaia$ and other existed data sets.  

Due to the line blending effect, a rigorous determination of stellar labels from low-resolution spectra needs a few steps. First, it requires an accurate, physics-sensible modelling of stellar spectra in high-dimension space that involves many labels. Second, when fitting the model to a given spectrum, one needs to adjust all the labels simultaneously to achieve an optimized convergence. This only became reality in the past few years thanks to the deep involvement of machine learning \citep[e.g.][]{TingYS2017b, TingYS2019, XiangMS2019, O'Briain2021}. In particular, the {\sc data-driven Payne} \citep[{\sc DD-Payne};][]{TingYS2017b, XiangMS2019} has been proposed for this specific purpose. {\sc DD-Payne} is hybrid method that takes the advantage of data-driven method \citep[e.g.][]{Ness2015} and model-driven method \citep{TingYS2019, XiangMS2022} for spectra modelling and parameter fitting. It builds up self-consistent spectra models in high-dimension parameter space by  training neural network models on the survey data set, while at the same time, it regularizes the training process with stellar atmospheric models to ensure the neural network models to be physically sensible. 

The {\sc DD-Payne} has been succeed in deriving $\sim15$ labels for a number of low-resolution survey spectra data sets. \citet{TingYS2017b} first applied it to derive the labels for giant stars in LAMOST DR3 as a demonstration of the key aspect of the method. \citet{XiangMS2019} implemented a detailed demonstration of the method and applied it to LAMOST DR5, yielding stellar parameters and abundances of 16 individual elements. \citet{WangZX2022} applied the {\sc DD-Payne} to the Multi-Unit Spectroscopic Explorer \citep[MUSE;][]{Bacon2010} spectra for stellar labels determinations. \citet{ZhangM2024} used it to derive stellar labels from the DESI EDR spectra. 


In this work, we further develop the {\sc DD-Payne} to derive abundances from LAMOST DR9 spectra. The current work has a few major improvements upon previous implementations. First, we succeed in determining the abundances for 22 elements, compared to 16 elements in \citet{XiangMS2019} for LAMOST DR5. In particular, for the first time we have achieved rigorous abundance determination for a large number of s- and r- process elements, including Sr, Y, Zr, Ba, La, Ce, Nd, Sm, and Eu, while there is only one such element, Ba, in \citet{XiangMS2019}. Second, we have updated the training sets for the modelling by using APOGEE DR17 and the GALAH DR3, upon earlier versions in previous work. We also improve the determination at the metal-poor regime by including a large set of metal-poor stars with uniform labels from literature \citep{LiHN2022} in our training set. This makes our method be valid for stellar parameter determination down to $\feh\simeq-4.0$. 

Because of uncertainties in stellar spectra modelling and artifacts in observations, e.g., line blending, it is a challenge to have homogeneous abundances determination for different types of stars. Also, systematic differences are common among different survey data sets \citep[e.g.][]{XiangMS2019, Soubiran2022}. These inhomogeneity may induce difficulties in the understanding and interpreting to results of Galactic archaeology. To minimize such effect, in the current work, we have carried out dedicated internal and external validation and calibration for the label determinations. We are ended up with a sample of $\simeq6$ million F/G/K stars with robust and homogeneous atmospheric parameters and abundances. In particular, non-local thermal equilibrium (NLTE) effect has been corrected for the [Fe/H] estimates. The effective temperature is calibrated to the widely used infrared flux method (IRFM) scale, while the $\log~g$ are validated with asteroseismic measurements. The elemental abundances are internally calibrated using wide binaries, eliminating systematic trend with effective temperature. 




The paper is organised as follows. Section~2 describes the LAMOST spectra data set. 
Section~3 introduces the {\sc DD-Payne} method to determine the stellar labels. 
The external calibration and validation of the label determinations are presented in Section~4.
Content and properties of the LAMOST abundance catalog are presented in Section~5. Section~6 present a comparison with existed LAMOST stellar parameter catalogs. Section~7 is a discussion on the zero point of the labels, followed by a summary in Section~8.
\section{Data}
\subsection{The LAMOST survey spectra}
The ninth data-released (DR9)\footnote{\url{http://www.lamost.org/dr9/v1.0/}}of LAMOST has been public since September 2023. A part of the release is 11,226,252 low-resolution ($R\simeq1800$) optical ($\lambda3700-9000${\AA}) spectra targeted by the LAMOST low-resolution spectroscopic surveys \citep{ZhaoG2012, DengLC2012, LiuXW2014, YanHL2022}  during the observation seasons between Oct., 2011 and July, 2019. 
The release also provides object classifications delivered by the LAMOST 1D pipeline \citep{LuoAL2015}, as well as stellar atmospheric parameter estimates including radial velocity, effective temperature, surface gravity, and metallicity, for F/G/K-type stars derived with the LAMOST stellar parameter pipeline \citep[LASP;][]{WuY2011, WuY2014, LuoAL2015}. The classification yields 10,907,516 stellar spectra for 7,906,319 unique stars. 

\begin{figure*}[htb!]
\centering 
\subfigure
{
	\begin{minipage}{0.48\linewidth}
	\centering     
	\includegraphics[width=0.97\columnwidth]{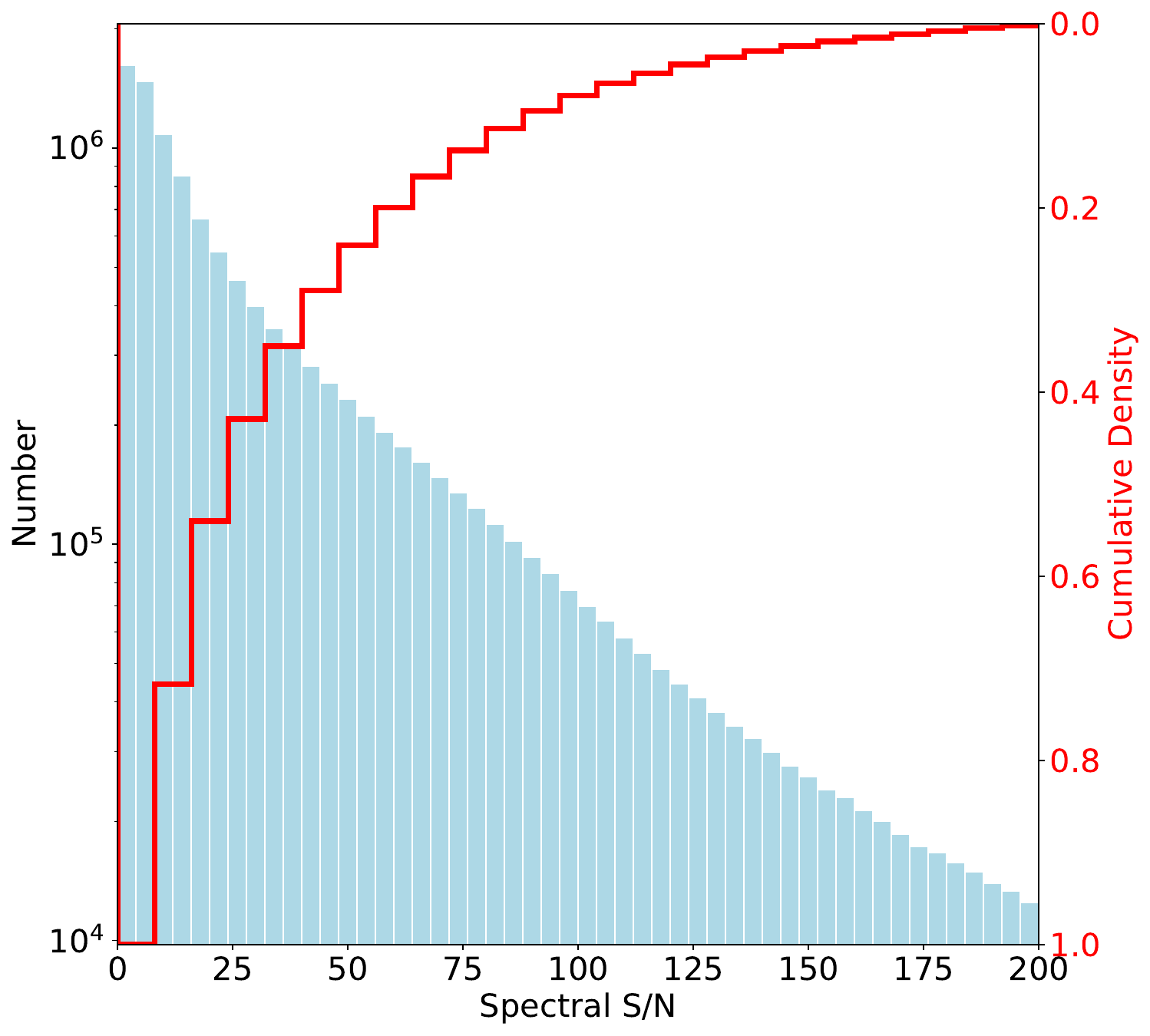}  
	\end{minipage}
}
\subfigure
{
	\begin{minipage}{0.48\linewidth}
	\centering     
	\includegraphics[width=0.95\columnwidth]{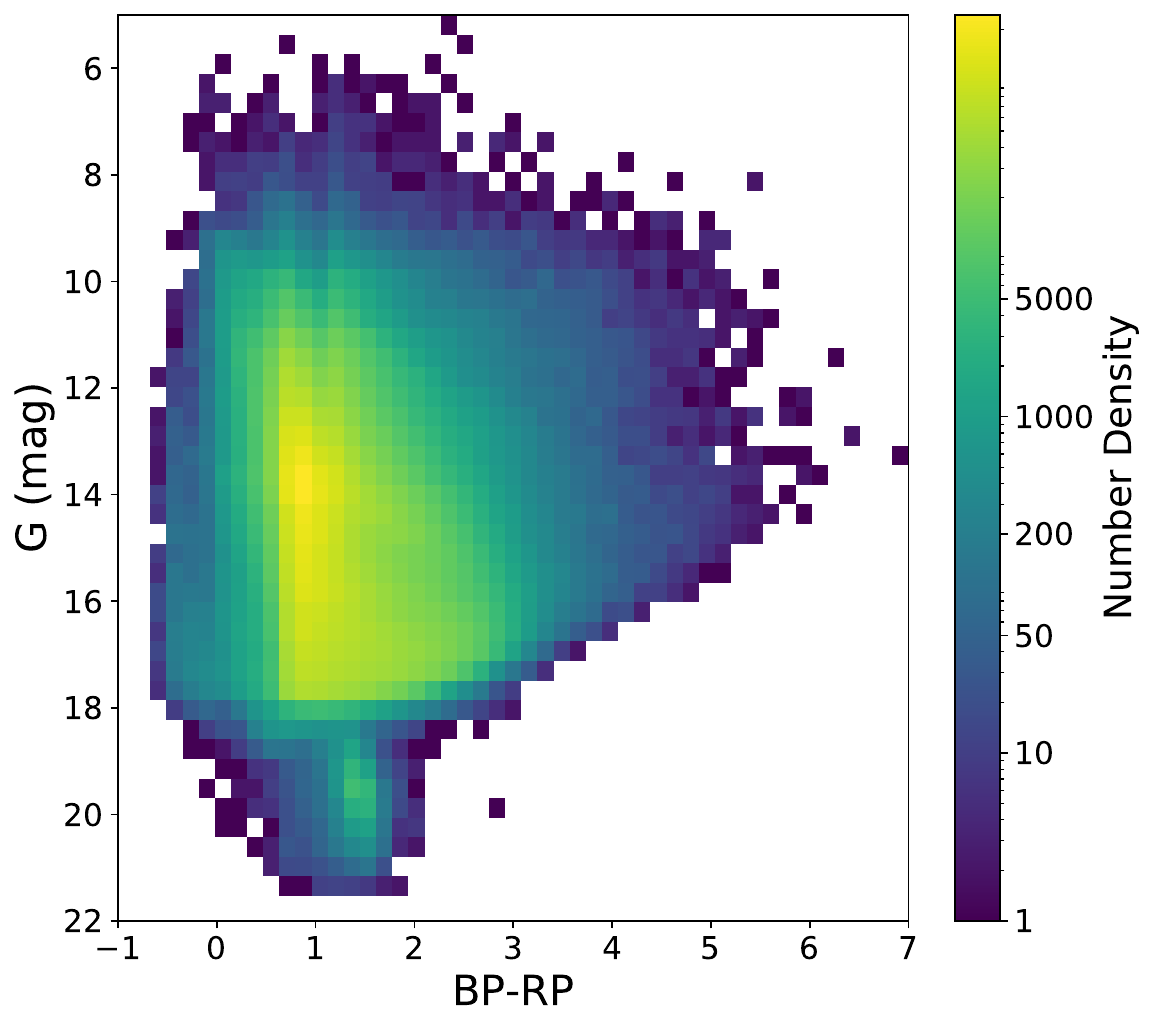}
	\end{minipage}
}
\caption{{$\it Left$}: Distribution of the LAMOST DR9 sample stars as a function of $g$-band $S/N$ , and {$\it right$}: in the $(BP-RP, G)$ color-magnitude diagram.
\label{figsnr}}
\end{figure*}

The spectral signal-to-noise ratio (S/N) distribution in the $(BP-RP, G)$ color-magnitude diagram of the full sample are shown in Figure~\ref{figsnr}. About 67.1 per cent and 24.7 per cent of the spectra have a g-band $S/N$ higher than 10 and 50, respectively. 
The vast majority of stars have a $Gaia$ G-band magnitude between 6 and 17.8\,mag, as a result of the surveys' target selection strategy. For most of the cases, the survey's target selection was implemented in  the ($g-r$, $r$) and ($r-i$, $r$) diagram \citep{Carlin2012, LiuXW2014, YuanHB2015}, where the value of SDSS r magnitude is almost identical to that of $Gaia$ G magnitude for FGK -type stars. There are also some stars fainter than 17.8~mag, mostly targeted during the pilot survey \citep[e.g.][]{YuanHB2015, XiangMS2017}. 

\begin{figure}[htb!]
\centering
\vspace{1.em}
\includegraphics[width=0.5\textwidth]{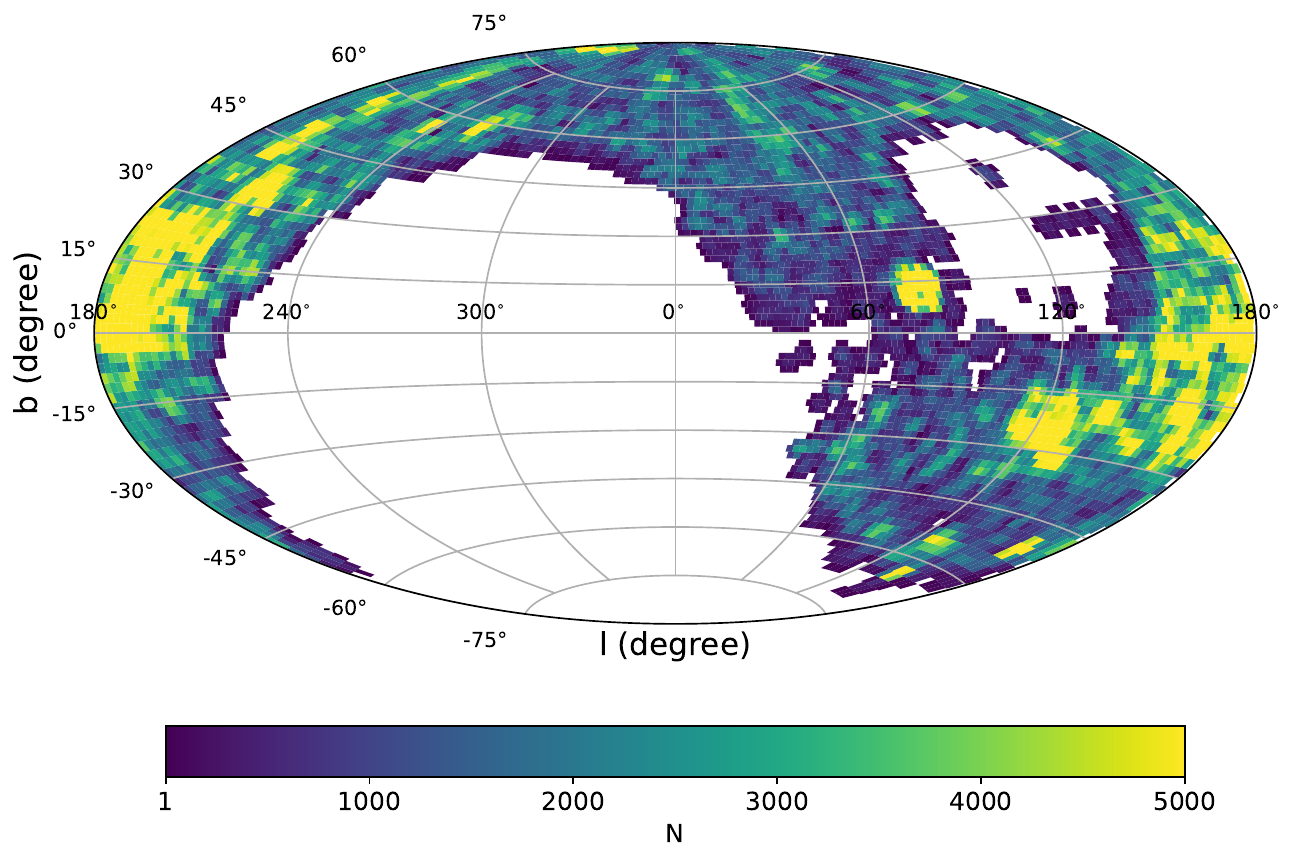}
\vspace{0.5em}
\caption{The LAMOST DR9 stellar density distribution in the Galactic coordinate. Each pixel of the plot represents a constant sky area of 4.0 square degrees ($2.0^\circ\times2.0^\circ$). 
\label{figlb}}
\end{figure}

Fig.~\ref{figlb} shows the sky coverage of the sample stars in Galactic coordinates. The sample stars have a nearly contiguous distribution in the Galactic latitude range $-10^\circ<b<60^\circ$. Particularly, the survey has a continuous and dense ($\gtrsim1000$ stars per square degree) sampling for stars in the anti-center disk ($150^\circ\lesssim l\lesssim210^\circ$, $|b|<30^\circ$) and the M31/M33 vicinity fields $(l,b)\simeq(120^\circ, -20^\circ)$, as a result of dedicated targeting by the LAMOST Spectroscopic Survey of the Galactic Anti-center \citep[LSS-GAC;][]{LiuXW2014, YuanHB2015, XiangMS2017}. At higher Galactic latitudes, the number density drops because the intrinsic stellar density of the Galaxy decreases with increasing latitude. The figure also shows a dense spot at the $Kepler$ field $(l,b)\simeq(70^\circ, 15^\circ)$, as a result of the dedicated LAMOST-$Kepler$ survey \citep{DeCat2015, FuJN2020, ZongWK2020}. 

\subsection{Spectra normalization}
In this work, the stellar abundance determination is implemented with a normalized LAMOST spectra. The normalization is implemented as follows:
\begin{itemize}
    \item First, the LAMOST spectra are converted to rest frame by correcting line-of-sight velocity $V_{\rm {los}}$ provided in the fits file header, which is derived with the LAMOST 1D pipeline. A linear interpolation is then adopted to sample the rest-frame spectra into a unified wavelength grid. 
    \item Second, for each spectrum $f(\lambda)$ we obtained a smoothed spectrum $\bar{f}(\lambda)$ by convolving a Gaussian kernal, 
    \begin{equation}
      \bar{f}(\lambda_i) = \frac{\Sigma_jf_j\omega_j(\lambda_i)}{\Sigma_j\omega_j(\lambda_i)},
    \end{equation}
    \begin{equation}
      \omega_j(\lambda_i) = \frac{1}{\sqrt{2\pi}L}\exp\left(-\frac{(\lambda_j-\lambda_i)^2}{2*L^2}\right),
    \end{equation}
where the Gaussian kernel width $L$ is adopted as a constant of 50{\AA}. The normalized spectrum $f_{\rm n}(\lambda)$ is then obtained via
 \begin{equation}
     f_{\rm n}(\lambda) = \frac{f(\lambda)}{\bar{f}(\lambda)}.
 \end{equation}
   \item Finally, we mask problematic pixels in the spectra $f(\lambda)$ by setting infinity values in their flux errors. For problematic pixels, we refer to those are flagged by the `andmask', `ormask' labels of the LAMOST pipeline, or those have unrealistic (zero) values of flux or invariance. These bad pixels are not used for the convolution by setting a zero value to their weights.  
\end{itemize}

\section{Method} 
In this work, we utilize the data-driven Payne ({\sc DD-Payne}) to derive 25 labels, including $T_{\rm eff}$, $\log\,g$, $v_{\rm mic}$, and abundances of 22 elements, namely, C, N, O, Na, Mg, Al, Si, Ca, Ti, Cr, Mn, Fe, Ni, Ba, Sr, Y, Zr, La, Ce, Nd, Sm, and Eu, from the LAMOST low-resolution spectra. The {\sc DD-Payne} is a hybrid approach that combines data-driven methods and physical priors for spectral modelling and fitting. We adopt LAMOST spectra that have precise label estimates from either APOGEE DR17 or GALAH DR3 as our training set to build spectra models, which are then used to infer stellar labels for the remaining vast sets of LAMOST spectra.
\subsection{The Data-Driven Payne}
The {\sc DD-Payne} is a tool for modeling low-resolution spectra and fitting stellar labels in high-dimensional space. It combines the advantages of both the flexibility of a data-driven machine learning approach and the physical interpretability of a model-driven method.
Specifically, it trains a data-driven neural network model between the survey spectra and their labels using a subset of stars in the survey that have independent label measurements.  
The neural network model can be expressed with a function form,
\begin{equation}
 f_{\lambda} = \mathbf{w}_m\cdot\sigma\left(\cdots \sigma\left(\mathbf{w}_1\cdot\sigma\left(\mathbf{w}_0\cdot\mathbf{l} + \mathbf{b}_0\right) +\mathbf{b}_1\right) \cdots\right)+\mathbf{b}_m
 \label{equ1}
\end{equation}
where $\sigma$ is the activation function, and $\mathbf{l}$ is the stellar labels. $\mathbf{w}$ and $\mathbf{b}$ are coefficient tensors to be optimized, the index $m$ is the number of network layers. In this work, we opt to adopt a 2-layer simple network, similar to previous implementations \citep{TingYS2017a, XiangMS2019}. The number of neurons is set to be 50. We adopt the sigmoid function, $\sigma(x)= 1/(1+e^{-x})$, as the activation function, . 

The flexible neural network algorithm enables precise spectral modeling in high-dimensional ($\gtrsim20$) label space.
Meanwhile, superior to trivial data-driven models, {\sc DD-Payne} regularizes the neural network model in the training process using differential spectra from physical stellar atmospheric models. This is implemented by inventing a loss function as follows,
\begin{equation}
\begin{aligned}
&  \mathcal{L}(\{f_{\rm obs}(\lambda)\} | \mathbf{w},\mathbf{b}) =  \frac{1}{N_s}\sum_{i=1}^{N_s} 
         \frac{(f(\lambda; \boldsymbol{\ell}_{{\rm obs},i}) - f_{{\rm obs},i}(\lambda))^2}{\sigma^2_{{\rm obs},i}(\lambda)}  \\
 &   +\sum_{j=1}^{N_r} \mathbf{D}_{\rm scale} \cdot \sum_{k=1}^{N_l} | f^{\prime}(\lambda; \boldsymbol{\ell}_{\rm ref}) - f^{\prime}_{\rm {physical}}(\lambda; \boldsymbol{\ell}_{\rm ref})|
\end{aligned} 
\end{equation}
where the $f^{\prime}_{\rm {physical}}$ is the differential (gradient) spectra from stellar physical models, $N_{s}$, $N_{r}$, $N_{l}$ represents the number of training spectra, reference stars, and stellar labels, respectively. The vector parameters $\mathbf{D_{scale}}$ set a weight for each label, and is empirically determined. In our case, we adopt a value 10 times larger for labels with weak features like [X/Fe] than those for labels with strong features like \teff, \logg, and \feh. The coefficient tensors in this multi-layer perception neural network, $\mathbf{w}$ and $\mathbf{b}$, will be optimized using the training sample.

Once the spectral model is built, {\sc DD-Payne} fits the full spectrum of a target star to derive all the stellar labels simultaneously. In low-resolution spectra, absorbing lines from different labels are seriously blended, fitting all the labels simultaneously is essential for properly recovering the individual labels. Using full spectra for the fitting makes full use of the spectra information. Also, because the spectral features (gradients) of different labels may overlap in wavelength, the label estimates are inevitably co-variant. Using the full spectra rather than a small wavelength window helps to reduce the covariance.

\subsection{The training set}
\begin{figure*}[htb!]
\centering
\vspace{1.em}
\includegraphics[width=\textwidth]{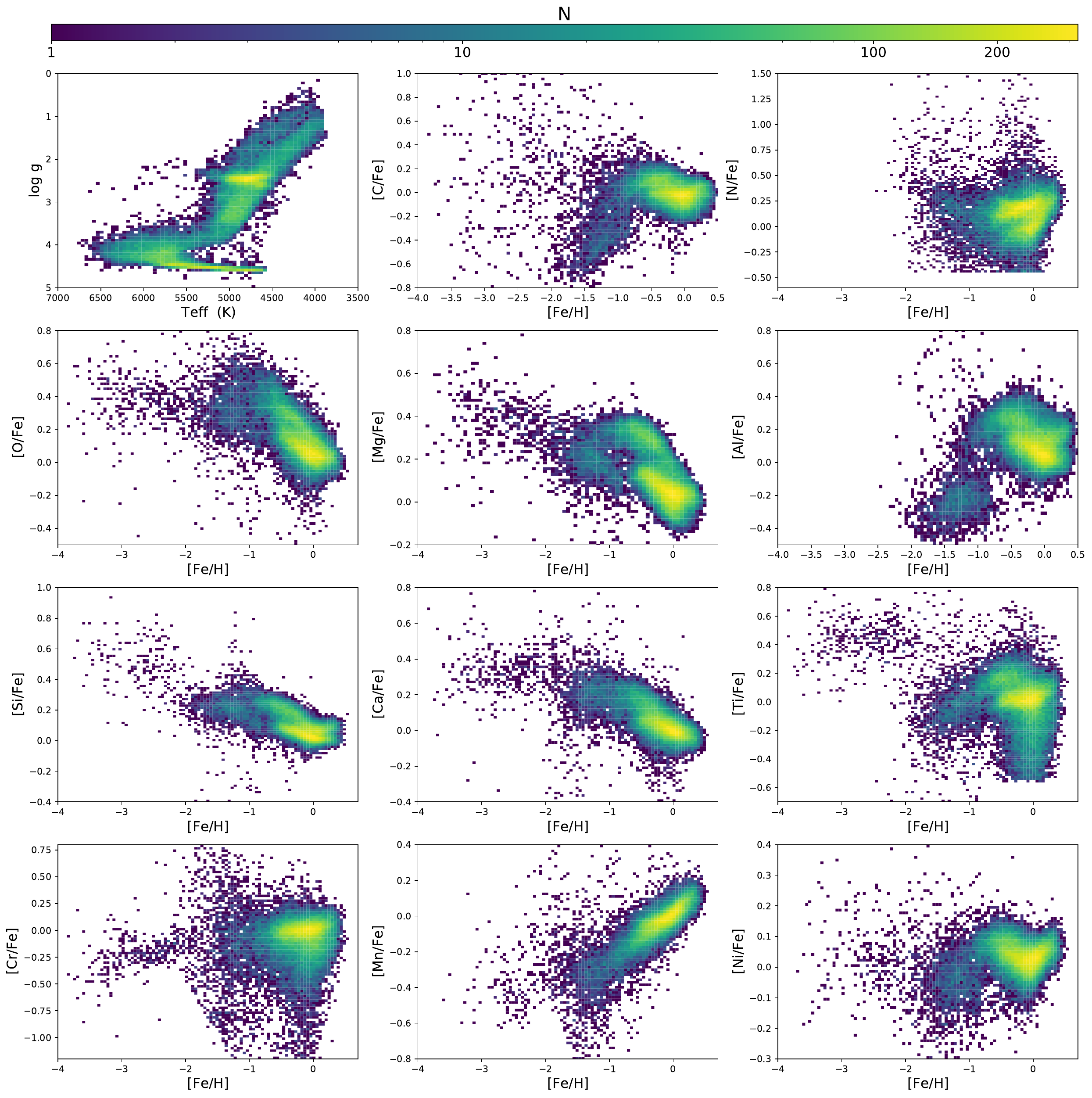}
\vspace{0.5em}
\caption{Stellar density distribution of the LAMOST-APOGEE training sample in the $T_{\rm eff}$--$\log\,g$ diagram and [X/Fe]-[Fe/H] planes. The labels shown are either from the APOGEE DR17 catalog or from the LAMOST VMP catalog of \citet{LiHN2022}. Some VMP stars have no abundance measurements for N, Al, Mn, Si, and we have assigned constant values to them in the training process.
\label{fig1}}
\end{figure*}
We adopted common stars both between LAMOST and APOGEE DR17, and between LAMOST and GALAH DR3 as our training sets. Common stars between LAMOST and APOGEE are used to train spectra models for the determination of 15 labels, namely, $T_{\rm eff}$, $\log\,g$, $v_{\rm mic}$, [C/Fe], [N/Fe], [O/Fe], [Mg/Fe], [Al/Fe], [Si/Fe], [Ca/Fe], [Ti/Fe], [Cr/Fe], [Mn/Fe], [Fe/H], and [Ni/Fe], while common stars between LAMOST and GALAH are used to train models for the determination of 23 stellar labels, i.e., $T_{\rm eff}$, $\log\,g$, $v_{\rm mic}$, [O/Fe], [Na/Fe], [Mg/Fe], [Al/Fe], [Si/Fe], [Ca/Fe], [Ti/Fe], [Cr/Fe], [Mn/Fe], [Fe/H], [Ni/Fe], [Sr/Fe], [Y/Fe], [Zr/Fe], [Ba/Fe], [La/Fe], [Ce/Fe], [Nd/Fe], [Sm/Fe], and [Eu/Fe]. These training sets are used to train spectra models separately, and the resultant LAMOST label estimates from these models will be provided, along with a recommended version by combining these results.   

We cross-match the stars' coordinates (RA, Dec) of LAMOST DR9 and APOGEE DR17 using a radius of 3", and find 193,156 common stars. From this common star set, we further select stars with good quality according to the following criteria,

\begin{align}
\left\{
\begin{aligned}
S/N_{\rm LAMOST} > 50, &\text{ if $\feh > -0.8$}, \\
S/N_{\rm LAMOST} > 30, &\text{ if $\feh < -0.8$}, \\
S/N_{\rm APOGEE} > 80, \\
{\rm APOGEE ~ Star\_flag} = 0, \\
{\rm APOGEE ~ X\_FE\_flag = 0}, &\text{ for X=Fe, C, O, Mg, Al}, \\
&\text{ Si, Ca, Ti, Cr, Mn, and Ni}, \\
\text{RUWE} < 1.4, \\
\dfrac{\omega_{\rm specphot} - \omega_{\rm Gaia}}{\sqrt{\sigma_{\omega, \rm specphot}^2+\sigma_{\omega, \rm Gaia}^2}} < 2, &\text{ if $\logg > 3.5$},
 \end{aligned}
\right.
\end{align}
 where `RUWE' refers to the $Gaia$ renormalised unit weight error \citep{Lindegren2021}, the RUWE cut discard binary stars that cause significant noise in their astrometric solution. $\omega_{\rm Gaia}$ refers to Gaia parallax, and $\omega_{\rm specphot}$ refers to spectrophotometric parallax estimates derived with the method presented in \citep{XiangMS2022}. The cut in the last line is adopted to discard binary stars that exhibit significant light excess with respect to single stars at the same distance \citep{XiangMS2021, XiangMS2022}. These criteria lead to a sample of 56,223 stars, from which we randomly draw 36,223 stars as our training sample, and the remaining 20,000 stars are adopted as a test sample. 

In order to increase the training sample at the very metal-poor (VMP; $\feh\lesssim-2$) side, we also make use of the VMP star abundance catalog of \citet{LiHN2022}. The \citet{LiHN2022} catalog contains abundance determinations for over 20 elements from high-resolution spectra taken by the Subaru Telescope, for a total of 385 VMP stars spreading a metallicity range from $\feh\simeq-1.7$ to $\feh\simeq-4.3$. These stars are initially selected from the LAMOST catalog, so that all of them have a LAMOST spectrum. We select those with a LAMOST spectral S/N higher than 20. This results in a sample of 345 stars, all of which are fed to the LAMOST-APOGEE training sample. 

Fig.~\ref{fig1} shows the distributions of the LAMOST-APOGEE training sample in the label spaces.
The training sample have a good coverage in the \teff-\logg plane, particularly for the red giant branch, red clump, and subgiant stars. The main-sequence dwarf and turn-off stars are limited within a temperature range of 4500--6800~K. These temperature boundaries are due to limitation of the APOGEE catalog, given the selection of data flags listed above. 
As for the elemental abundances, because the VMP stars of \citet{LiHN2022} do not contain measurements of N and Al abundances, we keep them in the training set, but set these abundances constant values of 0.5 and 0.0 respectively, to make the neural network work. Also, some of the VMP stars have no available Si, Mn, and Ni abundances, we set constant values of their abundances to be 0.5, $-0.2$, 0.0, respectively. The abundance estimates for some elements at the very metal-poor end will be discarded from our final results (Sect.\,5). 

To obtain measurements for some elements without artifacts, we also build up another training set (training set 2). The stars in training set 2 all have reliable N, Al, and Mn measurements. 

Similarly, we cross-match the LAMOST catalog with GALAH DR3, and find 40,377 common stars. However, not all of these stars have reliable (FLAG=0) abundance estimates for all of the 23 labels mentioned above. In particular, for some of the s- or r-process elements such as Sr and Eu, the GALAH catalog provide reliable estimates for only a small number of the stars. Therefore, rather than defining a uniform training set that meet a quality requirement in all dimensions, we choose to define a series of tailored training sets. These training sets are described in Table~\ref{table1}. Training set 3 is used to train {\sc DD-Payne} for the estimation of 17 labels, namely $T_{\rm eff}$, $\log g$, $V_{\rm mic}$, C, N, O, Mg, Na, Al, Si, Ca, Ti, Cr, Mn, Fe, Ni, and Ba. Each of the training sets 4-11 is used to determine a particular s- or r-process element, as listed in Table~\ref{table1}. Note that, even with this optimization, we still lack of N abundance measurements for all stars and C abundance measurements for most stars in the GALAH DR3 catalog. As C and N contribute significant lines to the LAMOST spectra, they cannot be omitted for constructing a precise spectra model, we thus adopt the LAMOST [C/Fe] and [N/Fe] derived with {\sc DD-Payne} using the LAMOST-APOGEE training set. 

\begin{figure*}[htb!]
\centering
\vspace{1.em}
\includegraphics[width=0.95\textwidth]{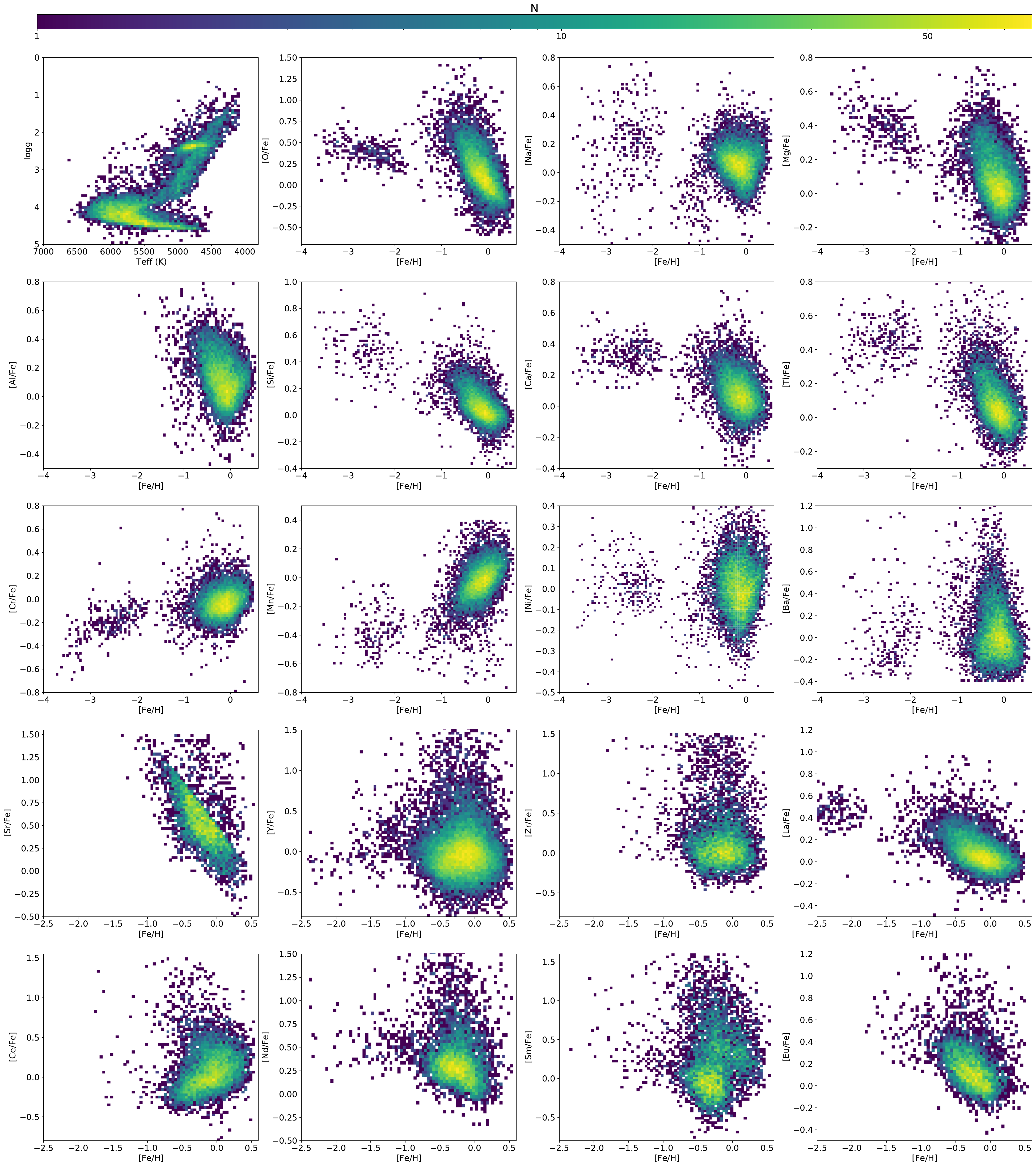}
\vspace{0.5em}
\caption{Stellar density distribution of the LAMOST-GALAH training samples in the $T_{\rm eff}$--$\log\,g$ diagram and [X/Fe]-[Fe/H] planes. The top three rows show distributions of the training set 3 as listed in Table\, \ref{table1}. The last two rows show distributions of the training sets 4-11 for s-/r- process abundance determinations. Some VMP stars have no abundance measurements for Al, Mn, Ni, and we have assigned constant values to them in the training process.
\label{fig1b}}
\end{figure*}

\subsection{Stellar model gradient spectra for regularization}
The physical spectra gradient $f^\prime_{\rm physical}(\lambda; \mathbf{l}_{\rm ref})$ used in Equation~(5) are computed using the Kurucz stellar atmospheric model \citep{Kurucz1970, Kurucz1993, Kurucz2005}. In doing so, we first build a set of 101 fiducial stars covering the parameter space of F/G/K stars with $3800<\teff<7500$~K, $0<\logg<5$, $-3<\feh<0.5$. The basic atmospheric parameters \teff, \logg, and \feh of these stars are randomly sampled from the MIST stellar isochrones \citep{Choi2016}. A plot shown their distribution in the $\teff$-$\logg$ plane can be found in \citet[][see their Fig.\,1]{ZhangM2024}. We assign a micro-turbulent velocity `$v_{\rm mic}$' for each star with a random value in the range of $0$--$5$~km/s. The elemental abundances [X/Fe] for each star are randomly assigned in the range tailored for different elements, from $-0.5$ to 1.5 for the majority of elements lighter than Zn, while $-2.0$ to 3 for the majority of s- and r-process elements. 
The model atmosphere for each star computed with the Kurucz {\sc ATLAS12} software \citep{Kurucz2005}, by dividing it into 72 layers. The spectrum at a resolution power of $R=500,000$ is then generated by solving the radiative transfer equation with the {\sc SYNTHE} software \citep{Kurucz1993, Kurucz2005}. We adopt the solar abundance scale of \citet{Asplund2009}, and a latest version of the Kurucz line list \citep{Kurucz2017}. We also incorporate the opacity of an extensive list of molecules in the calculation, including H$_2$, HD, CH, NH, OH, NaH, MgH, SiH, CaH, CrH, FeH, C$_2$, CN, CO, MgO, AlO, SiO, VO, TiO, and H$_2$O.

We derive the gradient spectrum $f^\prime_{\rm physical}(\lambda; \mathbf{l}_{\rm ref})$ for each of the 25 labels via 
\begin{equation}
f^\prime_{\rm Kurucz}(\lambda; l_{\rm ref}) = \frac{f_{\rm Kurucz}(\lambda; l_{\rm ref}+\Delta{l}) - f_{\rm Kurucz}(\lambda; l_{\rm ref})}{\Delta{l}}.
\end{equation}
To obtain the spectrum associated with $l_{\rm ref}+\Delta{l}$, we re-compute the {\sc ATLAS12} atmospheric model and then the {\sc SYNTHE} spectrum. For $\Delta{l}$, we adopt a value of 100~K for \teff, 0.2~dex for \logg, 0.2~dex for [Fe/H]. Finally, we computed a total of 2626 stellar model atmospheres and spectra, 102 for each star (for gradient spectra of 101 labels). This number is significantly increased with respect to the previous version of {\sc DD-Payne} by \citet{XiangMS2019}, who used only 16 stars (256 spectra for 15 labels), and all of which have solar-scaled abundance ratios (${\rm [X/Fe]}=0$). In addition, here we have utilized the CH molecule data of \citet{Masseron2015}, while \citet{XiangMS2019} adopted an old version \citep{Jorgensen1996}. We therefore expect a better performance of the current {\sc DD-Payne} implementation in constraining the gradient spectra of the data-driven neural network modelling. 
\begin{table*}
\caption{Descriptions for the LAMOST $DD$--$Payne$ training sets.}
\begin{tabular}{cccccc}

\hline
Training sets & Stellar labels & Label numbers & Label sources & Star numbers & Target labels\\
\hline         
\multirow{2}{*}{1}& $T_{\rm eff}$, $\log g$, $v_{\rm mic}$, C, N, O, Mg& \multirow{2}{*}{15} & APOGEE DR17 & \multirow{2}{*}{38,143} & \multirow{2}{*}{all }\\
 &Al, Si, Ca, Ti, Cr, Mn, Fe, Ni&  & LAMOST VMP \tablenotemark{1}  &  & \\
 \hline
 
 \multirow{2}{*}{2}& $T_{\rm eff}$, $\log g$, $v_{\rm mic}$, C, N, O, Mg& \multirow{2}{*}{15} & \multirow{2}{*}{APOGEE DR17} & \multirow{2}{*}{30,513} & \multirow{2}{*}{N, Al, Mn}\\
 &Al, Si, Ca, Ti, Cr, Mn, Fe, Ni&  &   &  & \\
 \hline
\multirow{2}{*}{3}&$T_{\rm eff}$, $\log g$, $v_{\rm mic}$, C, N \tablenotemark{2}, O, Na & \multirow{2}{*}{17} & GALAH DR3, & \multirow{2}{*}{10,916} &\multirow{2}{*}{all, except for C, N}\\
&Mg, Al, Si, Ca, Ti, Cr, Mn, Fe, Ni, Ba &  & LAMOST VMP  &  & \\
\hline    
\multirow{2}{*}{4}& $T_{\rm eff}$, $\log g$, $v_{\rm mic}$, C, N , O, Na & \multirow{2}{*}{18} & \multirow{2}{*}{GALAH DR3} & \multirow{2}{*}{4128}&\multirow{2}{*}{Sr}\\
& Mg, Al, Si, Ca, Ti, Cr, Mn, Fe, Ni, Ba, Sr & & & &\\
 \hline    
\multirow{2}{*}{5}& $T_{\rm eff}$, $\log g$, $v_{\rm mic}$, C, N, O, Na &\multirow{2}{*}{18} &\multirow{2}{*}{GALAH DR3}& \multirow{2}{*}{23,948} & \multirow{2}{*}{Y}\\
&Mg, Al, Si, Ca, Ti, Cr, Mn, Fe, Ni, Ba, Y  & & & & \\
 \hline    
\multirow{2}{*}{6}& $T_{\rm eff}$, $\log g$, $v_{\rm mic}$, C, N, O, Na &\multirow{2}{*}{18} &\multirow{2}{*}{GALAH DR3} &\multirow{2}{*}{6318} & \multirow{2}{*}{Zr}\\
&Mg, Al, Si, Ca, Ti, Cr, Mn, Fe, Ni, Ba, Zr &  &  & &\\
 \hline    
\multirow{2}{*}{7}& $T_{\rm eff}$, $\log g$, $v_{\rm mic}$, C, N, O, Na &\multirow{2}{*}{18}& \multirow{2}{*}{GALAH DR3}&\multirow{2}{*}{8130}&\multirow{2}{*}{La}\\
& Mg, Al, Si, Ca, Ti, Cr, Mn, Fe, Ni, Ba, La & &  &  & \\
 \hline    
\multirow{2}{*}{8}& $T_{\rm eff}$, $\log g$, $v_{\rm mic}$, C, N, O, Na &\multirow{2}{*}{18} & \multirow{2}{*}{GALAH DR3} &\multirow{2}{*}{7577} &\multirow{2}{*}{Ce}\\
& Mg, Al, Si, Ca, Ti, Cr, Mn, Fe, Ni, Ba, Ce & & &  & \\
 \hline    
\multirow{2}{*}{9}& $T_{\rm eff}$, $\log g$, $v_{\rm mic}$, C, N, O, Na &\multirow{2}{*}{18} & \multirow{2}{*}{GALAH DR3} &\multirow{2}{*}{5076}&\multirow{2}{*}{Nd}\\
&Mg, Al, Si, Ca, Ti, Cr, Mn, Fe, Ni, Ba, Nd & & & & \\
\hline    
\multirow{2}{*}{10}& $T_{\rm eff}$, $\log g$, $v_{\rm mic}$, C, N, O, Na &\multirow{2}{*}{18} & \multirow{2}{*}{GALAH DR3} &\multirow{2}{*}{ 5501}&\multirow{2}{*}{Sm}\\
&Mg, Al, Si, Ca, Ti, Cr, Mn, Fe, Ni, Ba, Sm & &  & & \\
\hline    
\multirow{2}{*}{11}& $T_{\rm eff}$, $\log g$, $v_{\rm mic}$, C, N, O, Na &\multirow{2}{*}{18} & \multirow{2}{*}{GALAH DR3} &\multirow{2}{*}{4108 }&\multirow{2}{*}{Eu}\\
&Mg, Al, Si, Ca, Ti, Cr, Mn, Fe, Ni, Ba, Eu & &  & & \\
\hline
\end{tabular}

\tablenotetext{1}{The LAMOST VMP sample from \citet{LiHN2022}.}
\tablenotetext{2}{The [C/Fe] and [N/Fe] values are estimated from LAMOST spectra using training set 1 with {\sc DD-Payne}.}
\label{table1}
\end{table*}

\subsection{Training the DD-Payne model}
\begin{figure}[htb!]
\centering
\includegraphics[width=0.48\textwidth]{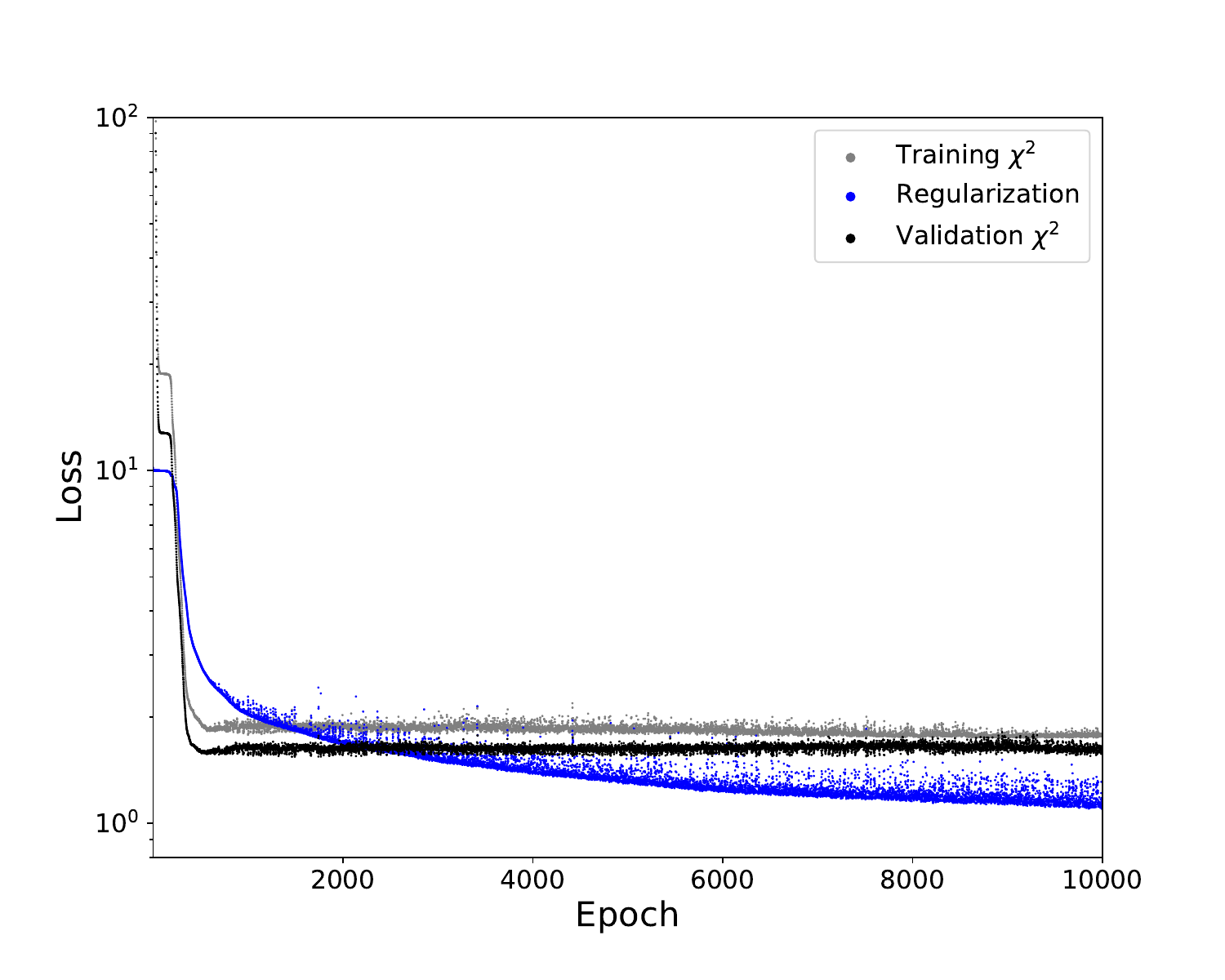}
\caption{Loss values as a function of training epoch for the modelling of a particular pixel of spectrum as an example. The grey curve shows the loss value calculated for the spectral flux of the training set, while the blue one shows the regularization term. The black curve shows the loss value for the spectral flux of the validation sample.  
\label{fig2}}
\end{figure}

For each training set, a neural network model of the {\sc DD-Payne} is trained for each pixel of the LAMOST spectra. The training process is implemented in the Pytorch environment, with the {\sc ADAM} method \citep{Kingma2014} adopted for model optimization. We adopt a learning rate varying from 0.01 at the beginning to 0.0001 at the end. To overcome overfitting and accelerate the convergence, the training sample is split into batches with a batch size of 2000. The training process is stopped until after 10000 iteration epochs. 

As an example to show the convergence, Fig.\,\ref{fig2} presents the loss function values for a typical pixel as a function of the iteration epoch. It shows that the loss function value contributed by the flux term in Equation\,(5), i.e., the $\chi^2$, drops quickly and reaches a plateau after about a few hundred iterations. However, it takes much more iterations for the converge of the regularization term. This demonstrates the necessarity to impose regularization of gradient spectra and to implement sufficient computation in order to build an effective spectra model. 

\begin{figure*}[htb!]
\centering
\vspace{0.5em}
\includegraphics[width=0.95\textwidth]{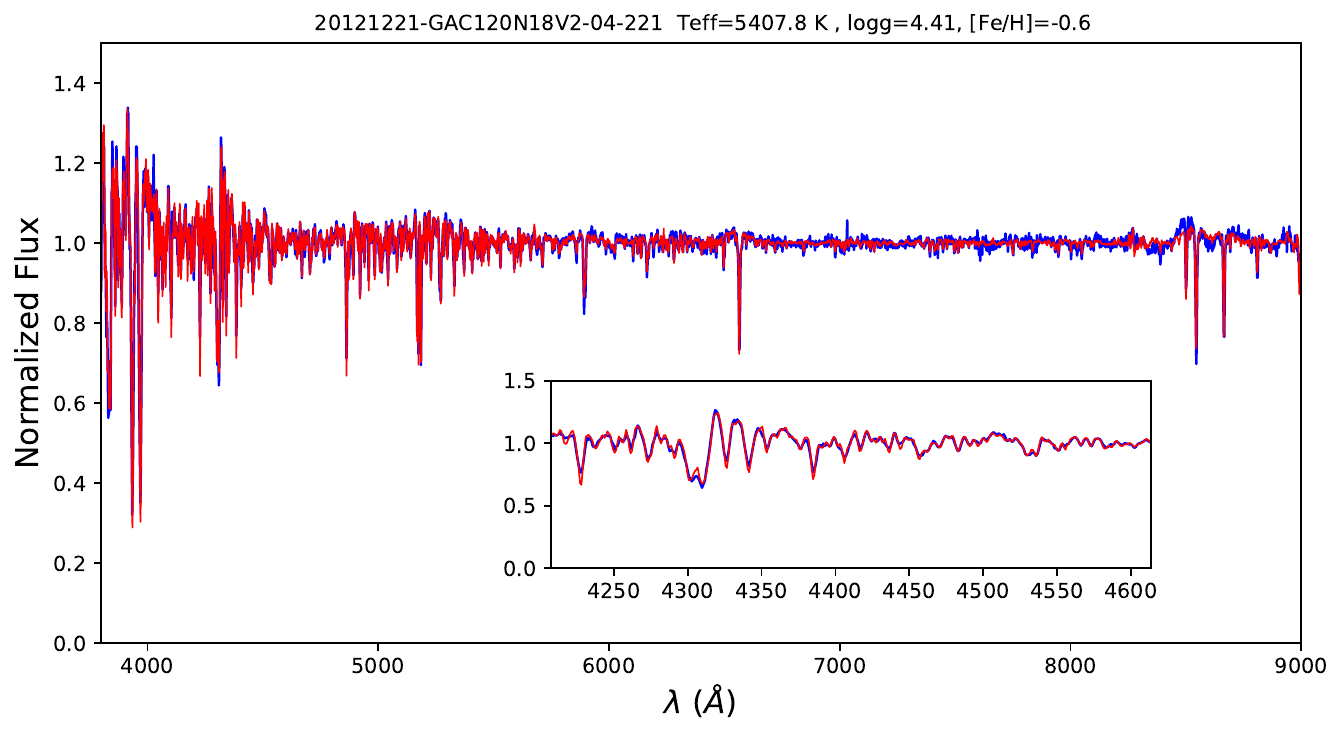}
\caption{The normalized LAMOST spectrum of a particular star as an example. The blue line is for the observed LAMOST spectra, while the red line is the best-fit {\sc DD-Payne} model, which parameters marked on top of the figure. The zoom-in plot shows the spectrum in a small window.
\label{fig4}}
\end{figure*}

\begin{figure*}[htb!]
\centering
\vspace{0.5em}
\includegraphics[width=0.87\textwidth]{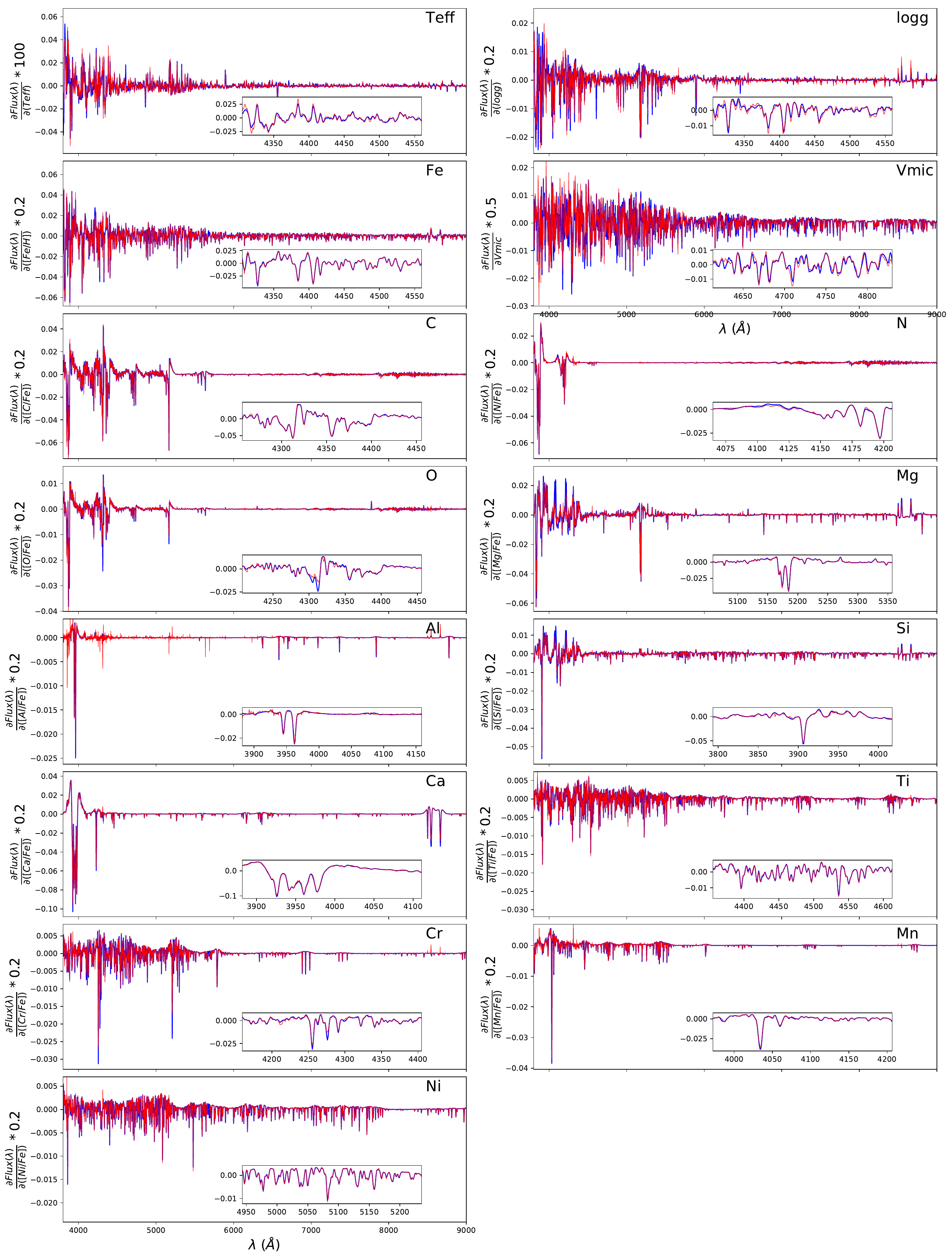}
\caption{Comparisons of differential spectrum for $T_{\rm eff}$, $\log\,g$, [Fe/H], $v_{\rm mic}$, and [X/Fe] for C, N, O, Mg, Al, Si, Ca, Ti, Cr, Mn, Ni between the {\sc DD-Payne} prediction (red) and the Kurucz $ab$ $initio$ model calculation (blue) for a fiducial star, with $T_{\rm eff} = 5117$\,K, $\log\,g = 3.29$, and  ${\rm [Fe/H]} = -0.37$. The {\sc DD-Payne} model used in this figure is trained on the LAMOST-APOGEE training set (training set 1 in Table\, \ref{table1}). The zoomed-in plot of each panel shows the gradient feature in a small window. The good consistency of differential spectrum between the {\sc DD-Payne} prediction and the Kurucz model suggests that the {\sc DD-Payne} abundance determination is rigorously physics-sensible.
\label{fig5}}
\end{figure*}

As an example, Figure~\ref{fig4} shows a normalized LAMOST spectrum and the {\sc DD-Payne} model spectrum for the best-fit labels. The {\sc DD-Payne} models the LAMOST spectrum flux well, with a $\chi^2$ value of 1.38. Moreover, the {\sc DD-Payne} can also correctly reproduce the gradient spectra for individual labels. Fig.\,\ref{fig5} shows the comparison of the {\sc DD-Payne} gradient spectra with gradient spectra generated from the Kurucz atmospheric model, for a fiducial star as an example. The Figure shows that some of the labels such as N have few but strong features, while some of them such as Ni only have weak but numerous features. As expected, {\sc DD-Payne} can correctly reproduce the Kurucz physical model gradient spectra for all these 25 labels, demonstrating its capability to catch up and make maximally use of these features for deriving stellar labels from the spectra. 

\begin{figure*}[htb!]
\centering
\vspace{0.5em}
\includegraphics[width=0.95\textwidth]{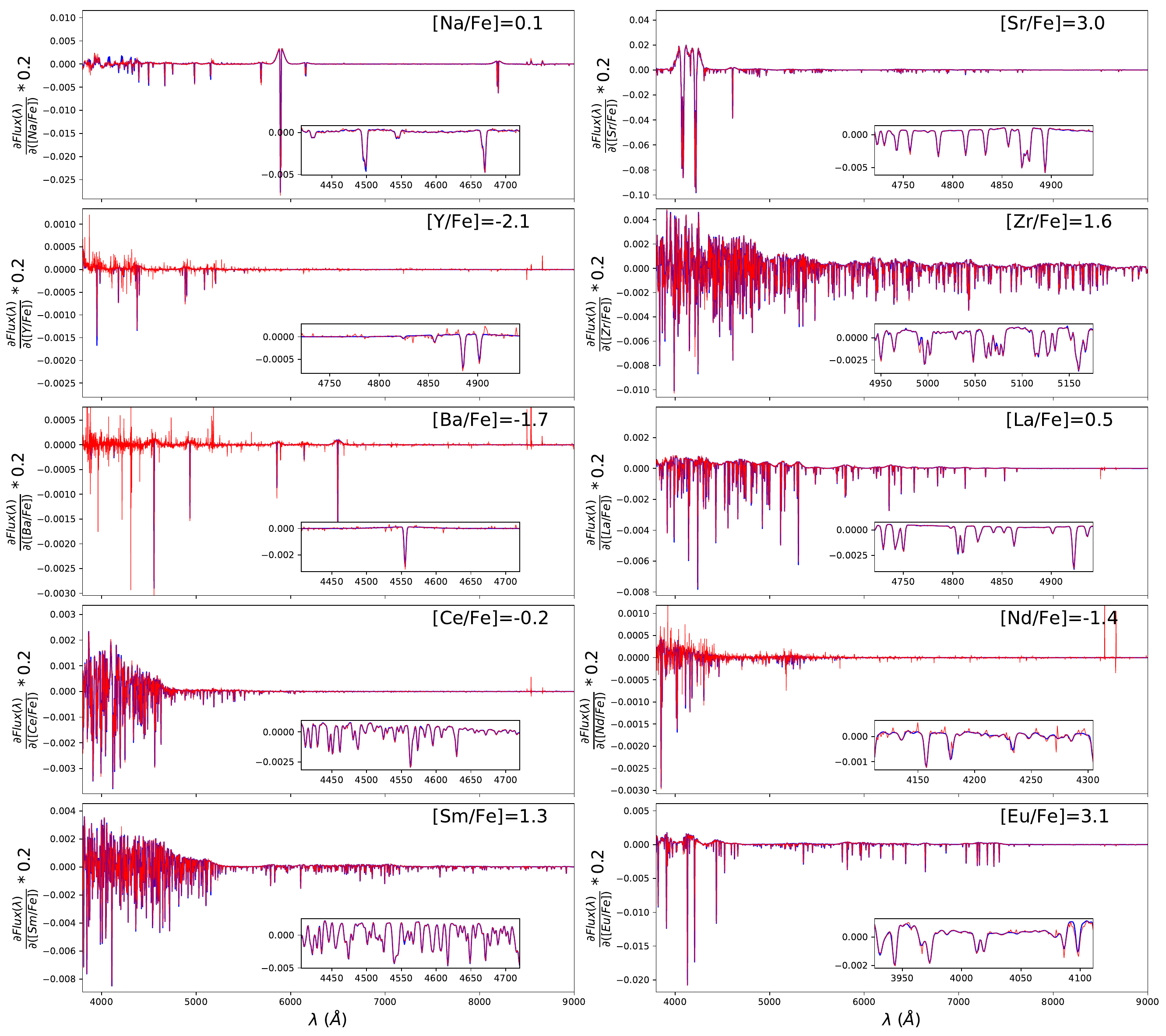}
\caption{Comparison of differential spectrum for Na, Sr, Y, Zr, Ba, La, Ce, Nd, Sm, and Eu between {\sc DD-Payne} prediction (red) and the Kurucz $ab$ $initio$ model calculation (blue) for the same star as in Fig.\,\ref{fig5}. The {\sc DD-Payne} models used here are trained on  LAMOST-GALAH training sets (training set 3-11 in Table\, \ref{table1}).
\label{fig6}}
\end{figure*}
\subsection{Validation with test set}
\begin{figure*}[htb!]
\centering
\vspace{1.em}
\includegraphics[width=\textwidth]{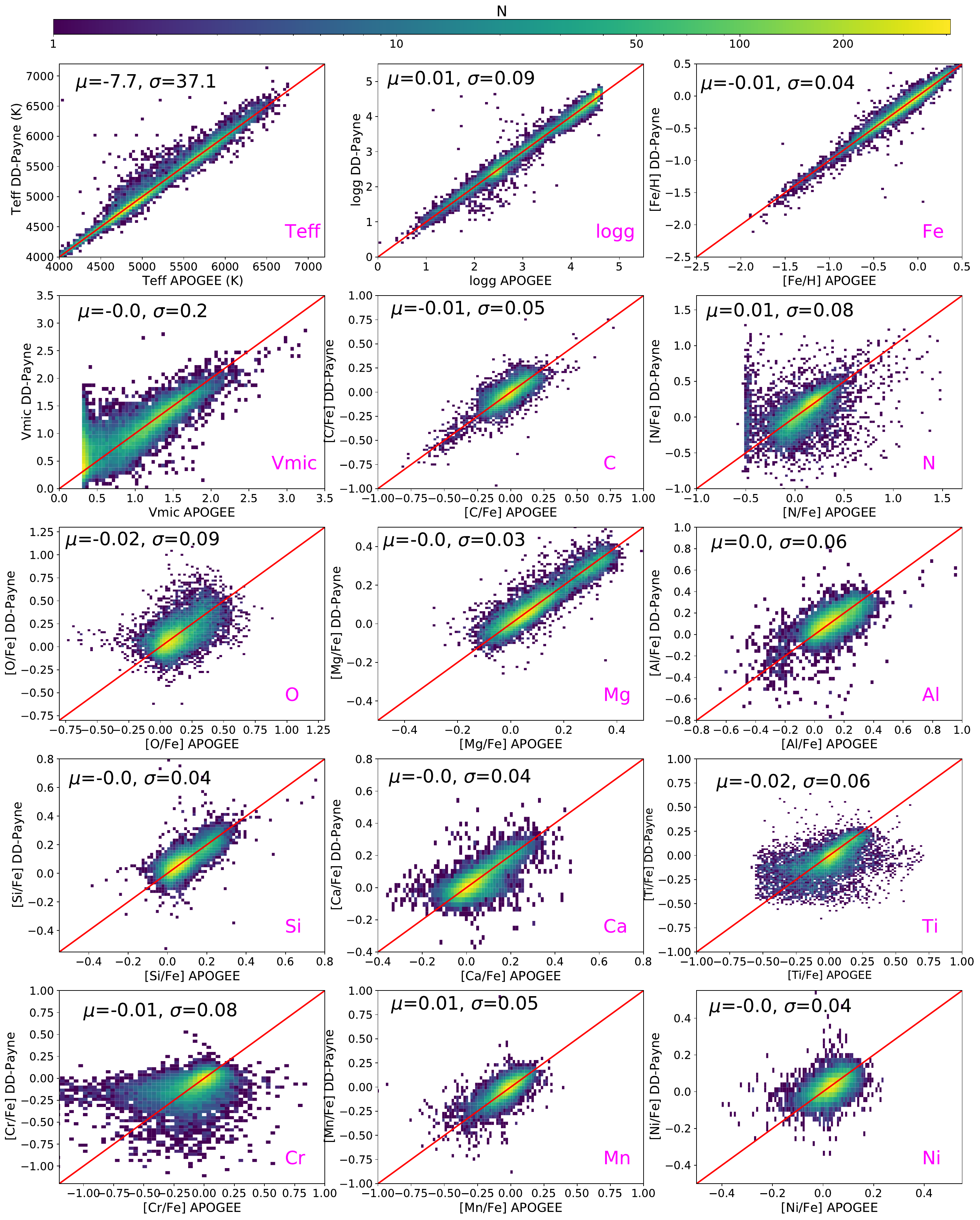}
\vspace{0.5em}
\caption{Comparisons of the stellar parameters ($T_{\rm eff}$, $\log\,g$, [Fe/H], $v_{\rm mic}$) and elemental abundances between {\sc DD-Payne} determinations and those from APOGEE DR17 catalog for the validation sample. Color represents the number density of stars. Solid lines show the 1:1 line of X and Y axes. The mean and dispersion, defined as the standard deviation after $3\sigma$ clipping, of the differences between {\sc DD-Payne} and the APOGEE measurements are marked in each panel.
\label{fig3}}
\end{figure*}

\begin{figure*}[htb!]
\centering
\vspace{1.em}
\includegraphics[width=\textwidth]{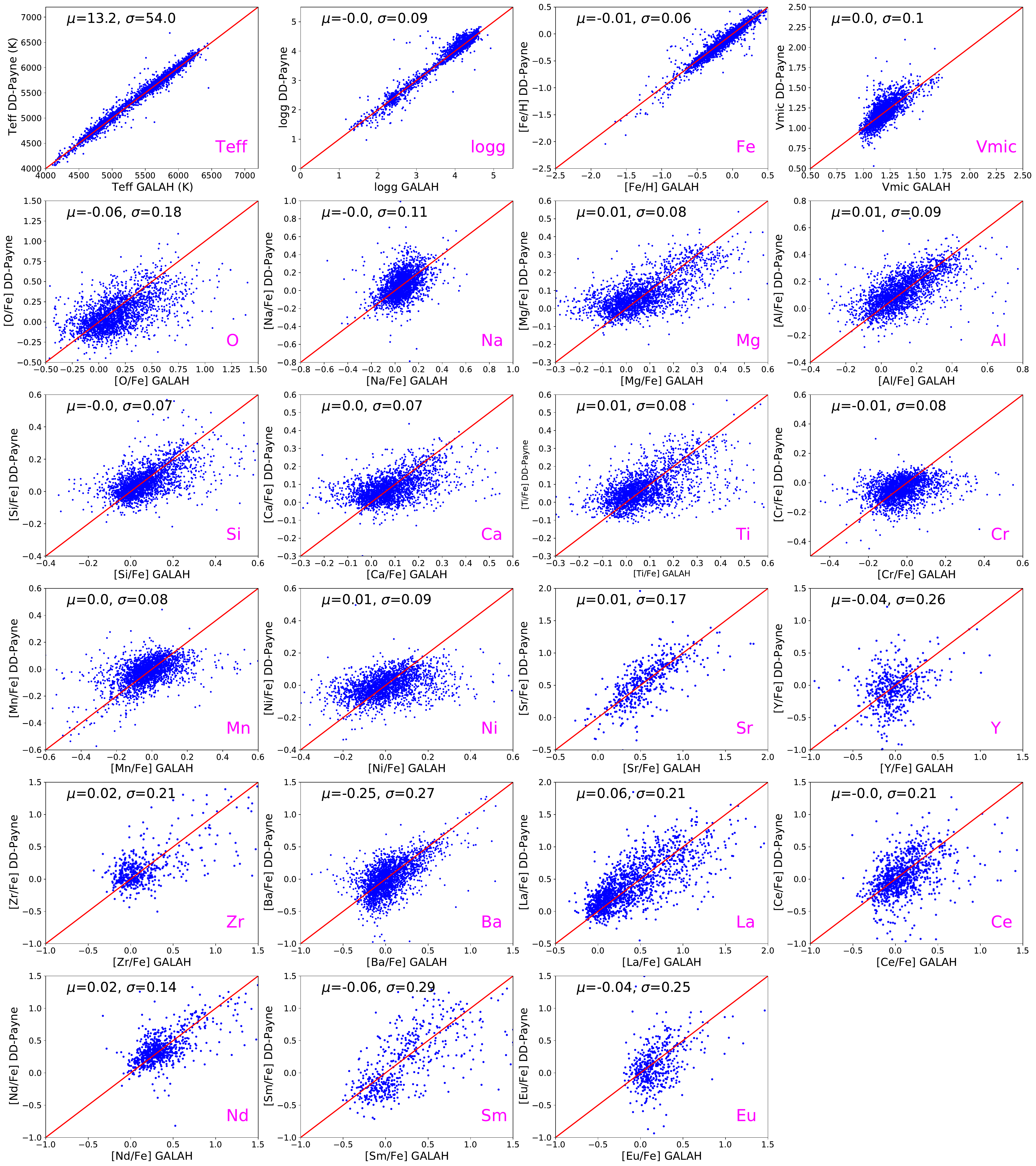}
\vspace{0.5em}
\caption{Comparison of the stellar parameters ($T_{\rm eff}$, $\log\,g$, [Fe/H], $v_{\rm mic}$) and elemental abundances between {\sc DD-Payne} determinations and those from GALAH DR3 catalog for the validation sample. Solid lines show the 1:1 line of X and Y axes. The mean and dispersion of the differences between {\sc DD-Payne} and the GALAH measurements are marked in each panel.
\label{fig3b}}
\end{figure*}
Fig.\,\ref{fig3} shows a comparison of the LAMOST {\sc DD-Payne} label estimates with APOGEE DR17 for 20,000 test stars. Here the {\sc DD-Payne} model is trained on the LAMOST-APOGEE training set. The figure shows good agreements, with a negligible mean deviation, for all the labels. The standard deviation in the differences is only 37~K for \teff, 0.09~dex for \logg, 0.03--0.05~dex for [Fe/H], [C/Fe], [Mg/Fe], [Si/Fe], [Ca/Fe], [Mn/Fe], [Ni/Fe], 0.06--0.09~dex for [N/Fe], [O/Fe], [Al/Fe], [Ti/Fe], and [Cr/Fe], 0.2~km/s for $v_{\rm mic}$. For [N/Fe], the APOGEE DR17 values exhibit a lower border with ${\rm [N/Fe]}\simeq-0.5$. However, most of the low-[N/Fe] stars at this border have a normal [N/Fe] value in the LAMOST {\sc DD-Payne} estimates. We found that these stars generally have a warning flag in the APOGEE catalog. 
 For [Cr/Fe], APOGEE DR17 exhibits a lower tail (${\rm [Cr/Fe]}\lesssim-0.5$), while our results show normal values for most of these stars, with unclear reason.
Note that Cr has weak but a massive set of features in the LAMOST spectra (Fig.\,\ref{fig5}), which allow a robust [Cr/Fe] estimate with {\sc DD-Payne}.

Similarly, Fig.\,\ref{fig3b} shows a comparison of the LAMOST {\sc DD-Payne} label estimates with GALAH DR3 for test stars. Here the LAMOST labels are estimated with {\sc DD-Payne} models trained on the LAMOST-GALAH training sets.
The figure show good agreements between the LAMOST {\sc DD-Payne} and GALAH DR3 labels. The standard deviation in the differences is only 54~K for \teff, 0.09~dex for \logg, 0.06~dex for [Fe/H], 0.07--0.09~dex for [Mg/Fe], [Al/Fe], [Si/Fe], [Ca/Fe], [Ti/Fe], [Cr/Fe], and [Mn/Fe], and 0.2 dex for [O/Fe] and [Ba/Fe]. The heavier elements, the standard deviation in the abundance difference is 0.14\,dex for Nd, $\sim0.2$\,dex for Ba, Sr, Zr, La, and Ce, 0.25-0.3\,dex for Y, Sm, and Eu. 
    
Note that the scatters in Fig.\,\ref{fig3} and Fig.\,\ref{fig3b} are due to errors in both the LAMOST {\sc DD-Payne} estimates and the APOGEE DR17 or GALAH DR3 labels. Therefore, the standard deviations are not measures of the {\sc DD-Payne} results. Actually, errors in the {\sc DD-Payne} label estimates can be significantly smaller than the standard deviations (see Sect.\,4).

\begin{figure*}[htb!]
\centering
\vspace{1.em}
\includegraphics[width=\textwidth]{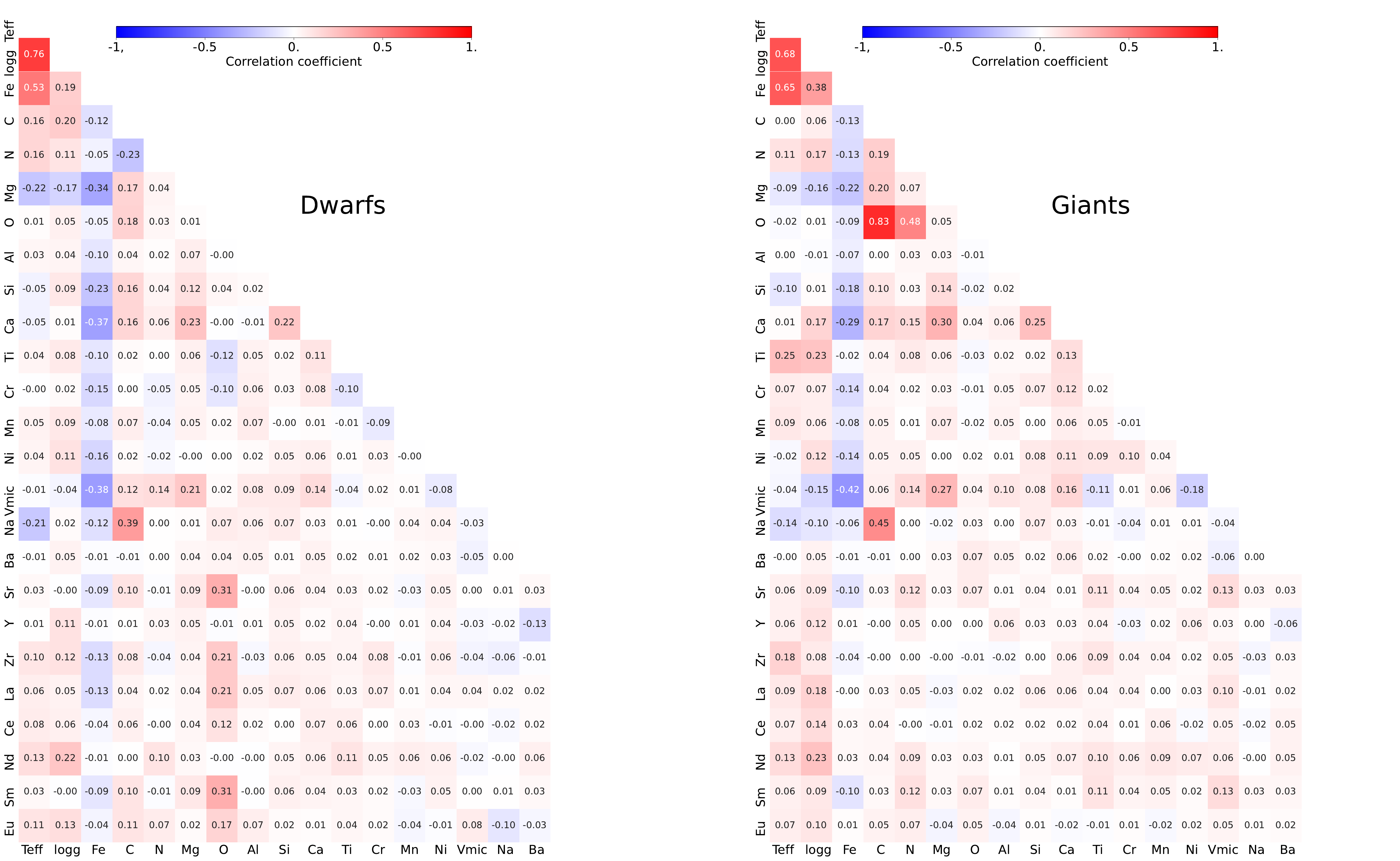}
\vspace{0.5em}
\caption{Covariance among different labels for {\sc DD-Payne} spectral fitting. Colors represent the median value of correlation coefficients, derived from the covariance matrix for dwarfs ($left$) and giants ($right$), separately. The numbers marked in the figure are identical to the colors.
\label{fig_correlation}}
\end{figure*}

In order to understand the covariance among the individual label estimates, we calculate their correlation coefficients from the covariance matrix of the spectra fitting.
Fig.\,\ref{fig_correlation} shows the median value of the correlation coefficients ($R$) for the LAMOST sample stars. The left panel shows that, there are strong covariance between $T_{\rm eff}$ and $\log\,g$ ($R=0.76$ for dwarfs, $R=0.68$ for giants), $T_{\rm eff}$ and [Fe/H] ($R=0.53$ for dwarfs, $R=0.65$ for giants).
For giants, there is a strong covariance between C and O abundances ($R=0.83$), as well as between N and O abundances ($R=0.48$). This is because the O abundance has a strong impact on the C features (e.g., CH\,4314\AA\,band) and N features (e.g., CN\,3880\AA\,band) in LAMOST spectra through the CNO network \citep{TingYS2017b}. There are moderate (anti-)covariance ($0.3\lesssim \vert{R}\vert \lesssim 0.4$) among some of the labels, such as Fe and Mg, Fe and Ca, Fe and $v_{\rm mic}$, C and Na for dwarfs. However, the covariance between $T_{\rm eff}$ and [X/Fe] are generally small. This is also the case for covariance between $\log g$ and [X/Fe], as well as for most of the cases among different abundances.

\section{External calibration and validation}
\subsection{Calibrating $T_{\rm eff}$ to IRFM scale}
 \begin{figure*}[htb!]
\centering   
\subfigure
{
	\begin{minipage}{0.48\linewidth}
	\centering  
	\includegraphics[width=0.98\columnwidth]{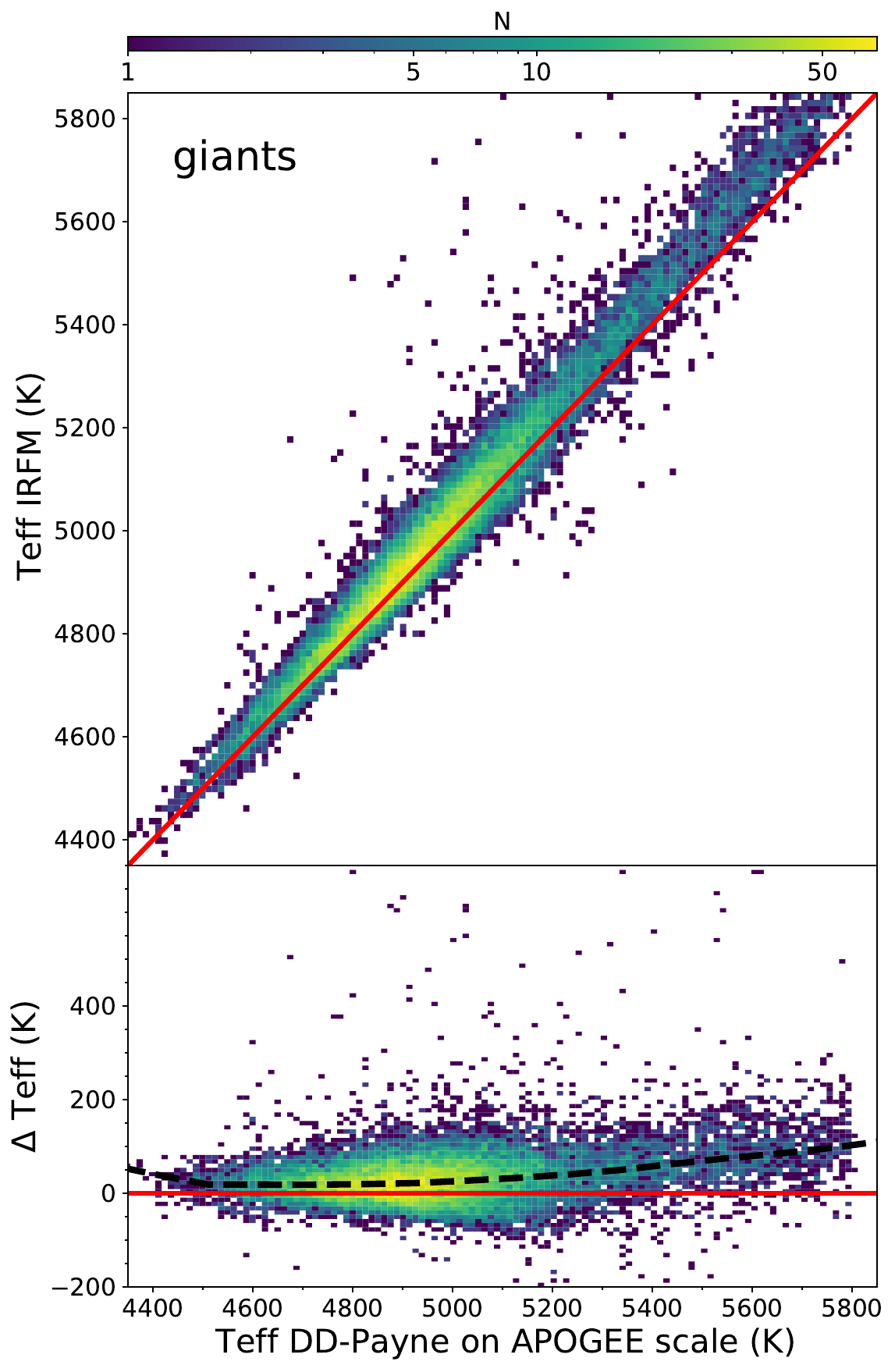} 
	\end{minipage}
} 
\subfigure
{
	\begin{minipage}{0.48\linewidth}
	\centering  
	\includegraphics[width=0.98\columnwidth]{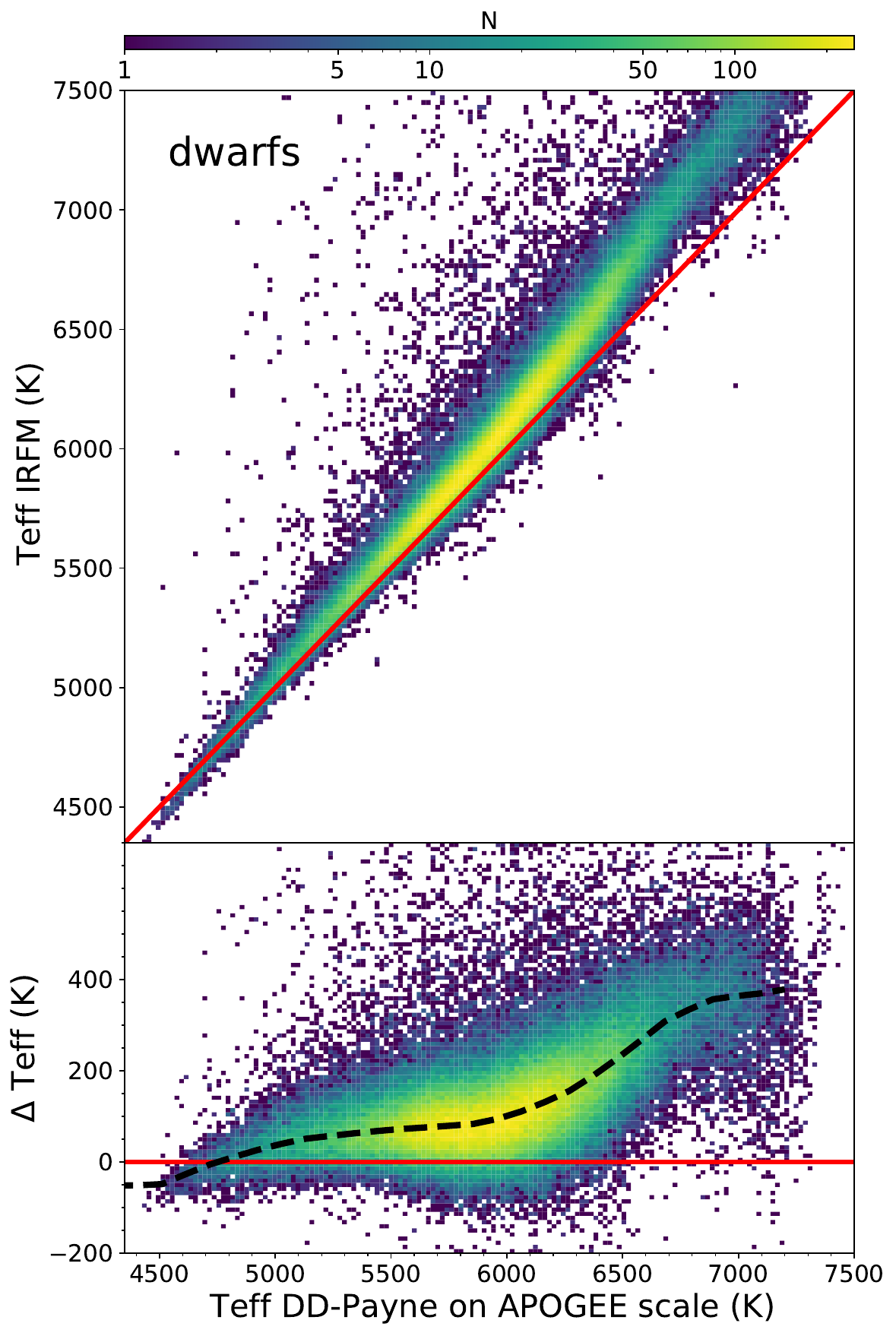} 
	\end{minipage}
} 
\caption{Comparison of effective temperature between {\sc DD-Payne} determination and the IRFM estimates for giants (left) and dwarfs (right). Color represents the number density of stars. Solid lines in the upper panels show the 1:1 line of X and Y axes. The dotted lines in the bottom panels show the median of systematic differences between the IRFM and {\sc DD-Payne} results.
\label{fig5a}}
\end{figure*}

The infrared flux method (IRFM) has been widely suggested to be an effective way for accurate temperature estimation \citep[e.g.][]{Gonzalez2009, Casagrande2010, Soubiran2024}. Alongside with the direct spectroscopic $T_{\rm eff}$ estimates, the APOGEE DR17 catalog also provided a set of $T_{\rm eff}$ calibrated to IRFM temperature, which is derived using the color-temperature relation of \citet{Gonzalez2009} for giant stars \citep[see details in][]{Abdurrouf2022}. This IRFM-scaled temperature is adopted as our training label for the LAMOST-APOGEE training set. However, since dwarf stars have a different color-temperature relation to giant stars, we found that the APOGEE IRFM-scaled temperatures for dwarf stars suffer significant systematics (Figure~\ref{fig5a}). We therefore opt to re-calibrate our temperature estimates to IRFM scale here.

We define a calibration data set by selecting stars with good photometry in both V band from the AAVSO Photometric All-Sky Survey \citep[APASS;][]{Henden2014, Henden2016} and K band from the Two Micron All Sky Survey \citep[2MASS;][]{Skrutskie2006}. We require the calibration stars to have a V-band magnitude from 12 to 16~mag, with a photometric error smaller than 0.02~mag, and the photometric error in 2MASS K band is set to be less than 0.03~mag. We derived the reddening $E(B-V)$ of the individual stars using the star-pair method as described in \citet{ZhangM2024}, and require the calibration stars to have an error in the $E(B-V)$ estimates to be smaller than 0.03~mag. We then derive the IRFM temperature of the calibration stars using the $T_{\rm eff}$ - $(V-K)_0$ relations of \citet{Gonzalez2009}, for giants ($T_{\rm eff}<5800$~K, $\log~g<3.8$) and dwarfs, separately.  

Figure~\ref{fig5a} shows the comparison between the DD-Payne and IRFM temperatures. For giant stars, we found a good agreement, with a typical mean difference of only 10-20~K for stars with $T_{\rm eff}<5200$~K. This good agreement reflects the fact that the training $T_{\rm eff}$ of DD-Payne, i.e., the APOGEE $T_{\rm eff}$, were calibrated to the same scale as our IRFM temperature estimates. The minor zero point offset (10-20~K) is possibly due to either different photometry or different extinction correction adopted. For stars with temperature higher than 5200~K, the deviation continuously increases to $\sim100$~K, similar to the case of dwarf stars. For dwarf stars, temperature difference between DD-Payne and IRFM scale increases from $\sim0$~K at 4800\,K $\sim400$~K at 7000~K, with a sudden increase of slope above 6200~K. 

For both dwarfs and giants, we subtract the mean trend shown by dashed lines in Figure~\ref{fig5a} to calibrate the DD-Payne temperature to the IRFM scale. This calibration is only done for DD-Payne temperature estimated using the LAMOST-APOGEE training set, as it is the temperature in our recommended catalog (Sect.\,5).  

\subsection{Validating $\log~g$ with asteroseismology}
\begin{figure}[htb!]
\centering
\includegraphics[width=0.48\textwidth]{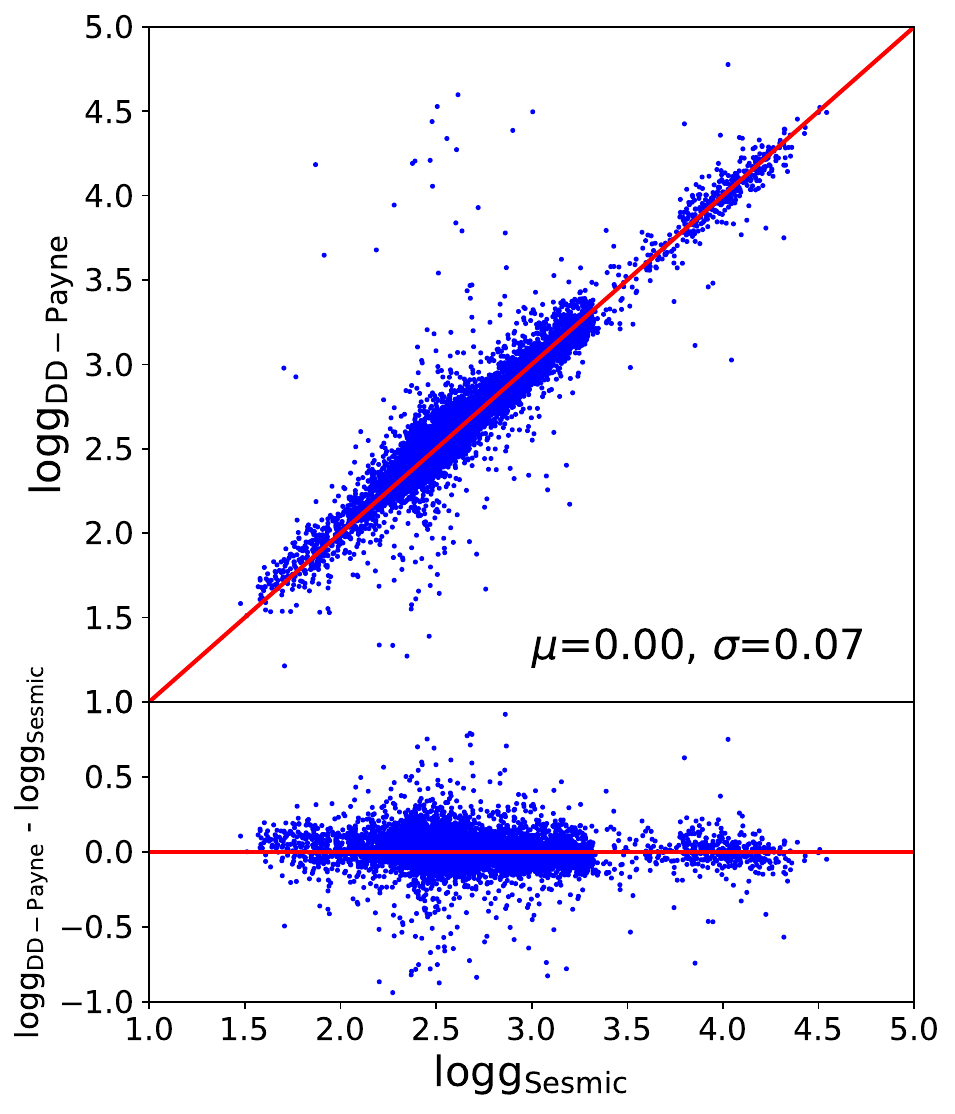}
\caption{The one-to-one comparison of $\log\,g$ values for 8981 red giants, and 426 subgiant and dwarf stars between the asteroseismic measurements and {\sc DD-Payne} determinations. The red solid line in the upper panel shows the 1:1 line of the X- and Y-axes. The mean and dispersion of the differences are marked in the figure.
\label{fig5b}}
\end{figure}

Accurate and precise $\log~g$ derived with Asteroseismology provide a golden calibration source for spectroscopic $\log~g$ estimates \citep[e.g.,][]{RenJJ2016}. Light curves from the $Kepler$ mission \citep{Borucki2010} allow precise determination of asteroseismic frequency for more than nearly 20 thousands red giant stars \citep[e.g.][]{YuJ2018} and a few hundred of main-sequence and subgiant stars \citep[e.g.][]{Mathur2022}. About half of the $Kepler$ stars are observed by LAMOST via a dedicated survey to the $Kepler$ field \citep{DeCat2015,FuJN2020}.

Following \citet{Kjeldsen1995}, we derive the asteroseismic $\log~g$ as follows,
\begin{equation}
\log g = \log g_\odot + \log \left(
{{\nu_{\rm max} \over \nu_{{\rm max}, \, \odot}}}
\right)
+ {1 \over 2}
\log \left({T_{\rm eff}\over T_{{\rm eff}, \, \odot}}\right),
\label{Eq.logg}
\end{equation}
where the $\nu_{\rm max}$ is frequency of maximum stellar oscillation power, which we adopted from \citet{YuJ2018}. For the solar frequency $\nu_{{\rm max}_{,\sun}}$, we adopt a value of 3090~$\rm{\mu Hz}$ \citep{Huber2011}. For the effective temperature, we use the DD-Payne estimates calibrated to the IRFM scale, as mentioned above. Typical error is in the $\log~g$ estimates is found to be $\sim0.01$~dex.

Fig.\,\ref{fig5b} shows the comparison between the DD-Payne $\log\,g$ estimates and the asteroseismic values for 8981 stars with LAMOST spectral S/N higher than 30. The DD-Payne estimates agree very well with the asteroseismic values, with no significant mean deviation in the whole range of $\log~g$ from 1.5 to 4.5. The difference has a dispersion of only 0.07~dex. Note that here the DD-Payne $\log\,g$ is derived using the LAMOST-APOGEE training set, which we adopted in the recommended catalog (Sect.\,5).
 
\subsection{Correcting ${\rm [Fe/H]}$ for NLTE effect}
\begin{figure}[htb!]
\centering
\includegraphics[width=0.48\textwidth]{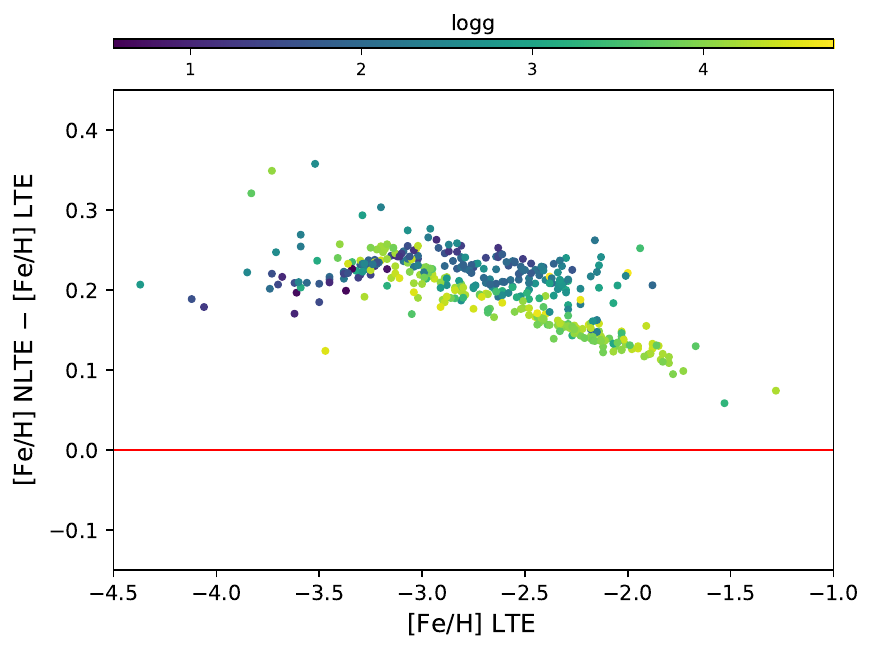}
\caption{The amount of NLTE correction for ${\rm[Fe/H]}$ of the metal-poor sample stars of \cite{LiHN2022}. Colors represent $\log~g$ of the stars.
\label{fig_nlte}}
\end{figure}
NLTE effects can have a large and complicated impact on the abundance determinations. For neutral iron and similar minority species, these effects are typically larger towards lower metallicities, as well as for hotter stars and for stars of lower surface gravity \citep[e.g.,][]{Mashonkina2011, Bergemann2012, Lind2012, Sitnova2015, Lind2024}. The abundances for both the APOGEE DR17 and the high-resolution sample of \citet{LiHN2022} are derived using model spectra computed under LTE assumption. While our determination taking these labels as training set has achieved a good internal precision, the resultant labels may suffer non-negligible systematic errors due to NLTE effects that are not properly considered in the training labels. In this work, we conduct a NLTE correction for the {\sc DD-Payne} [Fe/H] determination. 
 


In doing so, we first need a good reference sample with NLTE ${\rm [Fe/H]}$ measurements. The GALAH DR3 itself forms a large reference sample, as NTLE effects for Fe and a few other elements have been incorporated in its abundance determination \citep{Amarsi2016, Amarsi2020, Buder2021}. Nonetheless, the number of metal-poor stars in common between GALAH DR3 and LAMOST that have good abundance quality is too small to do the calibration at this regime. We therefore build up a metal-poor star sample with NLTE ${\rm [Fe/H]}$ from high-resolution spectroscopy. This is carried out by directly making a NLTE correction to the LTE abundance measurements of \citet{LiHN2022}. For each star in \citet{LiHN2022}, we estimate the NLTE correction for the Fe\,{\sc i} abundance measurements, using the 1D non-LTE model of \citet{Amarsi2022}. The model has corrections for 171 optical Fe\,{\sc i} lines; for a given star the correction was estimated as the median correction for lines having reduced equivalent widths between -6 and -5.
Fig.~\ref{fig_nlte} shows the amount of abundance correction as a function of ${\rm [Fe/H]}$. Typical correction varies from 0.1~dex at $\feh\simeq-1$ to 0.25~dex at $\feh\simeq-3$. The 1D NLTE Fe\,{\sc i} abundance is further validated by the Fe\,{\sc ii} abundance. Generally, NLTE effect for Fe\,{\sc ii} is small, but the 3D effect may be significant.We also tested calibrating the Fe\,{\sc ii} abundances of \citet{LiHN2022} to the 3D LTE scale (given that Fe\,{\sc ii} lines show negligible departures from LTE down to {\rm [Fe/H]}$\approx-3$; \citealt{Amarsi2022}). The 3D corrections were estimated from the grid of \citet{Amarsi2019}, taking for each star the median of 142 optical Fe\,{\sc ii} lines.  The 3D corrections for the  Fe\,{\sc ii} lines are smaller than the NLTE corrections for the Fe\,{\sc i} lines (around $+0.08$ dex to $+0.12\,$dex for Fe\,{\sc ii}, compared to $+0.10\,$dex to $+0.25\,$dex for  Fe\,{\sc i}); and we found improved consistency between the two species after applying corrections to both species.
Nonetheless, we ultimately only adopt the 1D NLTE Fe\,{\sc i} abundance as our reference for simplicity, because only a few sample stars have the measurements of Fe\,{\sc ii} lines. 

Note that, the abundance of \citet{LiHN2022} also has a slightly (0.05~dex) different zero point to GALAH DR3, as the former adopted the solar abundance of \citet{Asplund2009}, while the latter adopted \citet{Grevesse2007} (Sect.\,7). This difference has also been corrected for to ensure a uniform abundance scale. 

The calibration of {\sc DD-Payne} [Fe/H] to the NLTE reference is conducted separately for giants ($\log~g<3.5$, $\teff<5800$~K) and dwarfs. We found that the  correction is small for dwarfs with $\feh\gtrsim-1$, but non-negligible for giants or metal-poor stars. For elemental abundances other than [Fe/H], currently we do not make any NLTE correction. However, for Na and Ba, the abundance determinations are automatically in NLTE scale as inherited from GALAH DR3 training set (except for [Ba/Fe] at the $\feh\lesssim-2$ metal-poor regime where the abundance is tied to \citet{LiHN2022}). 


 
\subsection{Comparing [Fe/H] with the PASTEL catalog}

\begin{figure*}[htb!]
\centering   
\subfigure
{
	\begin{minipage}{0.48\linewidth}
	\centering  
	\includegraphics[width=0.98\columnwidth]{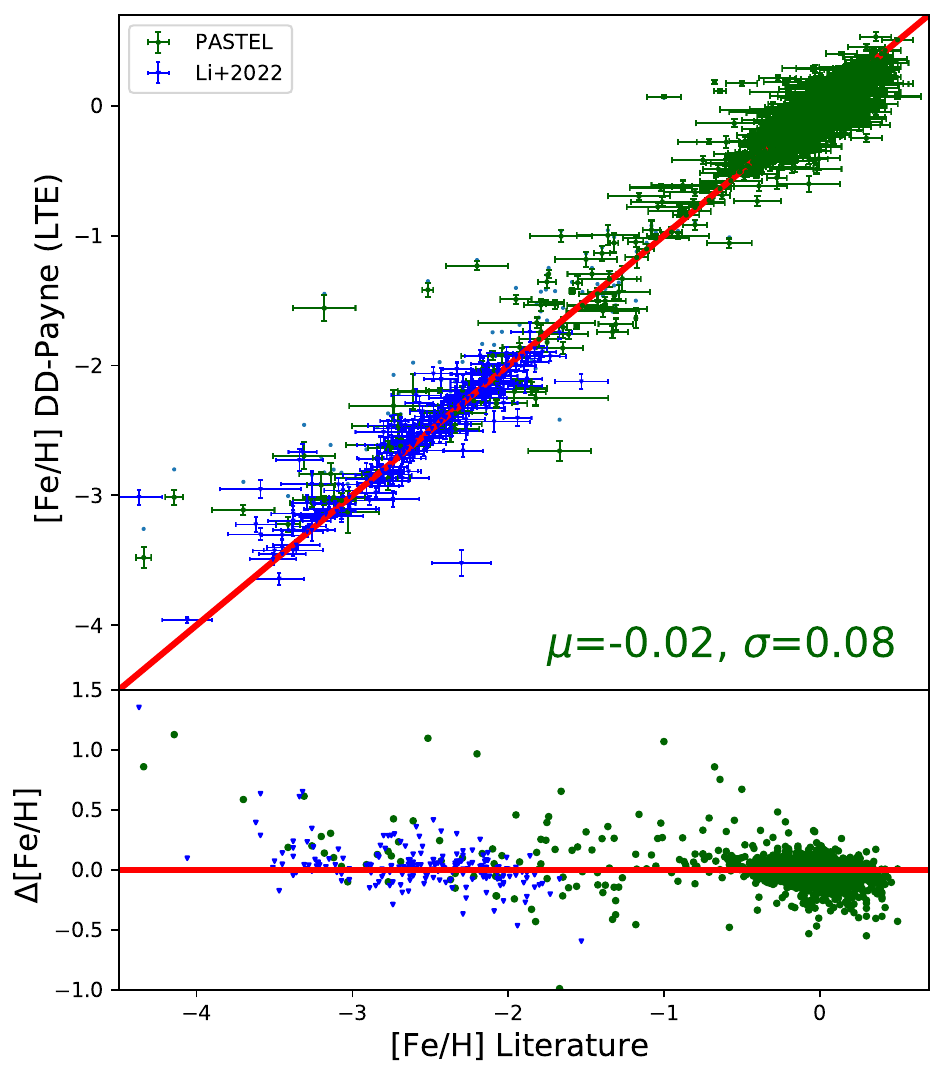} 
	\end{minipage}
} 
\subfigure
{
	\begin{minipage}{0.48\linewidth}
	\centering  
	\includegraphics[width=0.97\columnwidth]{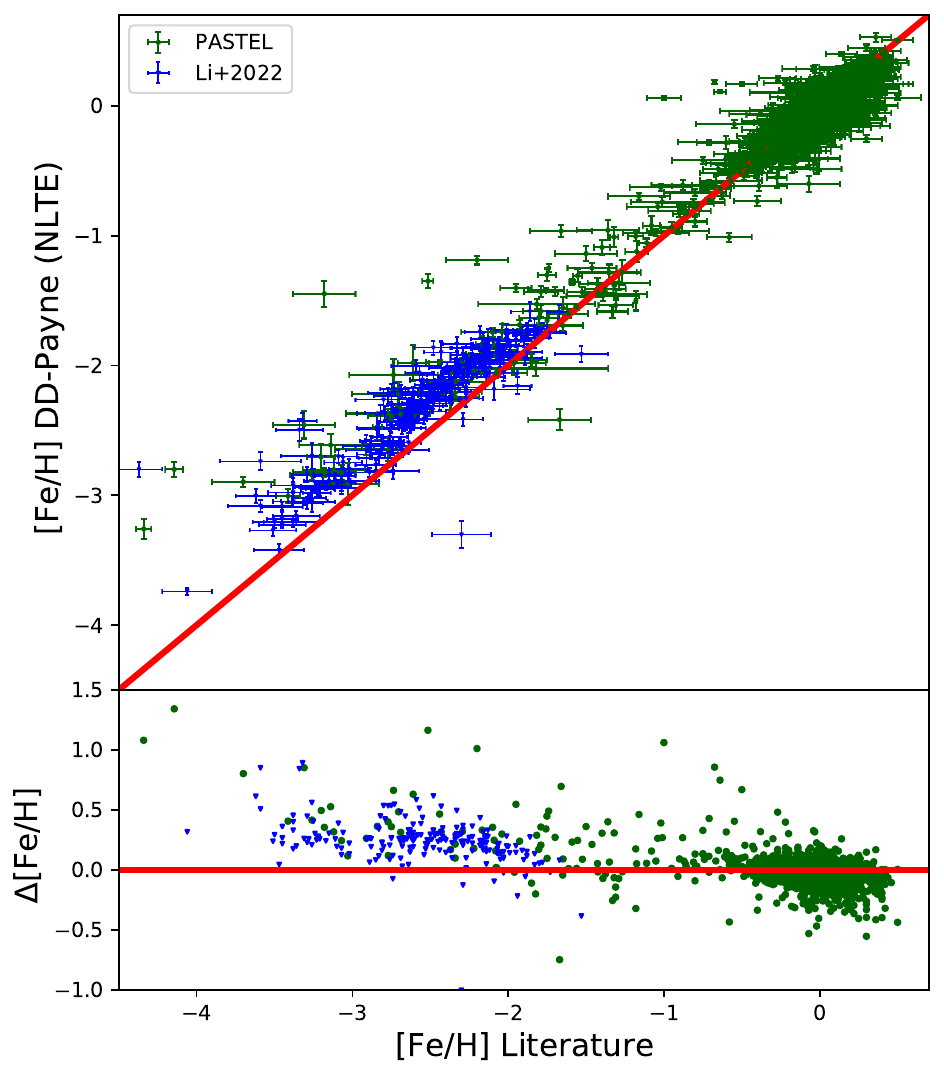} 
	\end{minipage}
} 
\caption{Comparing the {\sc DD-Payne} ${\rm [Fe/H]}$ determinations with high-resolution spectroscopy samples from the PASTEL catalog \citep{Soubiran2016} and the LAMOST VMP catalog of \citet{LiHN2022}. The left panel shows {\sc DD-Payne} ${\rm [Fe/H]}$ determination using the LAMOST-APOGEE training set (training set 1 in Tab.\ref{table1}), without external calibration, while the right panel shows the same sample but after NLTE correction (see text).
\label{fig5c}}
\end{figure*}

The PASTEL catalog is a compilation of stellar parameters derived from high-resolution and high-S/N spectra \citep{Soubiran2016}. It has been often used as a reference sample for validating the [Fe/H] estimates from spectroscopic surveys with lower spectra resolution \citep[e.g.][]{XiangMS2017, Soubiran2022}. The PASTEL catalog includes 31,401 stars with $T_{\rm eff}$, $\log\,g$ and [Fe/H] records. A cross-identify of the LAMOST DR9 with PASTEL with a 3" distance criterion yields 4902 common stars with a LAMOST spectral S/N higher than 20. Many of the stars have more than one record in the PASTEL catalog. To have an accurate reference [Fe/H] value, we discard all measurement before the year 1990. For the remaining stars that have multiple records, we further compute the weighted average value, taking the invariance errors as the weights. Ultimately, 1702 stars are used for our validation. 

The left panel of Fig.\,\ref{fig5c} shows the comparison of [Fe/H] between PASTEL and the {\sc DD-Payne} determinations using the LAMOST-APOGEE training set, without NLTE correction. The Figure shows a good consistency for [Fe/H] down to $\feh<-3$~dex. The difference for the overall sample exhibits only a minor ($-0.02$~dex) mean deviation, and a dispersion of only 0.08~dex. For stars with a PASTEL [Fe/H] below $-3.5$, the DD-Payne may yield higher value, up to $\sim1$~dex at the metal-poor end. 

The Figure also shows the comparison with high-resolution results of \citet{LiHN2022} for their LAMOST VMP star sample. This sample is adopted as a part of our training set. Nevertheless, as DD-Payne is not simply tied to the training sample as trivial data-driven approach may do, such a comparison can still be informative for the verification of the DD-Payne [Fe/H] estimates at the metal-poor end. The good consistency down to $-4$ shown in the Figure suggests that the DD-Payne [Fe/H] estimates are robust down to such a low metallicity. However, stars with ${\rm [Fe/H]}\lesssim-4$ may exhibit a complicate behaviour.   

However, the right panel shows that the {\sc DD-Payne} abundances after NLTE correction are systematically higher than the high-resolution results. This is because both the PASTEL and the \citet{LiHN2022} abundances are mostly derived with LTE models. They thus may suffer a significant underestimate for metal-poor stars. 

\subsection{Calibrating elemental abundances with wide binaries}
\begin{figure*}[htb!]
\centering
\includegraphics[width=0.98\textwidth]{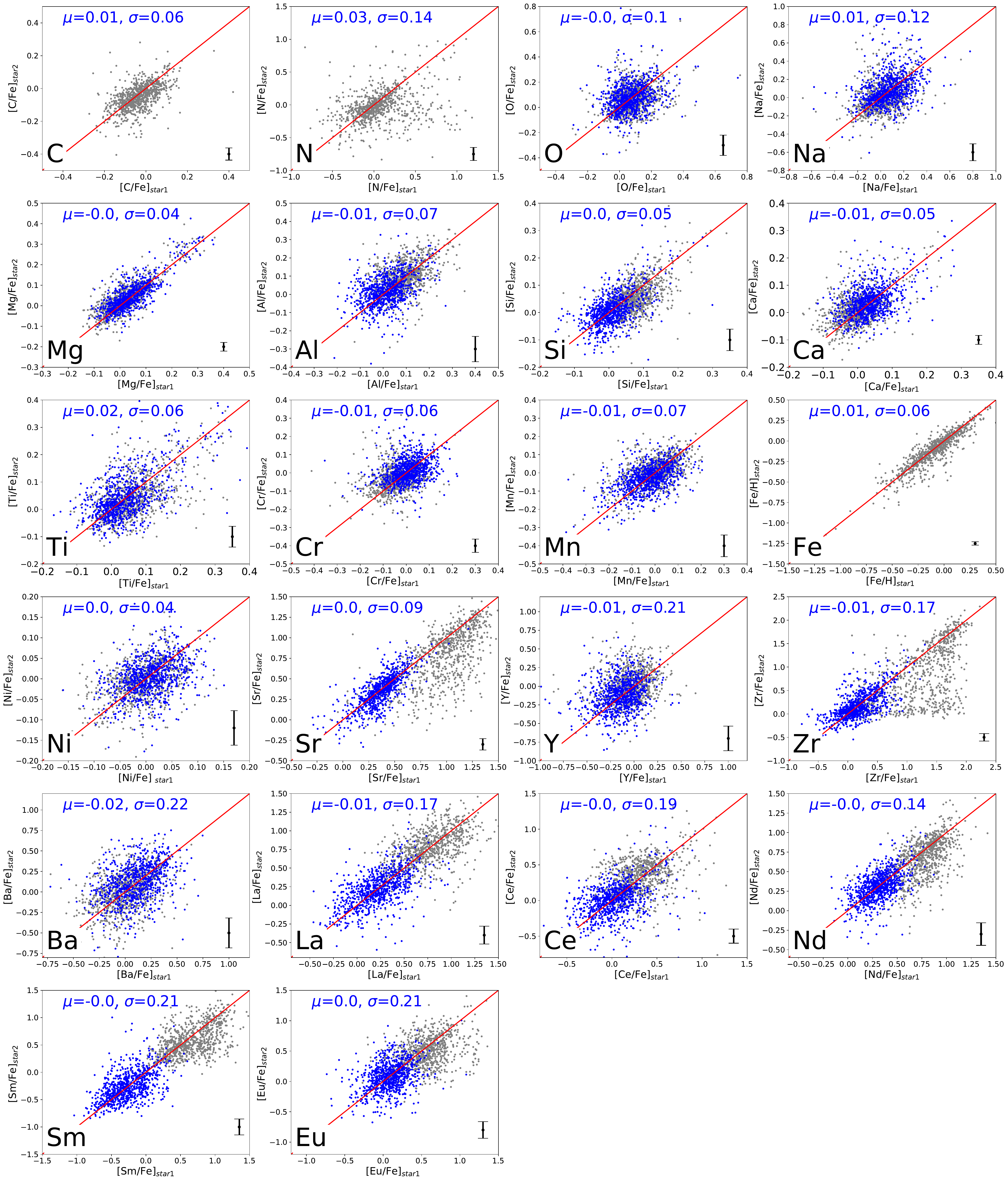}
\caption{Comparison of stellar abundances between the pair components of wide binaries. The grey dots show the abundance determinations before calibration. The blue dots show the abundances after calibration as a function of $T_{\rm eff}$ values. The median and dispersion of the difference for each abundance after calibration is marked in top-left corner. For [C/Fe], [N/Fe] and [Fe/H], no calibration is implemented. The error bar in the bottom-right corner delineates the typical error in the abundance determinations. 
\label{fig_wb}}
\end{figure*}

\begin{figure*}[htb!]
\centering
\includegraphics[width=0.98\textwidth]{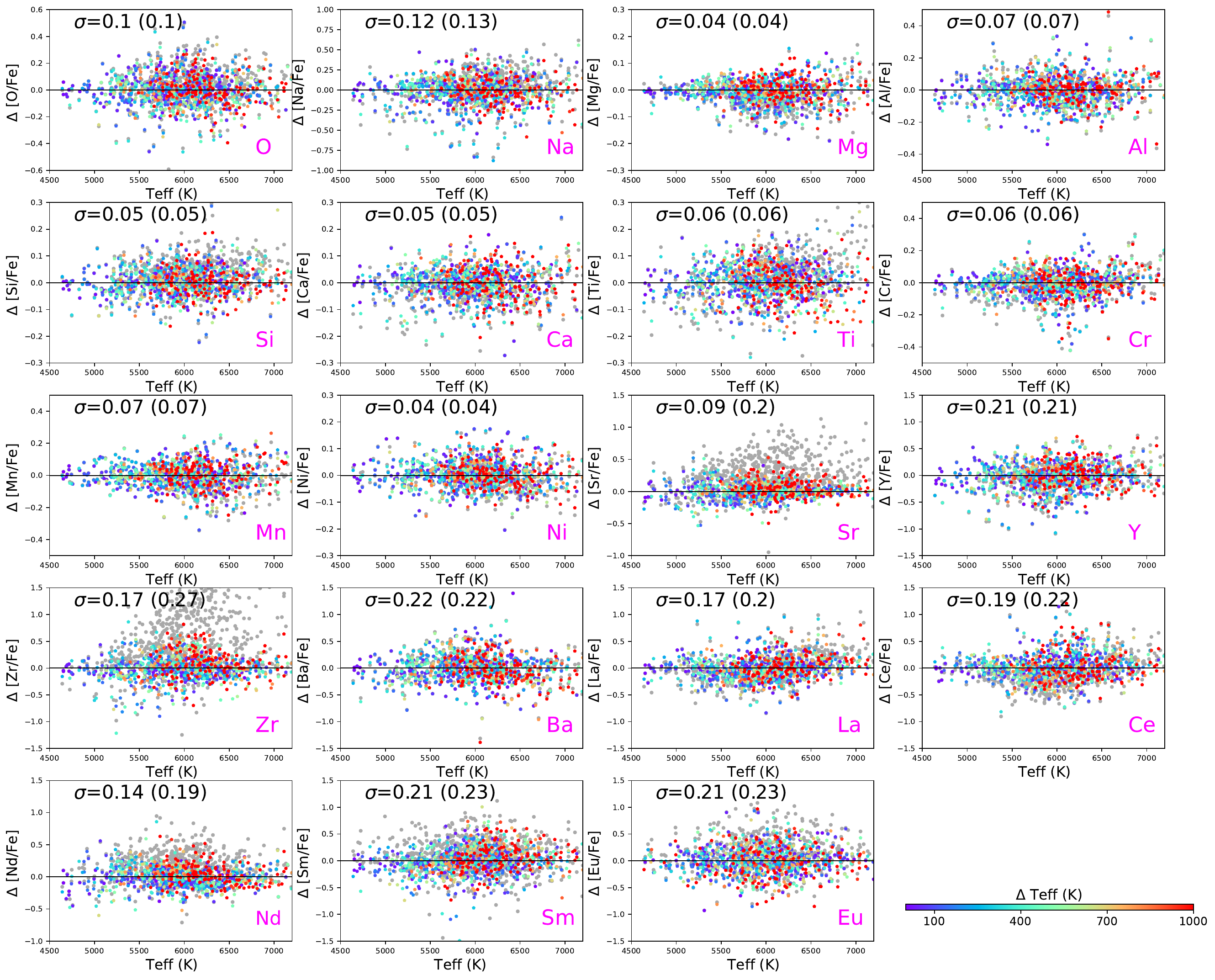}
\caption{The differential abundance between the wide binary component stars, $\Delta{\rm [X/Fe]}$, as a function of $T_{\rm eff}$ of the primary. The grey dots show the results for abundances before calibration, while the dots in colors are results for abundances after calibration, and the colors represent the temperature differences between the primary and the secondary of the wide binaries.
\label{fig_wb_corre}}
\end{figure*}
We use wide binaries to calibrate systematic trend as a function of $T_{\rm eff}$ in our abundance estimates for dwarf stars. Stars in a wide binary system are generally believed to be born simultaneously from the same molecular cloud and thus share the same initial elemental abundances, but they evolve independently following the evolutionary track of single stars. For unevolved main-sequence stars, their surface abundances are almost the same as the initial values, so that the abundances of the binary components should keep almost identical. Depending on the masses, the temperature of the binary components can be very different, so that wide binaries provide a golden source for calibrating possible temperature trend of the abundance determinations. Note that in the current work we do not make any calibration for giant stars due to a number of reasons. First, there are very few giant stars in our wide binary sample to allow for a robust calibration; Second, as giant stars span only a narrow temperature range, any temperature-dependent systematics are expected to be small compared to dwarf stars. Indeed, by looking into the training set, we realize that abundances for dwarf stars suffer much more severe systematics than giants, especially for heavy elements, which can be also seen in the GALAH DR3 paper \citep{Buder2021}.  

\citet{El-Badry2021} built a catalog of 1 million spatially resolved binary stars within about 1\,kpc of the Sun from the $Gaia$ early data release 3 \citep[Gaia EDR3][]{Gaia2021}. A cross-match of this catalog with LAMOST DR9 yields 1793 wide binary systems that both components have a LAMOST spectrum with $S/N>40$. We then select our wide binary calibration sample according to the following criteria,
\begin{align}
    \left\{
    \begin{aligned}
 &S/N_{1} > 40~ {\rm and}~ S/N_{2} > 40,\\
 &{\rm \chi^2\_flag}_{1} <3~{\rm and}~ {\rm \chi^2\_flag}_{2} <3, \\
&   \logg_{1}> 3.8~{\rm and}~ \logg_{2}> 3.8,\\
&  4000\,{\rm K}<T_{\rm eff,1}<6800\,{\rm K}~{\rm and}~ 4000\,{\rm K}<T_{\rm eff,2}<6800\,{\rm K}, \\
&   \dfrac{|RV_{1} - RV_{2}|}{\sqrt{\sigma_{RV,1}^2+\sigma_{RV,2}^2}} < 3,\\
    \end{aligned}
      \right.
\end{align}
where $\chi^2\_{\rm flag}$ is a quality flag of the parameter determination based on $\chi^2$ values of the spectra fitting (Sect.\,5.1). These criteria leave nearly 1000 wide binary systems in our abundances calibration set. 
Figure~\ref{fig_wb} presents a comparison of the {\sc DD-Payne} abundances between the binary component stars. While for the vast majority of elements the binary abundances show a good agreement, there are large deviations in the results for some elements, such as Sr and Zr. A correction of the temperature trend as described below can effectively remove such deviations. 

For the calibration, we first build a temperature-dependent model of the systematic errors, 
\begin{equation}
\begin{aligned}  
\delta_{\rm [X/Fe]}(T_{\rm eff}) =a_{1}T_{\rm eff}^3+a_{2}T_{\rm eff}^2+a_{3}T_{\rm eff}+c,\\
\end{aligned}
\label{E11}
\end{equation}
which is a cubic polynomial function. 
The coefficients {$a_1$, $a_2$, $a_3$} in the model can be estimated using the differential abundances of the binary stars,
\begin{equation}
\begin{aligned}  
 \Delta{\rm [X/Fe]} & := {\rm [X/Fe]}_1 - {\rm [X/Fe]}_2 \\
 & = \delta_{\rm [X/Fe]}(T_{\rm eff,1}) - \delta_{\rm [X/Fe]}(T_{\rm eff,2}) \\
 & = a_{1}(T_{\rm eff,1}^3-T_{\rm eff,2}^3)+a_{2}(T_{\rm eff,1}^2-T_{\rm eff,2}^2)\\
 & +a_{3}(T_{\rm eff,1}-T_{\rm eff,2}).\\
\end{aligned}
\label{E2}
\end{equation}
We adopt the Markov Chain Monte Carlo (MCMC) method in Python {\sc emcee} package \citep{Foreman-Mackey2013} to obtain a best estimate of \{$a_1$, $a_2$, $a_3$\}. The constant term $c$ serves as an absolute zero point, and need external calibration. 

Currently, we are still lack of a good benchmark set of elemental abundances for an accurate determination of $c$. However, since our main goal is to have a self-consistency abundance determination across different temperatures, as a compromise, we opt to tie the abundances of dwarfs to the abundances of giants using the M67 member stars. Specifically, we take the average DD-Payne abundances of M67 giant stars as a reference, and derive $c$ by tying the average abundances of M67 dwarfs to the reference values. In doing so, we assume that dwarfs and giants in M67 should have the same abundance values. This is a good assumption for most of the elements but not for C and N, which may suffer from non-negligible alternation due to dredge-up processes of giant stars. We therefore decide not to do any calibration for [C/Fe] and [N/Fe], which is reasonable as we do not observe strong temperature-trend for these elements. Similarly, stellar evolution in the main-sequence phase may suffer non-negligible atomic diffusion effect \citep[e.g.,][]{Korn2007,Bertelli_Motta2018, GaoXD2018, Souto2018, Souto2019, Semenova2020}. This effect may cause an intrinsic difference of 0.1~dex level between the wide binary component stars, depending on their stellar masses. We therefore opt not to do any calibration for [Fe/H] as well. 


Figure~\ref{fig_wb_corre} shows the differential abundance between the wide binary component stars, $\Delta{\rm [X/Fe]}$, as a function of $T_{\rm eff}$ of the primary, i.e., the one with higher $T_{\rm eff}$.  
For many of the elements, including O, Al, Si, Ca, Ti, Cr, Mn, Ni, and Y, $\Delta{\rm [X/Fe]}$ shows little trend with temperature. The calibration thus only cause a negligible change to the abundance estimates, and the overall dispersion of $\Delta{\rm [X/Fe]}$ before and after the calibration is almost identical. For some elements, such as Sr, Zr, and Nd, $\Delta{\rm [X/Fe]}$ shows strong trend with temperature. The calibration significantly improves the abundance estimates. The dispersion of $\Delta{\rm [Sr/Fe]}$ decreases to 0.1~dex after the calibration from a value of 0.2~dex before the calibration. The dispersion of $\Delta{\rm [Zr/Fe]}$ decreases from 0.28~dex to 0.21~dex, while that of $\Delta{\rm [Nd/Fe]}$ decreases from 0.20~dex to 0.13~dex for results before and after the calibration. The Figure also illustrates that our cubic polynomial model of $\delta_{\rm [X/Fe]}(T_{\rm eff})$ is a good description of the temperature-trend in the abundance estimates, as wide binaries with all temperature differences (indicated by colors in the Figure) between the components stars show a flat $\Delta{\rm [X/Fe]}$ distribution after the calibration.  

However, as already shown in Figure~\ref{fig_wb}, for many of the elements such as Al, Si, Sr, Zr, Ba, La, Ce, Nd, Sm, Eu, the calibration significantly changed the mean value of the abundance estimates through the zero point coefficient $c$. Many of these elements especially for heavy ones such as Sr, Zr, La, Nd, Sm, Eu have too high abundance values before the calibration, we suspect it is due to systematic difference in the GALAH DR3 training labels for dwarfs. This is also presented in the abundances of solar twins (Sect.\,4.5). Note that the abundances shown in this section refers to that in the recommend data set (Sect.\,5).

\subsection{Abundances of solar twins}
\begin{figure*}[htb!]
\centering
\includegraphics[width=0.9\textwidth]{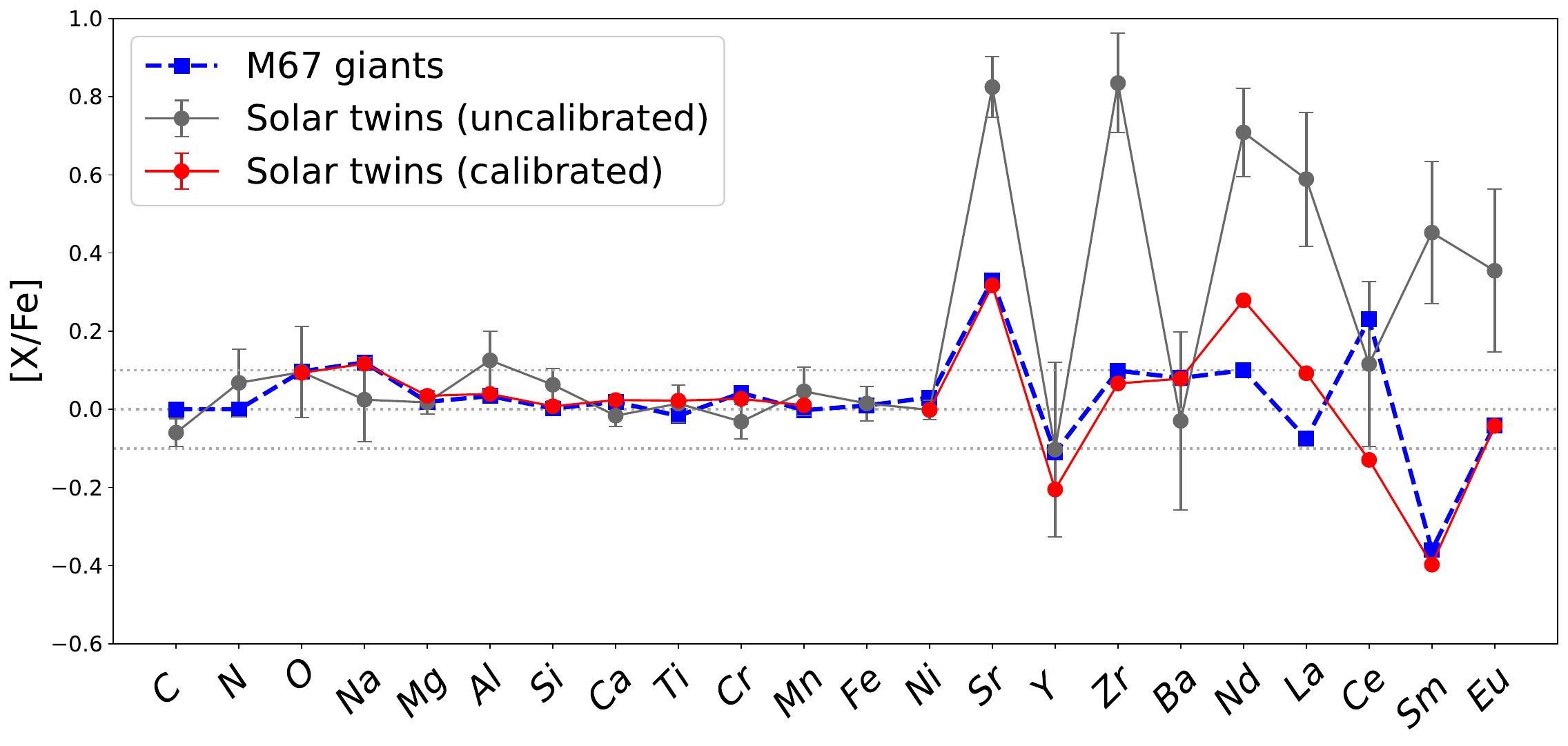}
\caption{Abundance patterns of solar twins for {\sc DD-Payne} abundances before (grey) and after calibration (red). Solar twins are selected based on the atmospheric parameters using criteria listed in Equation (12). The error bar delineates the median measurement error of the individual solar twins. The blue dotted line shows the median abundance values for red giant stars in M67, which are adopted as the zero points of our abundance calibration (see text).
\label{fig_solar_twin}}
\end{figure*}
We define solar twins in LAMOST DR9 based on their atmospheric parameters using the criteria below
\begin{align}
    \left\{
    \begin{aligned}
&  5727\,{\rm K}<T_{\rm eff}<5827, \\
&  4.39<\logg<4.49,\\
&  -0.05 < {\rm [Fe/H]} < 0.05,\\
    \end{aligned}
      \right.
\end{align}
All stars selected are required to have spectral S/N higher than 40. Fig.\,\ref{fig_solar_twin} shows the abundance patterns of these solar twins. 
For all elements with atomic number no larger than that of Ni, the [X/Fe] values are within $\pm0.1~dex$ from zero, even for the uncalibrated values. However, for heavier elements, the uncalibrated estimates may show a huge scatter, with particularly high values for Sr, Zr, Nd, La, Sm, and Eu. The abundance calibration using wide binaries significantly reduced the scatter. For both the s-/r-processes elements and relatively lighter ones such as Al, Si, Cr, and Mn, the abundance values after calibration are much closer to zero compared to the uncalibrated ones.    
\section{The {\sc DD-Payne} catalog for LAMOST DR9}
We apply {\sc DD-Payne} to the LAMOST DR9 low-resolution spectra data set. Due to the limitation of the training set, we only determinate labels for FGK stars and M giants. By doing so, we make use of the classification of the LAMOST pipeline \citep{LuoAL2015}, and select spectra classified as F, G, and K types, as well as M giants, but discarding spectra of other types such as O, B, A types or M dwarfs. Ultimately, our sample for LAMOST DR9 contains  spectra of 6439,359 unique stars. We derive their stellar labels using DD-Payne models trained on all the 11 training sets listed in Table\,\ref{table1}. 

We then combine label estimates derived from these 11 training sets. The training set used for each label in the recommended catalog is listed Table~\ref{table2}. Results from the LAMOST-APOGEE training set (Training set 1) are adopted for the atmospheric parameters $T_{\rm eff}$, $\log~g$, $v_{\rm mic}$, as well as most elements with atomic number no larger than Ni. Two exceptions are Ti and Cr, which we adopt estimates from the LAMOST-GALAH training set (Training set 3) as the APOGEE abundances for elements exhibit some artificial trend in the [X/Fe]-[Fe/H] plane (Figure~3). For the other elements, only results trained on GALAH labels are available. Furthermore, we discard stars that have large differences in $T_{\rm eff}$ between results derived using the LAMOST-APOGEE (Training set 1) and LAMOST-GALAH (Training set 3) training sets. We only reserve stars that the two temperature estimates are consistent within $3\sigma$, where $\sigma$ is the square root of the quadratic sum of temperature errors.  

Finally, we obtain a catalog of 25 stellar labels, including atmospheric parameters $T_{\rm eff}$, $\log\,g$, ${\rm [Fe/H]}$, and $v_{\rm mic}$, as well as elemental abundance ratio [X/Fe], with X being C, N, O, Na, Mg, Al, Si, Ca, Ti, Cr, Mn, Ni, Sr, Y, Zr, Ba, La, Ce, Nd, Sm, and Eu. The atmospheric parameters and abundance ratio for 13 elements, namely C, N, O, Na, Mg, Al, Si, Ca, Ti, Cr, Mn, Ni, and Ba, are valid for 6439,359 stars ( 9245,463 spectra) from LAMOST DR9. While the abundance ratios for Sr, Y, Zr, La, Ce, Nd, Sm, and Eu are provided only for about 3.6 million stars with S/N higher than 20.

For convenience, we provide two separate tables in {\sc .FITS} format, one for recommended values of the labels, after applying for the calibrations in Sect.\,4. The other is for original abundance measurements without calibrations. Table~\ref{table3} and Table~\ref{table4} present a description of the columns in the these catalogs. The catalogs are available in Zenodo, \url{https://zenodo.org/records/15254859}, in fits format. They can be found in the NADC, \url{https://nadc.china-vo.org/res/r101600}. Future updates to the catalogs, including new application to LAMOST DR12 with the same {\sc DD-Payne} pipeline sets, are presumed to made public via this NADC link.

\begin{table*}
\caption{Training Sets for the Recommended DD-Payne Stellar Labels}
\label{table2}
\begin{tabular}{ccccccccccccccccc}
\hline         
 Label &$T_{\rm eff}$ & $\log\,g$ & Fe & $v_{\rm mic}$ & C & N &O & Na & Mg & Al& Si& Ca &Ti &Cr & Mn & Ni  \\
\hline
Training set No. & 1 & 1  &1 & 1  &1  &2 &1&3&1&2& 1 &1 &3&3 &2 & 1 \\
\hline 
 Label & Sr & Y &Zr & Ba & La& Ce & Nd & Sm & Eu& & & & & & \\
\hline 
Training set  No. &4 & 5 & 6 & 3 &7  &8 &9&10 &11 & & & &  & &\\
\hline 
\end{tabular}
\end{table*}

\begin{table*}
\caption{Descriptions of the main recommended catalog.}
\label{table3}
\begin{tabular}{lc}
\hline
 Field &    Description  \\
\hline        
SPECID & LAMOST spectral ID \\
RA &   Right ascension from the LAMOST catalog (deg; J2000)\\
Dec  &  Declination from the LAMOST catalog (deg; J2000)  \\
UNIQFLAG & Flag of the repeat visit spectra, 1 means unique star; $>1$ means repeat visits\\
S/N\_{u/g/r/i/z}&  median spectral signal-to-noise ratios in u/g/r/i/z bands \\
$T_{\rm eff}$ &  Effective temperature (K)   \\
$T_{\rm eff}$\_{err}  & Uncertainty in $T_{\rm eff}$ (K)\\
$\log g$ &  Surface gravity  \\
$\log g$\_err & Uncertainty in $\log g$  \\
{\rm [Fe/H]} & Iron abundance \\
{\rm [Fe/H]\_{LTE}} & Iron abundance without NLTE correction \\
{\rm [Fe/H]}\_{err} &  Uncertainty in {\rm [Fe/H]} (dex)  \\
{\rm [X/Fe]} &  Element-to-iron abundance ratio of the recommended values\\
{\rm [X/Fe]}\_{err} & Uncertainty in {\rm [X/Fe]} (dex) \\
$V_{\rm mic}$  & Micro-turbulent velocity (km/s)  \\
$V_{\rm mic}$\_{err} &  Uncertainty in $V_{\rm mic}$ (km/s)  \\
Chisq\_LAM\_APO  &  The $\chi^2$ of the spectral fit using LAMOST-APOGEE training set\\
Chisqflag\_LAM\_APO  & A quality flag describing the $\chi^2$ of the spectral fit\\
     & using LAMOST-APOGEE training set, the smaller the better \\
Chisq\_LAM\_GAL  &  The $\chi^2$ of the spectral fit using LAMOST-GALAH training set\\
Chisqflag\_LAM\_GAL  & A quality flag describing the $\chi^2$ of the spectral fit \\
                    & using LAMOST-GALAH training set, the smaller the better \\ 
Flag\_X\_Fe & A flag describing the X\_Fe estimation quality based on the adopted training set\\ 
Chisq\_X & The $\chi^2$ of the spectral fit using different training sets in  Table\, \ref{table1}\\
\hline
\end{tabular}
\end{table*}

\begin{table*}
\caption{Descriptions of the abundance catalog without calibration.}
\begin{tabular}{cc}
\hline
Field                              &          Description   \\
\hline        
SPECID                             &              LAMOST spectral ID \\
{\rm [X/Fe]}\_${\rm uncali}$        &                 Element-to-iron abundance ratio without correcting for the trend with $T_{\rm eff}$       \\
\hline
\label{table4}
\end{tabular}
\end{table*}





\subsection{data flag}
In our sample there are some spectra with poor model fits due to either data artifacts or anomaly spectral features. In order to identify these outliers conveniently, we introduce a quality flag ``qflag\_$\chi^2$" based on the $\chi^2$ of the spectral fitting, defined as 
\begin{align}
     {\rm qflag}\_\chi^2 = \frac{\chi^2 - \chi^2_{\rm median}}{\chi^2_{\rm mad}},
\end{align}
where the $\chi_{\rm median}^2$ is the median value of $\chi^2$ for stars of similar spectral S/N, while $\chi_{\rm mad}^2$ is the mean absolute deviation from the median value. A larger value of qflag\_$\chi^2$ means a less good fit. 

In the following analysis, we take ${\rm qflag}\_\chi^2\leq3.0$ as a criterion to select label estimates with good quality, utilizing  ${\rm qflag}\_\chi^2$ derived from the LAMOST-APOGEE and LAMOST-GALAH training sets (Training sets 1 and 3). We eliminate labels with ${\rm qflag}\_\chi^2>3.0$ for qflag\_$\chi^2$ from either Training set 1 or Training sets 3. 

For many elements, the abundance measurements for metal-poor stars of ${\rm Fe/H]}\lesssim -2.0$ are inaccurate due to the lack of training sample (Sect.\,5.3). We therefore set a quality flag for each of the elemental abundance (Table~\ref{table3}) to mark this effect. The abundance should be used with caution if the flag\_X\_Fe value is none zero.


\subsection{$T_{\rm eff}$--$\logg$ and $T_{\rm eff}$--$\feh$ diagrams}
\begin{figure*}[htb!]
\centering
\includegraphics[width=\textwidth]{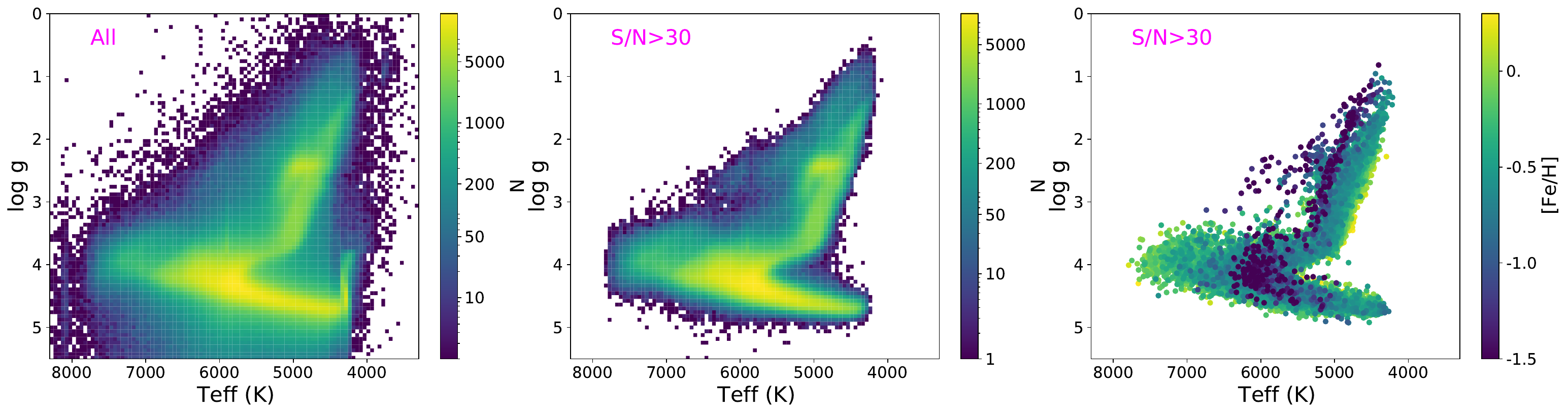}
\caption{Stellar number density distribution in the $T_{\rm eff}$--$\log\,g$ diagram for the {\sc DD-Payne} results. The left panel is for all the LAMOST DR9 sample stars, while the middle panel is for stars with $S/N>30$. The right panel shows the $T_{\rm eff}$-$\log~g$ distribution of a randomly selected sub-sample of 25,000 stars, color-coded by their metallicity.
\label{fig6a}}
\end{figure*}

\begin{figure*}[htb!]
\centering
\includegraphics[width=1.\textwidth]{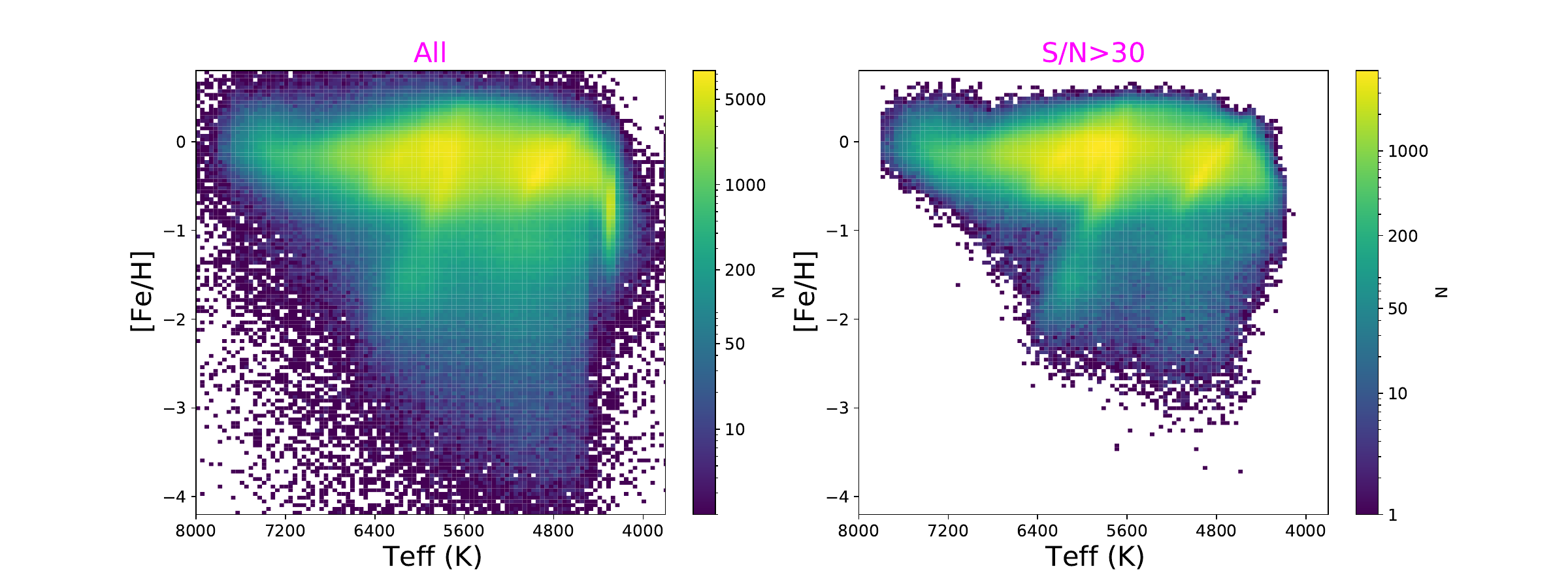}
\caption{Stellar number density distribution in the $T_{\rm eff}$--{\rm [Fe/H]} plane for the {\sc DD-Payne} results. The left and right panels are all sample stars and sample stars with $S/N>30$, respectively.
\label{fig6b}}
\end{figure*}

The left panel of Fig\,\ref{fig6a} shows the stellar number density distribution in the $T_{\rm eff}$--$\log~g$ diagram for our sample stars of all spectral S/Ns. The sample stars span the parameter space of $3500$~K $\lesssim T_{\rm eff}\lesssim 8000$\,K, $0 \lesssim \log~g \lesssim 6$. For stars with $\log~g \gtrsim 5$, their $\log~g$ should be erroneously estimated as a consequence of large measurement error due to due to low spectral S/N. These stars are discarded after setting a cut of $S/N>30$ (the middle panel). Also, for stars with $T_{\rm eff}<4300$~K and $\log~g>3$, their $\log~g$ values are systematically underestimated due to a lack of training set in this regime. Nonetheless, the main-sequence stars, main-sequence turn-offs, subgiants, core-helium-burning sequence stars (red clumps and red horizontal branch stars) are clearly visible in the figures as robust features. The right panel of Fig\,\ref{fig6a} shows a clear {\rm [Fe/H]} trend that the more metal-poor giant stars exhibit higher temperature, which is expected. 
Fig.\,\ref{fig6b} shows the stellar distribution in the $T_{\rm eff}$--[Fe/H] plane. The full sample stars cover a metallicity range of $-4 \lesssim \feh \lesssim0.5$~dex (the left sample). Most of the very metal-poor stars of $\feh\lesssim-2.5$~dex are faint halo stars with low spectral S/N, especially for dwarfs. They would thus escape from the sample if a $S/N>30$ cut is set (the right panel). The figures show two prominent ridged sequences. For both them, stars with lower [Fe/H] exhibit higher temperature. The ridge with hotter temperature of $5600\lesssim \teff \lesssim6400$~K spans the full [Fe/H] range. This ridge is mainly composed of main-sequence turn-off stars of the high-$\alpha$, old disk. The tight $\teff$ range at a fixed metallicity is due to the fact that the stars have almost an identical age of about 11~Gyr \citep{XiangMS2022}. The ridge with cooler temperature of $4600\lesssim \teff \lesssim5100$~K occurs mainly for stars with $\feh\gtrsim-1$. This ridge is mainly composed of red clump stars of the Galactic disk. 


\subsection{Stellar elemental abundances}
\begin{figure*}[htb!]
\centering
\includegraphics[width=0.99\textwidth]{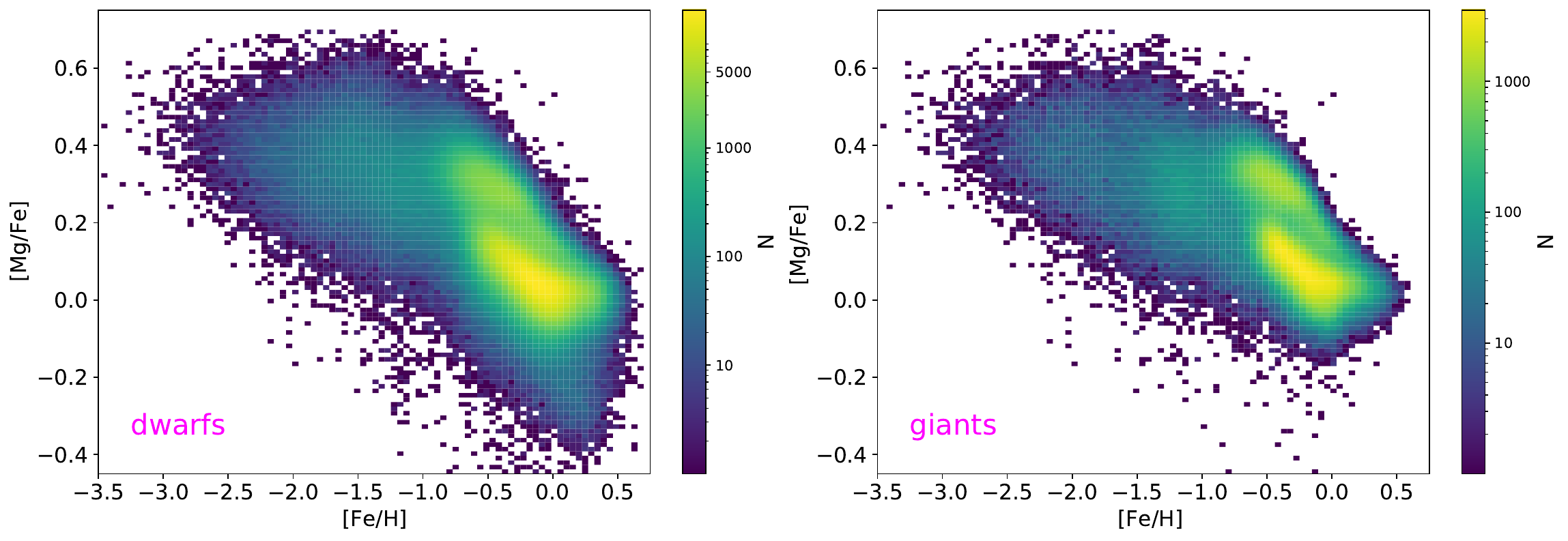}
\caption{Stellar number density distributions in the [Fe/H]--[Mg/Fe] plane for dwarf stars (left) and for giant stars (right). In both panels only stars with spectral $S/N>30$ are shown.
\label{fig6c}}
\end{figure*}
The [Mg/Fe] has been most frequently used to distinguish different major Galactic stellar populations. The LAMOST DD-Payne has achieved a high precision for the [Mg/Fe] estimate, with a median measurement error of only 0.03~dex for the sample stars with $S/N>50$. Fig.\,\ref{fig6c} shows the stellar density distribution in the [Mg/Fe]--[Fe/H] plane for our sample stars of $S/N>30$ and $\vert b \vert>30^\circ$. The latter selection criterion is set to balance the relative stellar number density of the Galactic low-$\alpha$ and high-$\alpha$ disks. At the metal-rich part of $\feh \gtrsim -1$, the high-$\alpha$ sequence is clearly distinguishable from the low-$\alpha$ sequence, confirming the high precision of the [Mg/Fe] estimates. For dwarf stars (the left panel), there is also a tile of Mg-depleted stars with ${\rm [Mg/Fe]}<0$ that is not presented in the giant sample (the right panel). These stars are mostly chemically peculiar A/F type stars with intermediate stellar mass \citep[e.g.][]{XiangMS2020, ZhangM2023}. 

At the metal-poor side, the giant sample clearly shows a low-$\alpha$ sequence in the range of $-1.8\lesssim \feh\lesssim-0.6$, which is well known as the accreted halo population of Gaia-Enceladus-Sausage \citep[GES;][]{Belokurov2018, Helmi2018, Helmi2020}. This GES sequence is, however, not presented in the dwarf sample. We believe this is mainly because our sample contains only a small number of dwarf stars from GES, as they are too faint to be targeted by the survey. The metal-poor side of $\feh\lesssim-1$ also contains a high-$\alpha$ population, which is an extension of the high-$\alpha$ disk to lower metallicity \citep[e.g.][]{XiangMS2024}. Finally, stars with $\feh\lesssim-2$ may be a complicate mixing population of various merger galaxies \citep[e.g.][]{Myeong2018} as well as in-situ stars \citep{Belokurov2022, Rix2022}.   



\begin{figure*}[hb!]
 \centering
 \includegraphics[width=0.9\textwidth]{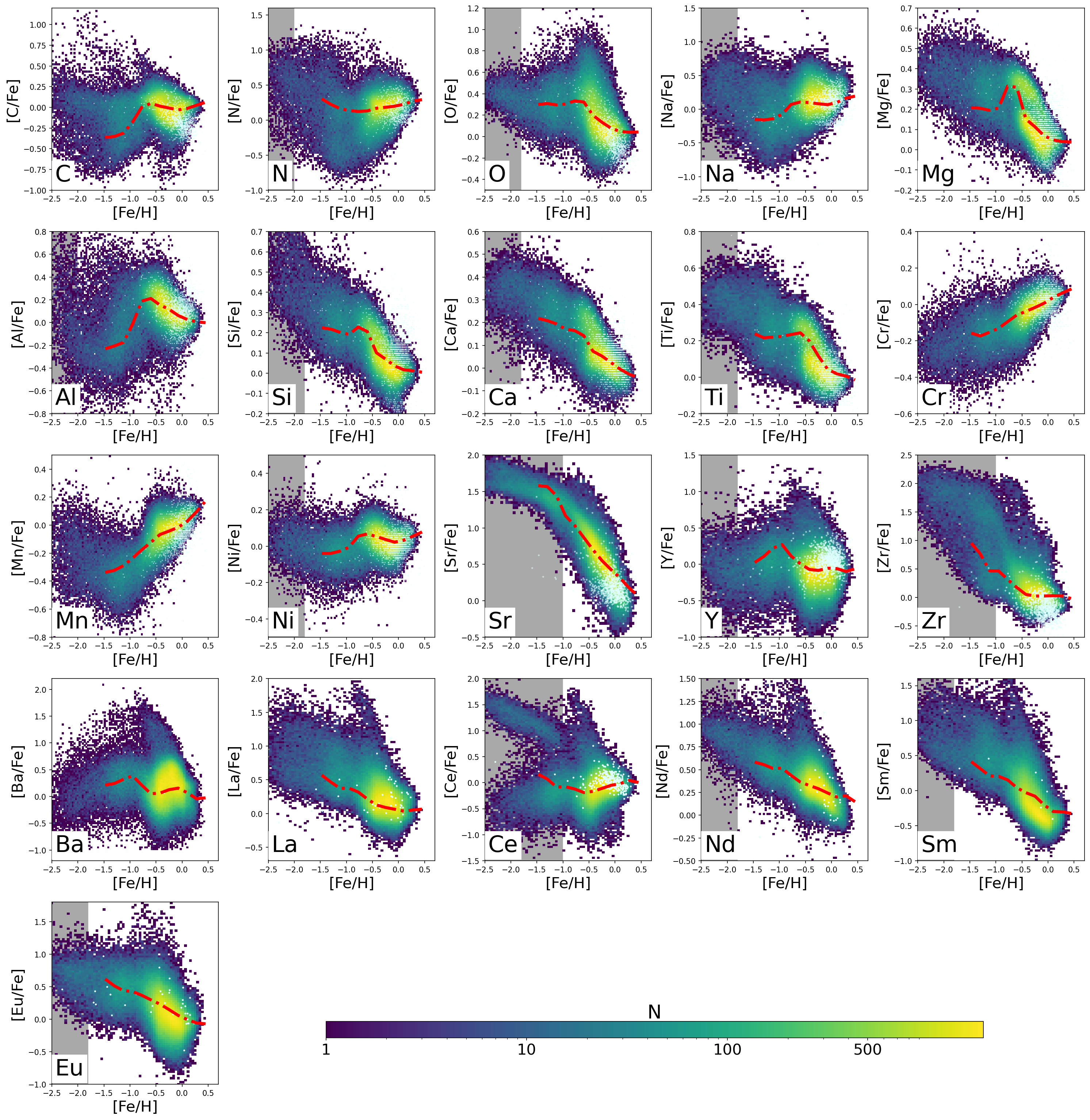 }
 \caption{Stellar number density distributions in the [X/Fe]--[Fe/H] planes for LAMOST giants with $S/N>50$. The red dotted line shows the mean trend giant stars in either APOGEE DR17 or GALAH DR3, depending on which is adopted as the source of the training set.  For elemental abundances of N, O, Na, Al, Si, Ca, Ti, Mn, Ni, Sr, Y, Zr, Ce, Nd, Sm, and Eu, the metal-poor regime is marked with grey shadow as there are few metal-poor stars in the training sample. Also shown in white dots are labels from the Hypita \citep{Hinkel2014} high-resolution database. 
 \label{fig7b}}
 \end{figure*}
 
\begin{figure*}[hb!]
 \centering
 \includegraphics[width=0.9\textwidth]{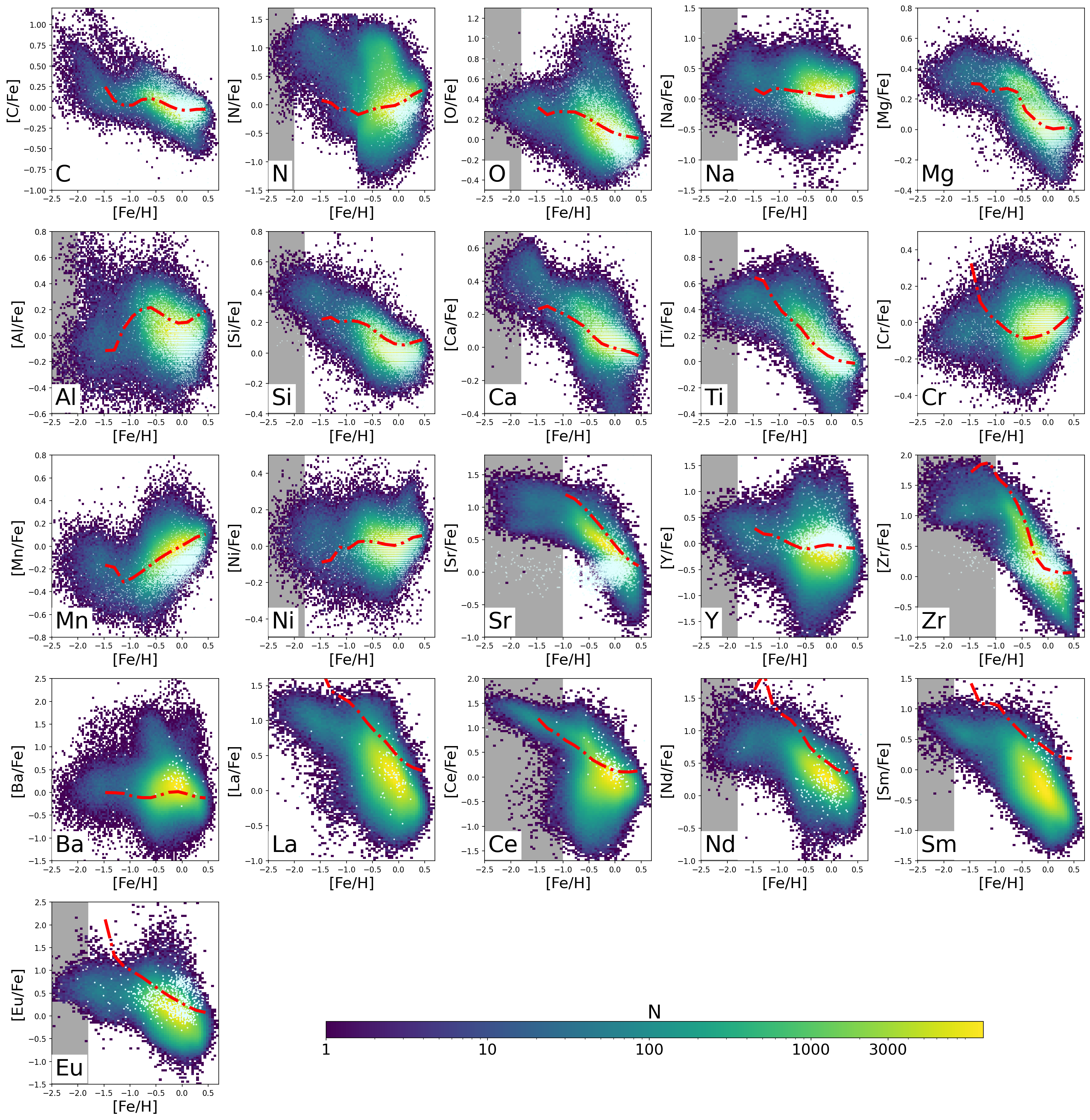 }
 \caption{Stellar number density distributions in the [X/Fe]--[Fe/H] planes for LAMOST dwarfs with $S/N>50$. Except for C and N, the abundances refer to those after calibration. The red dotted line shows the mean trend of dwarf stars in either APOGEE DR17 or GALAH DR3, depending on which is adopted as the source of the training set. For some elements such as Sr, La, Nd, Sm, and Eu, there are large offset between the red dotted line and the density peak of the LAMOST sample. This is attribute to the abundance calibration implemented in the current work. The grey shaded regimes mark the area where there are few metal-poor stars in the training set. Also shown in white dots are labels from the Hypita \citep{Hinkel2014} high-resolution database. 
 \label{fig7a}}
 \end{figure*}
 
Figs\,\ref{fig7b} and Fig\,\ref{fig7a} show, for giants and dwarfs, respectively, the stellar density distribution in the [X/Fe]-[Fe/H] plane for all the elements. For giant stars, the overall [X/Fe]-[Fe/H] trend is consistent well with the high-resolution (APOGEE/GALAH) results for most of the elements. For dwarfs, however, our [X/Fe] values for a number of heavy elements such as Sr, Zr, La, Md, Sm, and Eu can be substantially lower than GALAH DR3, which is due to the abundance calibration as presented in Sect.\,4.  

Beyond the overall trend, intriguing features are immediately visible for most of the individual elemental abundances. The [C/Fe] for the relatively metal-poor giant stars ($\feh\lesssim-1$) exhibit a broad distribution. While the high-[C/Fe] stars are presented for both giants and dwarfs,  the low-[C/Fe] stars are only giants. This is because the latter is a consequence of the dredge-up process of stellar evolution only for evolved stars. The [N/Fe] for giants are systematically higher than for dwarfs, which is likely also a consequence of the dredge-up process. However, there is a notable population of high-[N/Fe] dwarf stars in the range of  $\feh\gtrsim-1$. We also found a $T_{\rm eff}$-dependent trend of [N/Fe] values. Currently we are not clear about their nature, but we note that such stars also presented in the APOGEE DR17.

The abundances of $\alpha$ elements Mg, Si, Ca, and Ti show bimodal features for giant stars. Among them, the bimodal feature for [Mg/Fe] of giant stars is especially clear, probably an intrinsic feature reflecting different enrichment history for these individual $\alpha$ elements, as well as for dwarfs and giants. Although oxygen is also an $\alpha$ element, we do not observe a clear bimodal feature for [O/Fe], possibly because of its larger measurement error compared to other $\alpha$ elements.         

The Figures show a clear s-process enhanced giant star population at the metallicity range of $-1\lesssim\feh\lesssim-0.5$, with ${\rm [X/Fe]}\gtrsim1$ for many of the s-process elements, such as Y, Zr, Ba, La, Ce, and Nd. Such s-process enhanced stars also occur in the results of dwarfs for Y, Ba, and La. S-process enhanced stars in LAMOST have been studied in previous work utilizing their [Ba/Fe] \citep[][]{XiangMS2020, Norfolk2019, ZhangM2024}. Here our results provide abundances for many more elements, thus allow a more detailed study on their nature. The Zr-enhanced population is especially a strong feature for giant stars,  while there is no counterpart of dwarfs in the range of $\feh\gtrsim-1$. This is due to the contribution of a substantial population of S-type stars \citep{Smith1990,Keenan1954} in the giant sample. 

For dwarf stars, there is also a Ba-enhanced population with $\feh\gtrsim-0.2$, which has been found to be consequence of stellar internal element transport due to radiative acceleration in intermediate-mass stars \citep{XiangMS2020}. For these stars, their [C/Fe], [Mg/Fe], and [Ca/Fe] can be substantially lower, but their [Ni/Fe] can be higher \citep{Michaud2015}. All of these predicted features are seen in our data. Note that as mentioned above (Sect.\,3), for many of the elements such as O, Na, Al, Si, Ti, Sr, Y, Ce, Nd, Sm, and Eu, our abundance estimates at the metal-poor end are problematic due to the lack of appropriate training labels. We have marked those regime in the Figures with shaded area. For stars in these shaded regimes, we set a non-zero value in their abundance flags (Sect.\, 5.1).

\subsection{Abundance dispersion examined with star clusters}
 \begin{figure}[htb!]
 \centering
 \includegraphics[width=0.48\textwidth]{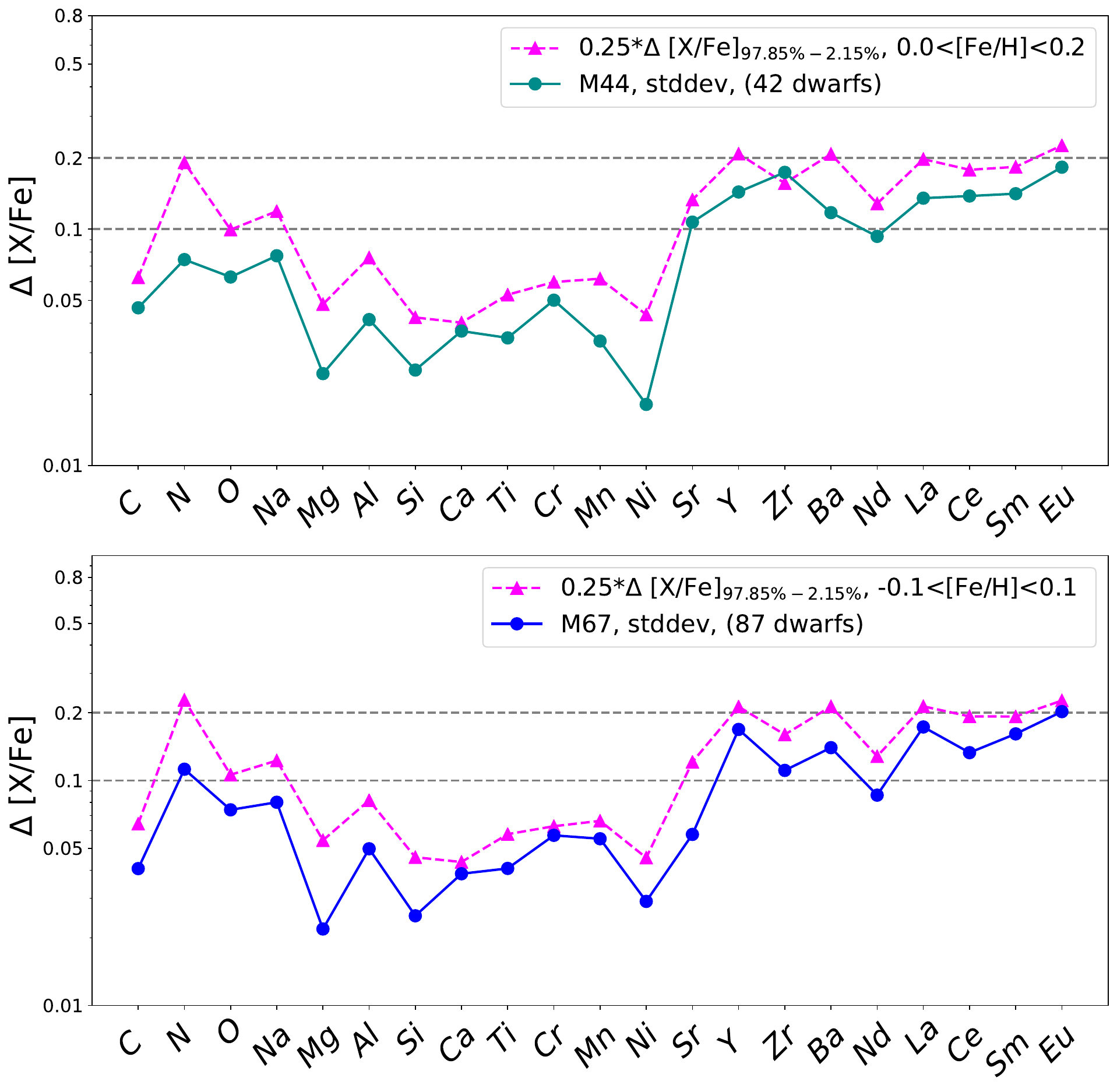}
 \caption{Comparison of abundance dispersion between member stars of open cluster and field stars with the same metallicity as the cluster. The top and bottom panels show results for M44 (42 dwarfs) and M67 (87 dwarfs), respectively. As marked in the figure, the abundance dispersion of the cluster member stars is computed as the standard deviation, after clipping outliers using $3\sigma$ criterion, while the dispersion of the field stars is computed from their abundance distribution by taking a quarter of the 97.5\%-2.5\% width. 
 \label{fig8}}
 \end{figure}

The capacity of the elemental abundances for disentangling different stellar populations depends on the relative scale of the measurement error to the intrinsic abundance scatter. To examine this capacity, we compare the dispersion of the [X/Fe] distribution between the field stars and cluster member stars for the same metallicity. Stars in a cluster share similar initial elemental abundances as they were most likely born from same molecular cloud, their abundance scatter is thus a good measure of the measurement error.

We select cluster member stars utilizing celestial coordinates, proper motion, and parallax from the Early data release of the third $Gaia$ mission \citep[Gaia eDR3;][]{Gaia2021}. By requiring a LAMOST spectral S/N higher than 40, we find large samples of member stars for two open clusters, namely M44 ($\feh\simeq0.1$) and M67 ($\feh\simeq0$). To compute the abundance scatter of cluster member stars, we first discard measurement outliers that deviate from the median abundance by more than $3\sigma$, where $\sigma$ is the dispersion of a Gaussian fit to the abundance distribution. We then take the standard deviation of the [X/Fe] for the remaining member stars as the scatter value. 

For the dispersion of the field stars in our sample, we first define a quantity $\Delta[X/Fe]_{\rm 97.5\%-2.5\%}$, which is the abundance difference between the 97.5th percentile and the 2.5th percentile of the abundance distribution. We the take a quarter of the $\Delta[X/Fe]_{\rm 97.5\%-2.5\%}$ value as a measure of the abundance dispersion of the field stars. This dispersion is strictly the same as the 1$\sigma$ value for a Gaussian distribution.

Fig.\,\ref{fig8} shows that for the M44, and M67, their abundance scatters are substantially smaller than the field stars for almost all the elements. For some elements, the intrinsic scatter, estimated as the square root difference in the dispersion between the field stars and cluster member stars, is at a level comparable to the scatter of the cluster member stars, i.e., the measurement error. But the scatters of member stars for many of the heavier elements such as Y, La, Ce, and Sm, are also smaller than the intrinsic dispersion of the field stars. 



\subsection{Comparison measurement errors with Cr\'amer-Rao bounds}
\begin{figure*}[htb!]
\centering
\subfigure
{
	\begin{minipage}{0.97\linewidth}
	\centering  
	\includegraphics[width=0.97\columnwidth]{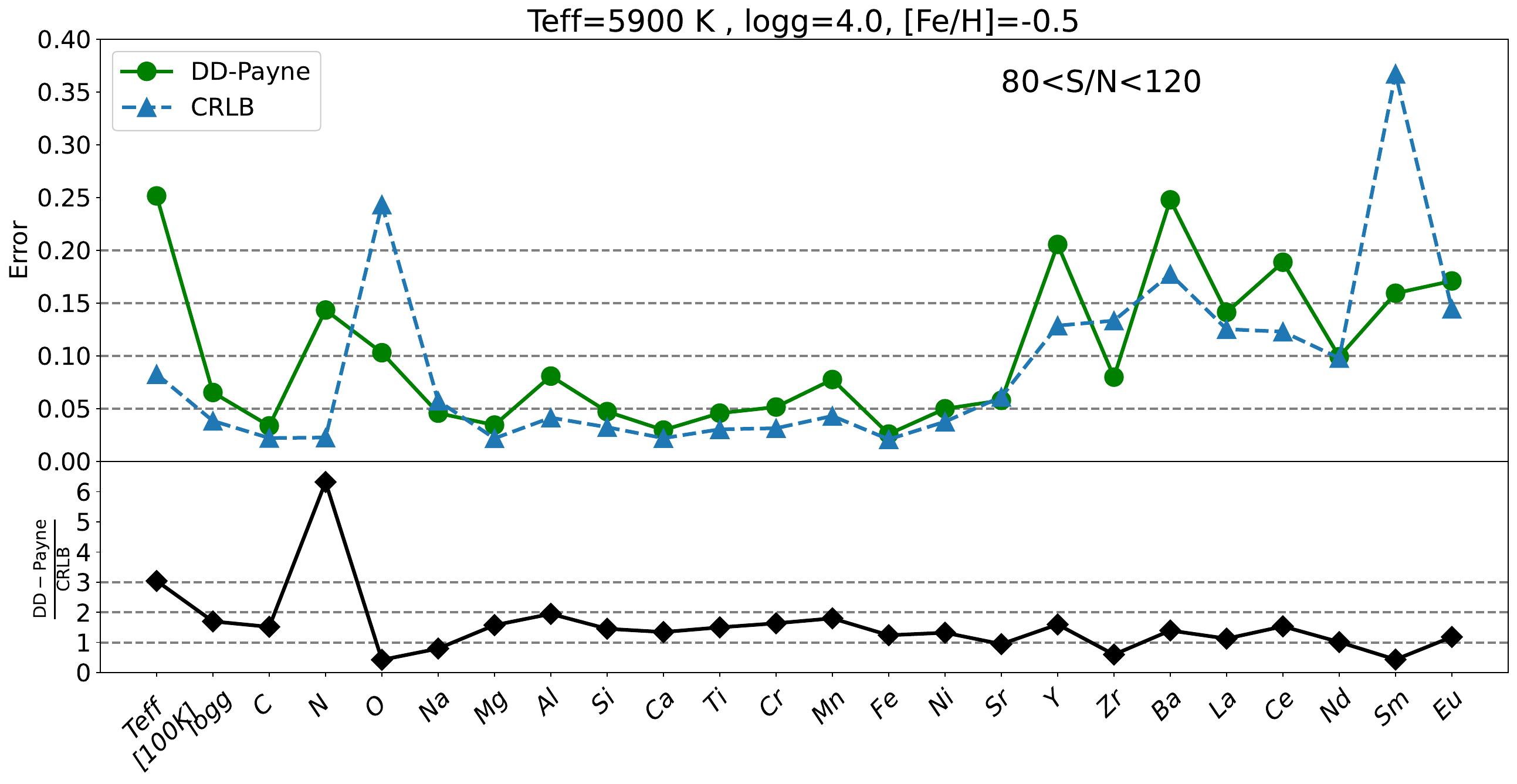}
	\end{minipage}
} 
\subfigure
{
	\begin{minipage}{0.97\linewidth}
	\centering     
	\includegraphics[width=0.95\columnwidth]{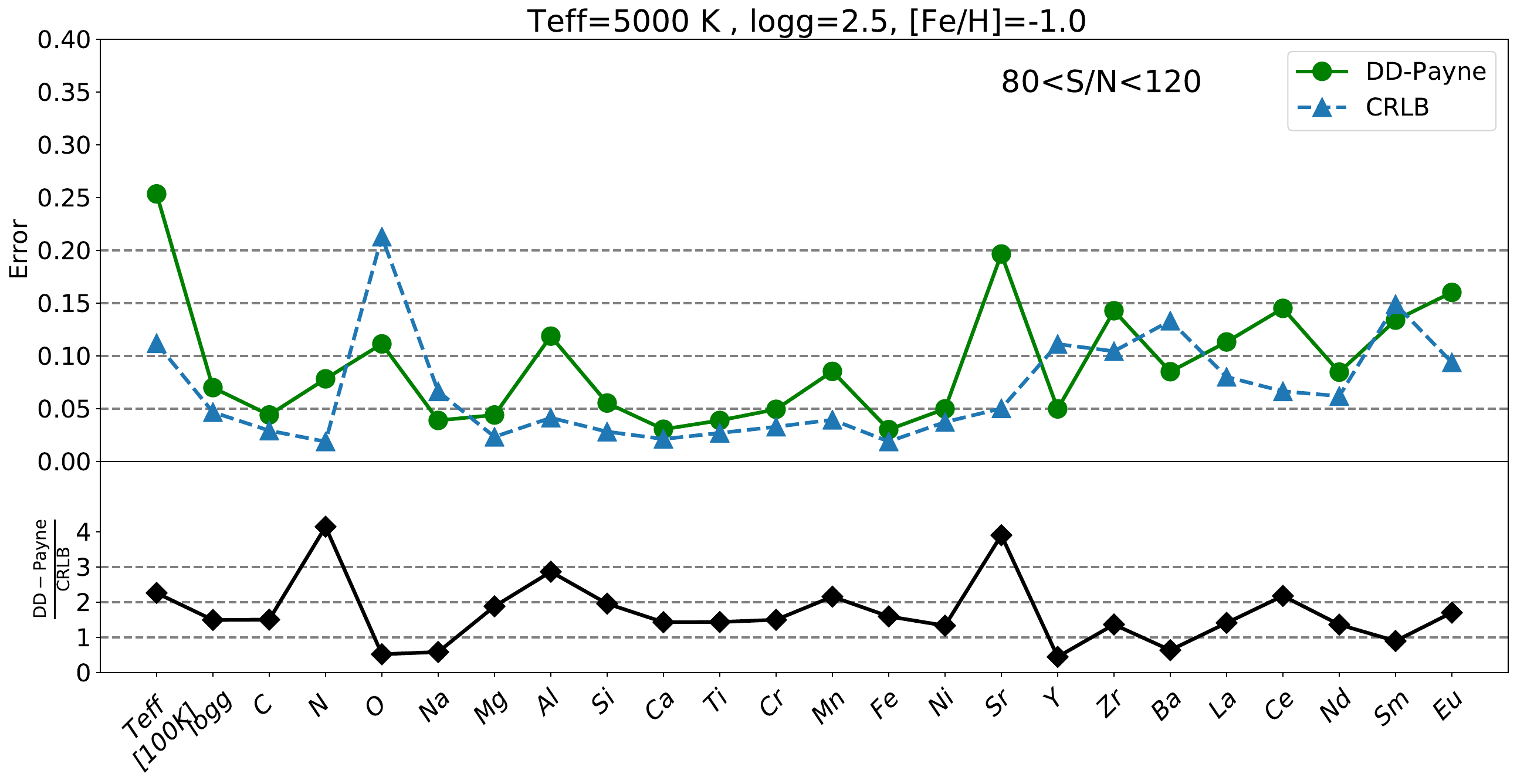}  
	\end{minipage}
}
\caption{Measurement errors of the {\sc DD-Payne} stellar labels compared to theoretical precision limit given by the CRLB, i.e., Cr\'amer-Rao lower bound. The upper panels show the case for a fiducial dwarf with $T_{\rm eff}=5900$~K, $\log~g=4.0$, and [Fe/H]=$-0.5$, while the lower panels shows the case for a giant star with $T_{\rm eff}=5000$~K, $\log~g=2.5$, and [Fe/H]=$-1.0$, at a spectral $S/N$ of 100. For $T_{\rm eff}$, the value in the vertical axis has an unit of 100~K. The bottom part of both cases shows the ratios between the {\sc DD-Payne} measurement errors and the CRLB values.
\label{fig11}}
\end{figure*}

Given the resolution, the wavelength range, and the S/N of a spectrum, a theoretical lower bound of measurement error for the underlying stellar labels can be estimated with the Cram\'er-Rao inequality \citep{Rao1945, Cramer1946}, as has been well demonstrated in \citet{TingYS2017a}. A comparison of the {\sc DD-Payne} measurement errors with this Cram\'er-Rao lower bound (CRLB) provides a validation to the former. 

Given the observation condition, the CRLB of a stellar label is determined by its strength of spectral gradient. As a result, it varies across the stellar parameter space. Generally, the CRLB for dwarf is larger than for giant, given the same temperature and metallicity. For same is temperature and surface gravity, the CRLB is larger for more metal-poor stars. 
Utilizing the Kurucz model spectra, here we calculate the CRLB for two stars, a dwarf with $T_{\rm eff} \simeq 5900$~K, $\log~g \simeq 4.0$, ${\rm [Fe/H]} \simeq -0.5$, and a giant with $T_{\rm eff} \simeq 5000$~K, $\log~g \simeq 2.5$, ${\rm [Fe/H]} \simeq−1.0$. The CRLB of these stars are shown in Fig.\,\ref{fig11}. The top panel shows that given a spectral S/N of 100, many of the elements, including, C, N, Na, Mg, Al, Si, Ca, Ti, Cr, Mn, Fe, Ni, and Sr, have a CRLB between 0.02 and 0.05, for both the dwarf and giant shown. For the heavier elements Y, Zr, Ba, La, Ce, Nd, and Eu, the CRLB values are 0.1--0.2~dex for the dwarf and 0.05--0.15 for the giant. The O and Sm exhibit larger CRLB for dwarfs, as their features in the LAMOST spectra are very weak. The giant has considerable amount of features for Sm, which leads to a CRLB of 0.15~dex. 

The Figure shows that the {\sc DD-Payne} abundance errors are in good agreement with the CRLB, as the ratio between the two are around 1-2 for most of the elements. The Nitrogen shows an exceptionally large ratio for both dwarf and giants. This is similar to the {\sc DD-Payne} results on DESI spectra \citep{ZhangM2024}, but different to the {\sc DD-Payne} estimates for LAMOST DR5 \citep{Sandford2020}, which shows a ratio of $\sim$1. The reason is unclear. We suspect our current estimates of {\sc DD-Payne} Nitrogen abundance may suffer relatively large uncertainty, compared to the LAMOST DR5 results. For giant stars, the Sr and Ce also exhibit relatively large measurement errors compared to the CRLB. This is possibly owing to the imperfect modelling of {\sc DD-Payne} for these elements due to a lack of training set at this metal-poor regime (Sect.\,3).    

The measurement errors in effective temperature and surface gravity are agreed with the CRLB with a factor of 3 and 2, respectively. This is remarkable, as the CRLB is only 8~K for dwarf and 12~K for giant, suggesting a measurement error of 20-30~K in effective temperature. This is consistent with previous work \citep[e.g.][]{XiangMS2019}.   

\section{Comparison with existed LAMOST stellar parameter catalogs}
\subsection{Comparison with LASP stellar parameters}
\begin{figure*}[htb!]
\centering   
\subfigure
{
	\begin{minipage}{0.3\linewidth}
	\centering  
	\includegraphics[width=0.97\columnwidth]{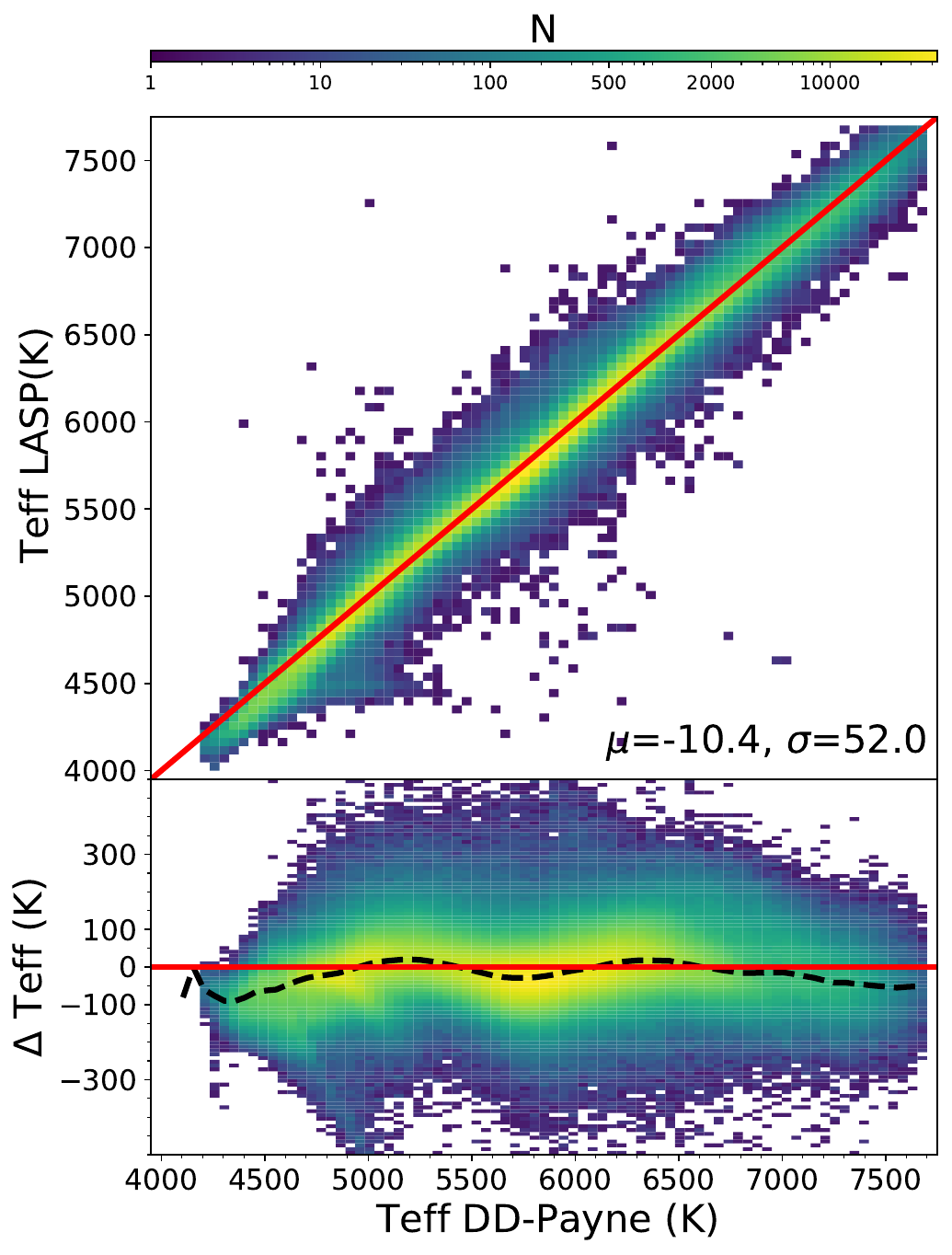} 
	\end{minipage}
} 
\subfigure
{
	\begin{minipage}{0.3\linewidth}
	\centering  
	\includegraphics[width=0.97\columnwidth]{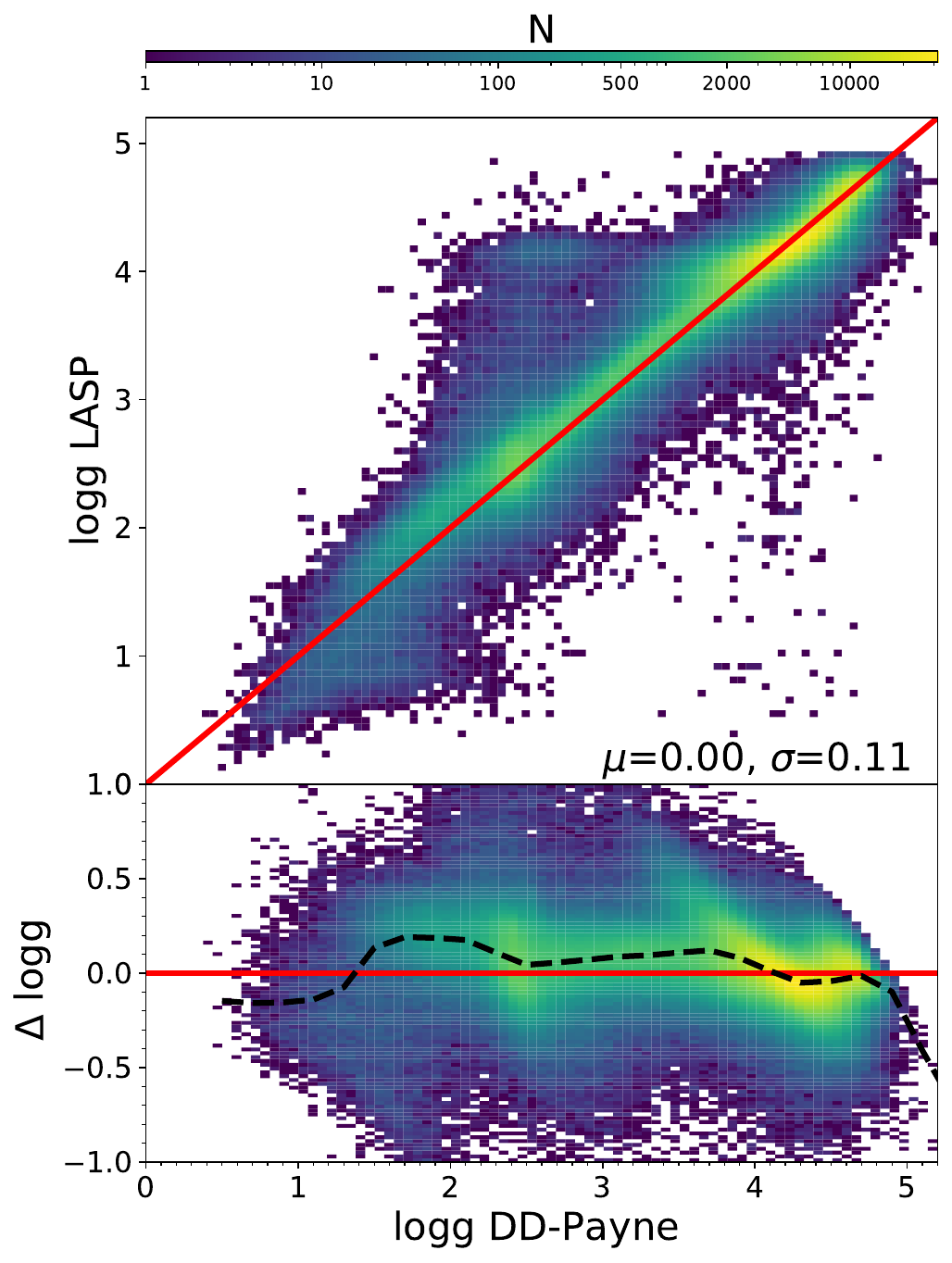} 
	\end{minipage}
} 
\subfigure
{
	\begin{minipage}{0.3\linewidth}
	\centering  
	\includegraphics[width=0.97\columnwidth]{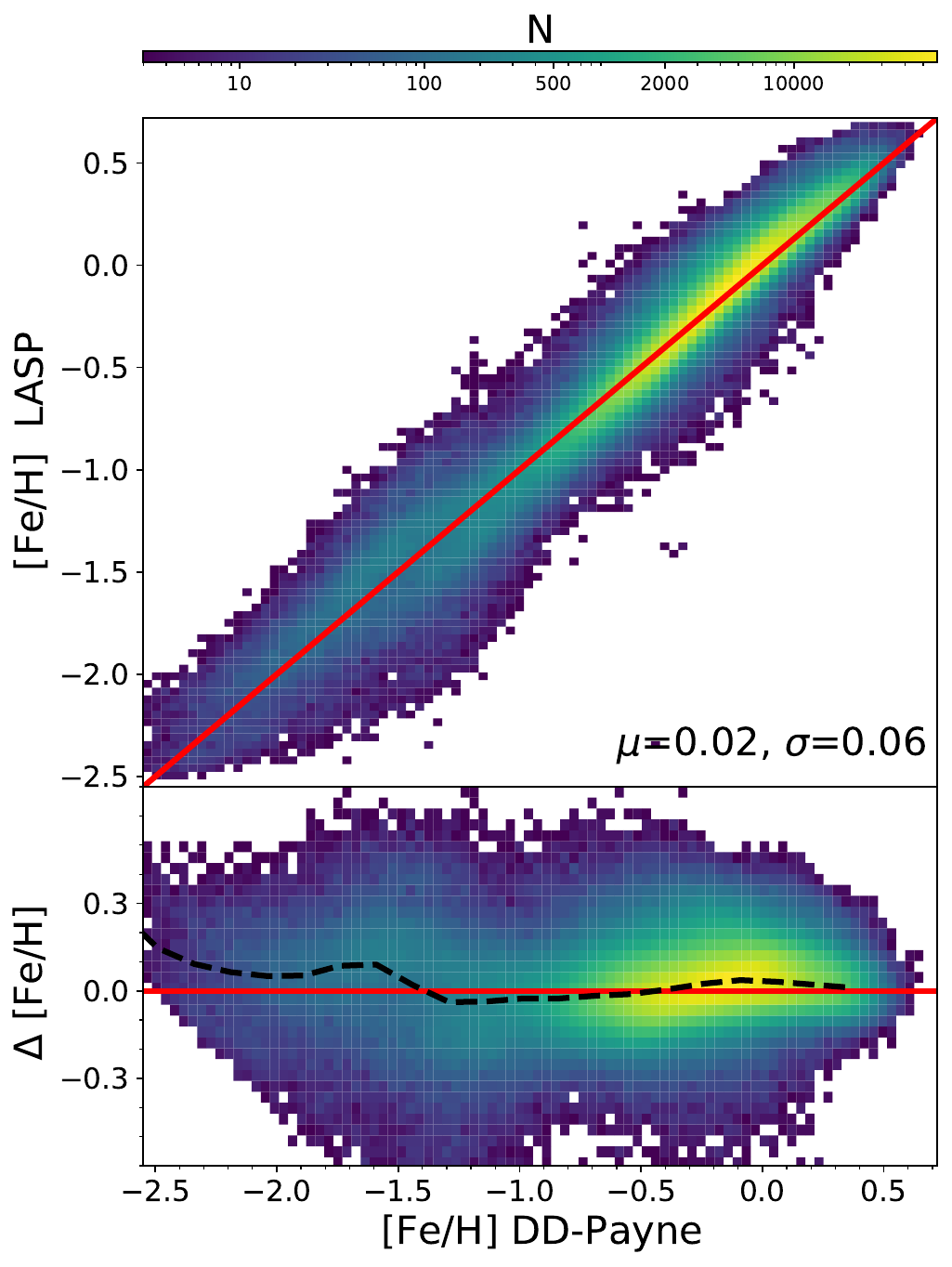} 
	\end{minipage}
} 
\caption{Comparison of basic atmospheric parameters between {\sc DD-Payne} determinations and the LAMOST DR9 official releases, from left to right are $T_{\rm eff}$, $\log~g$, and [Fe/H], respectively. Color represents the number density of stars. Here only stars with spectral $S/N>30$ are shown. Red solid lines in the upper panels show the 1:1 line of X and Y axes. The dotted lines in the bottom panels show the median values of the differences in different bins along X-axis. The mean and dispersion of the differences are marked in the figure.
\label{fig10}}
\end{figure*}

The LAMOST data release also contains basic stellar parameters, $T_{\rm eff}$, $\log~g$, and [Fe/H], derived from the LAMOST stellar parameter pipeline \citep[LASP;][]{WuY2011, LuoAL2015}. Fig.\,\ref{fig10} presents a comparison between our {\sc DD-Payne} estimates for these basic parameters with the LASP results for stars with spectral S/Ns higher than 30. The Figure shows a very good consistency for all these parameters. Remarkably, the effective temperature of LASP is different to our estimates by only 10~K, and the dispersion is only 52~K. The difference is larger for stars at both the cool ($T_{\rm eff}<4700$~K) and the hot ($T_{\rm eff}>7300$~K) end, but the agreement is still within 100~K. 

The surface gravity also shows good agreement, with a mean difference of only 0.004, and a dispersion of 0.11~dex. Nonetheless, for stars with $\log~g<4$, mostly subgiants and giants, the LASP estimates seems to be systematically higher than our estimates, by $\sim$0.1~dex for subgiants, and $\sim$~0.2~dex for giants with $\log~g\simeq2$. The mean difference in [Fe/H] is only 0.02~dex, and the dispersion is 0.06~dex. Remarkably, there is no significant systematic differences in the full metallicity range of $-2.0$ -- 0.5~dex. The LASP [Fe/H] estimates exhibit a lower border at a value of $-2.5$, but failed to provide results for more metal-poor stars.
 
\subsection{Comparison with {\sc DD-Payne} stellar labels on LAMOST DR5} 

\begin{figure*}[htb!]
\centering
\includegraphics[width=1.0\textwidth]{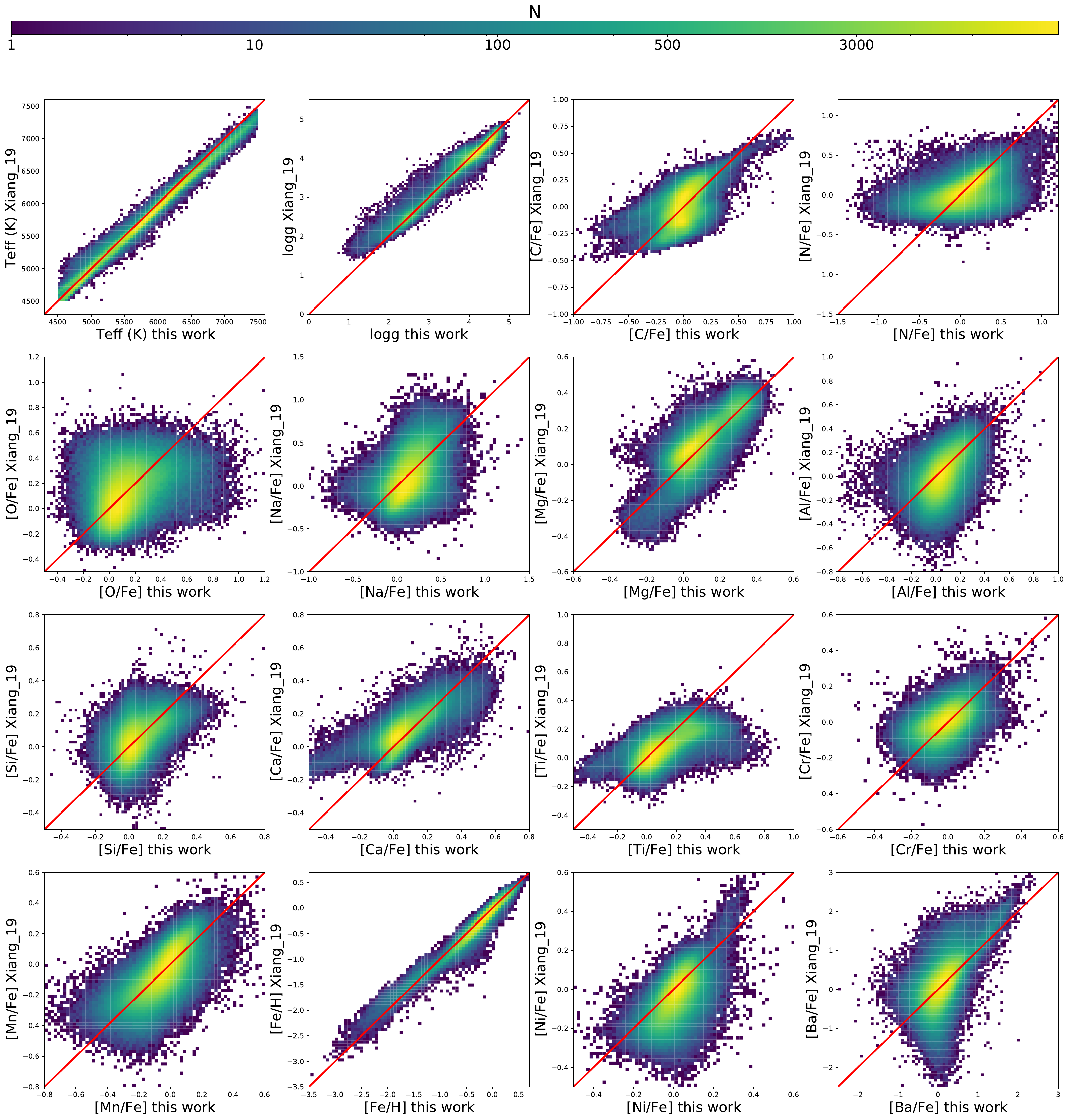}
\caption{Comparison of stellar parameters and abundances between {\sc DD-Payne} determination in this work and the LAMOST DR5 {\sc DD-Payne}  catalog of \citet{XiangMS2019}.
\label{fig_dr5}}
\end{figure*}

We also compare the current label estimates with the LAMOST DR5 abundance catalog of \cite{XiangMS2019}. As shown in Fig.\,\ref{fig_dr5}, good agreements are found to most of the labels for the vast majority of stars. The effective temperature shows some slight difference for the relatively hot ($\teff\gtrsim6500$~K) stars. The DR5 catalog gives $\sim0.3$~dex larger $\log~g$ value for bright giants of $\log~g\simeq1$. The [Fe/H] shows good agreement down to $-3.0$. The agreement of [Mg/Fe] is remarkably good in the full range, from very Mg-poor stars (${\rm [Mg/Fe]}\simeq-0.4$) to Mg-enhanced stars (${\rm [Mg/Fe]}\simeq0.4$). For some of the elements such as C, N, Na, Al, Ca, Ti, and Mn, the current estimates show a population of low-abundance stars, for which the DR5 catalog give higher values. This is possibly an imperfection of the DR5 estimates, due to imperfect training labels adopted by \cite{XiangMS2019}. The agreement of [O/Fe] is less good, reflecting the fact that it is difficult to determine [O/Fe] precisely from the LAMOST low-resolution spectra. Nonetheless, both sets give a [O/Fe] value of $\simeq$0.1 for the majority of stars.      

\subsection{Comparison with MEASNet estimates} 

\begin{figure*}[htb!]
\centering
\includegraphics[width=1.0\textwidth]{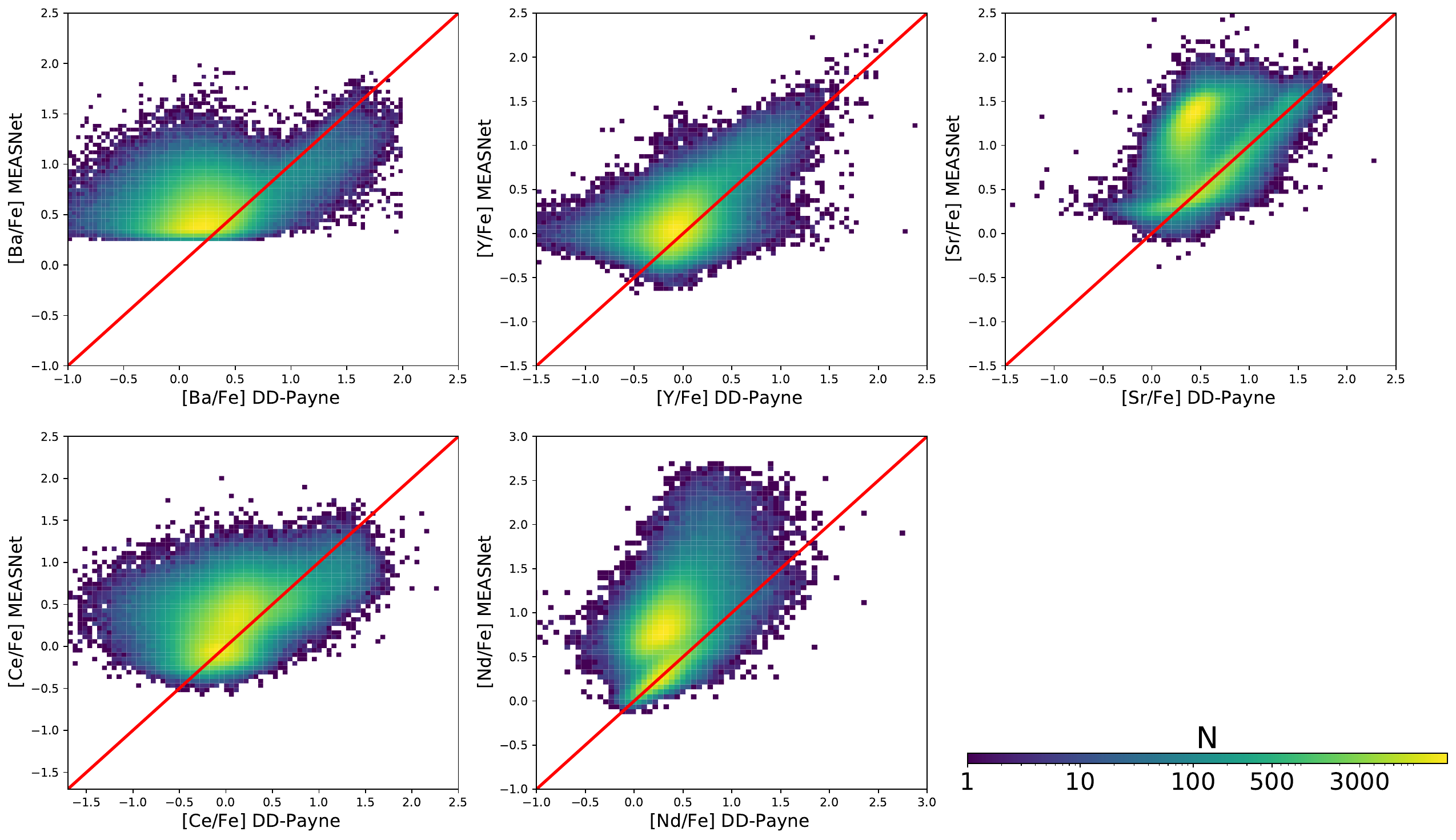}
\caption{Comparison of s-process elemental abundances between {\sc DD-Payne} determination in this work with \citet{Song2024}, who derived abundances for Barium stars with a data-driven approach MEASNet. The hard cut at ${\rm [Ba/Fe]}=0.3$ in the MEASNet estimates corresponds to the definition of Barium stars in \citet{Song2024}. For Sr, Ce, and Nd, most of the stars showing large discrepancy are dwarfs, for which our calibration has lead to a major improvement upon literature values.   
\label{fig_s}}
\end{figure*}

\cite{Song2024} has adopted a data-driven method named the memory-enhanced adaptive spectral network (MEASNet) to search for barium star candidates and estimate their abundances of Sr, Y, Ba, Ce, and Nd from LAMOST DR10 low-resolution spectra. They estimate the [Sr/Fe], [Y/Fe], [Ba/Fe], [Ce/Fe], and [Nd/Fe] based on a training set whose labels are from GALAH DR3. Fig.\,\ref{fig_s} shows the comparison between {\sc DD-Payne} determination and MEASNet estimates for common stars. The figure shows a good agreement in [Ba/Fe] for stars that the {\sc DD-Payne} determination gives a high [Ba/Fe] value (${\rm [Ba/Fe]}\gtrsim1$), while for a large portion of the sample, the {\sc DD-Payne} suggests a normal Ba abundances (${\rm [Ba/Fe]}\lesssim0.5$).

The [Y/Fe] shows a good overall consistency. Nonetheless, the {\sc DD-Payne} results show a tail of Y-poor stars with negative [Y/Fe] values (${\rm [Y/Fe]}\lesssim-0.5$), for which the MEASNet gives ${\rm [Y/Fe]}\simeq0$. These stars can be either intrinsic Y-poor stars or not. As {\sc DD-Payne} determines [Y/Fe] strictly based on the spectral features of Y itself, which are rather weak for Y-poor stars, it is possible that the extremely low-[Y/Fe] values are due to their relatively large measurement errors from {\sc DD-Payne}.

The [Sr/Fe], [Ce/Fe], and [Nd/Fe], however, shows a double-branch feature. The branch with good consistency is mainly composed of giants, while the branch with large deviation is mostly composed of dwarfs. The large deviation for dwarfs is due to a strong temperature-dependent systematics suffered in the MEASNet \citep{Song2024} but has been corrected in the current results.

\section{Absolute zero point of the label estimates}
The relative calibration presented in Sect. 4 lead to a massive data set with remarkable internal precision. Nonetheless, the absolute zero point of the label estimates remains unclear.

For effective temperature, extensive studies have suggested that the IRFM method is a state-of-the-art temperature scale. Typical zero point uncertainty as measured by offset in different implementations of the IRFM relation is within 50~K \citep[e.g.,][]{Blanco2014,Soubiran2024}. For surface gravity, as the comparison with asteroseismic estimates shows a very good agreement (Sect.\,4.2), we believe the zero point offset in $\logg$ estimates is negligible. 

Zero point of the abundances is much harder to be determined directly from our data, as the LAMOST survey is still lack of high-quality data of golden standard stars, such as the Sun and other well-measured stars. Fortunately, as our labels are tied to either GALAH or APOGEE scale via the training sets, they inherit the zero points of these surveys, which have been investigated in literature. 

For [Fe/H], GALAH DR3 suggests a zero point of A(Fe)=7.38, as measured from the solar spectrum of sky flat and from solar twins \citep{Buder2021}.  This is a lower value than the frequently used solar abundances of \citet[A(Fe)=7.45;][]{Grevesse2007}, \citet[A(Fe)=7.50][]{Asplund2009}, and the latest 3D non-LTE value of \citet[A(Fe)=7.46][]{Asplund2021}.  On the other hand, this 1D NLTE GALAH result is quite close to the average 1D non-LTE result presented in \citet{Asplund2021}, namely A(Fe)=7.40.On the other hand, \citep{Buder2021} also illustrated that the GALAH DR3 [Fe/H] estimates agree perfectly with the Gaia Benmark stars \citep[GBS][]{Jofre2014}, which have a zero-point of A(Fe) = 7.45, as of \citet{Grevesse2007}. This is confirmed by our own independently with a similar excercise via comparing the GALAH [Fe/H] with the GBS values. Considering there is many potential sources of uncertainty in the determination of solar abundances from the sky flat, arguably we are more inclined to suggest that the underlying zero-point of GALAH DR3 (thus our estimates) has a zero point of A(Fe) = 7.45. Notably, this is the same zero point as adopted by the APOGEE DR17 \citep{Abdurrouf2022}.

 The abundance estimates of [C/Fe], [N/Fe], [O/Fe], [Mg/Fe], [Al/Fe], [Si/Fe], [Ca/Fe], [Mn/Fe], and [Ni/Fe] should have similar zero point with APOGEE DR17 \citep{Abdurrouf2022}. The [X/Fe] zero point of APOGEE DR17 is determined by requiring the solar-metallicity stars at the solar neighborhood to have ${\rm [X/M]}=0.0$. The abundance estimates of [Na/Fe], [Ti/Fe], [Cr/Fe], [Sr/Fe], [Y/Fe], [Zr/Fe], [Ba/Fe], [La/Fe], [Ce/Fe], [Nd/Fe], [Sm/Fe], and [Eu/Fe] should be sharing the same zero point as GALAH DR3 \citep{Buder2021}. The GALAH DR3 abundance zero points have multiple sources of determination. Some of them are determined from solar spectra, while some of them are determined using stars in the Solar circle or adopted from literature \citep{Buder2021}.


\newpage
\section{Summary}
Stellar abundances are key fossils for unraveling the formation and evolution history of our Galaxy. While deriving stellar abundances from low-resolution spectra is very challenging, the {\sc DD-Payne} has paved a way for rigorously determining many stellar abundances from low-resolution spectra by using a data-driven and model-driven hybrid approach. In this work, we further develop the {\sc DD-Payne} method for stellar label determination by implementing it to higher dimensions of stellar labels. For the first time, we have achieved rigorous determination for a large number of s- and r-process elements, namely Sr, Y, Zr, Ba, La, Ce, Nd, Sm, and Eu, from the LAMOST low-resolution spectra.      

We adopt common stars between LAMOST and APOGEE DR17, as well as common stars between LAMOST and GALAH DR3, as the training sets for {\sc DD-Payne}. These result in the determination for stellar labels in 25 dimensions, including effective temperature, surface gravity, micro-turbulent velocity, and elemental abundances (ratios) for 22 elements (C, N, O, Na, Mg, Al, Si, Ca, Ti, Cr, Mn, Fe, Ni, Y, Sr, Zr, Ba, La, Ce, Nd, Sm, and Eu). We also make use of literature abundance tables from high-resolution spectroscopy, for improving the coverage of parameter space of the training sets at the metal-poor side, down to ${\rm [Fe/H]}\simeq-4$. The training process of the {\sc DD-Payne} is regularized with a large set of synthetic spectral gradients computed with the Kurucz ATLAS 12 model atmosphere. It turns out that the spectra model thus trained can properly reproduce the physical features of all the labels of concern, ensuring a physical rigorous determination of stellar labels with the {\sc DD-Payne}.      

An application of the {\sc DD-Payne} to the LAMOST DR9 spectra data set results in a massive stellar label catalog for 6 million stars. Both internal and external calibration and validation for these label estimates are implemented. The $T_{\rm eff}$ are calibrated to IRFM scale, while the $\log~g$ are validated with asteroseismic measurements. The [Fe/H] scale is tied to high-resolution spectroscopic measurements with NLTE correction. The elemental abundances are internally calibrated using wide binaries, eliminating systematic trend with effective temperature. The distribution of the sample stars in the [X/Fe]--[Fe/H] planes exhibit similar trends to high-resolution spectroscopy. 

A comparison of errors in the label estimates with the theoretical precision limits shows good consistency, as the error estimates for most of the labels agree with the CRLB within a factor of 2. Given a spectrum with $S/N$ higher than 50, typical measurement errors in the abundance estimates are about 0.05~dex for most of the elements with atomic number smaller than Sr, while the measurement errors for heavier, s- and r-process elements are typically 0.1--0.2~dex. 

By examining the abundance dispersion between field stars and star clusters, we have shown that the current abundance estimates are informative to tell the intrinsic abundance scatter, shedding light on unraveling the history of our Galaxy in a high-dimensional space with an unprecedented large data set. 

In this work, we are dedicated to measuring stellar abundances of stars from a large survey in an uniform way. As expected, there must be outliers and poor estimates due to complicate reasons, e.g. poor data quality, exotic parameter space, etc. For the sake of completeness and statistics, we prefer not to induce complex selection bias in the catalog. In this meaning, we urge users to be very careful to further choose their own sample from the catalog based on their own interests. 

\newpage
\noindent {\bf Acknowledgments}
We thank the referee for the suggestions that have improved the clarity of the manuscript.
This work has made use of data products from the Guo Shou Jing Telescope (the Large Sky Area Multi-Object Fibre Spectroscopic Telescope, LAMOST).
LAMOST is a National Major Scientific Project built by the Chinese Academy of Sciences.
Funding for the project has been provided by the National Development and Reform Commission. LAMOST is operated and managed by the National Astronomical Observatories, Chinese Academy of Sciences.

We acknowledges financial support from the National Key R\&D Program of China Grant (No.2022YFF0504200), the National Natural Science Foundation of China (NSFC; Grant No.2022000083, No.12303025, No.12588202), and the science research grants from the China Manned Space Project with No. CMS-CSST-2021-A08. J.-R.S. acknowledges support from NSFC No. 12090044.H.-L. Yan acknowledges support from the Youth Innovation Promotion Association of the Chinese Academy of Sciences No.12373036.

\newpage
\bibliographystyle{aasjournal}
\bibliography{lamost.bib}
\end{document}